\documentclass[reprint,prl,twocolumn,longbibliography,superscriptaddress]{revtex4-2}

\usepackage{graphicx}% Include figure files
\usepackage{dcolumn}% Align table columns on decimal point
\usepackage{bm}% bold math
\usepackage[mathlines]{lineno}% Enable numbering of text anddisplay math
\usepackage{siunitx}
\usepackage{amsmath, amsfonts, amssymb}
\usepackage{graphicx,epsfig}
\usepackage{longtable}
\graphicspath{{./Figures/}}

\begin{document}

\title{Visualization of Oscillatory Electron Dynamics on the Surface of Liquid Helium
}

\author{Hala Siddiq}
\affiliation{Department of Physics, Lancaster University, Lancaster LA1 4YB, United Kingdom}
\affiliation{Jazan University, Faculty of Science, Department of Physics, Jazan, Saudi Arabia}
\author{Kostyantyn Nasyedkin}
\affiliation{Quantum Condensed Phases Research Team, RIKEN CEMS, Wako, Saitama 351 0198, Japan}
\affiliation{Neutron Scattering Division, Oak Ridge National Laboratory, Oak Ridge, TN 37831, USA}
\author{Kimitoshi Kono}
\affiliation{Quantum Condensed Phases Research Team, RIKEN CEMS, Wako, Saitama 351 0198, Japan}
\affiliation{International College of Semiconductor Technology, National Yang Ming Chiao Tung University, Hsinchu 300, Taiwan}
\author{Dmitry Zmeev}%
\affiliation{Department of Physics, Lancaster University, Lancaster LA1 4YB, United Kingdom}
\author{Peter V. E. McClintock}%
\affiliation{Department of Physics, Lancaster University, Lancaster LA1 4YB, United Kingdom}
\author{Yuri A. Pashkin}
\affiliation{Department of Physics, Lancaster University, Lancaster LA1 4YB, United Kingdom}
\author{Aneta Stefanovska }
\affiliation{Department of Physics, Lancaster University, Lancaster LA1 4YB, United Kingdom}

\date{\today}% It is always \today, today,
             %  but any date may be explicitly specified

\begin{abstract}

We have measured signals induced in 5 Corbino electrodes by spontaneous oscillations of 2D surface electrons on liquid helium at $\sim$0.3\,K, with a perpendicular magnetic field and microwave radiation. Analysis using multi-scale, time-resolved, methods yields results consistent with magnetoplasmons modulated by slow surface gravity waves, with the latter requiring consideration of the 3rd dimension. Calculation of phase differences and phase coherences between signals from differently-positioned pairs of electrodes enables reconstruction of the electron dynamics on the helium surface.

\end{abstract}

\maketitle

The two-dimensional electron system (2DES) formed by electrons above the surface of liquid helium facilitates the exploration of 2D non-equilibrium phenomena in an almost perfectly clean environment \cite{Monarkha:04}, complementary to the 2DES in GaAs heterostructures \cite{Zudov:01}. Following the prediction \cite{Cole:69} and observation \cite{Williams:71} of electrons on helium, research highlights have included magnetoplasmons \cite{Grimes:76}, Wigner crystallization \cite{Grimes:79,Fisher:79}, the ripplonic Lamb shift \cite{Dykman:17}, coupling of Rydberg states to Landau levels \cite{Yunusova:19}, quantum information processing \cite{Platzman:99,Schuster:10,Koolstra:19}, incompressible electronic behaviour \cite{Konstantinov:12,Chepelianskii:15,Monarkha:16}, zero-resistance states \cite{Konstantinov:10,Monarkha:11} , a plethora of important results on many-electron phenomena and non-equilibrium physics \cite{Konstantinov:09,Konstantinov:09b,Ikegami:12,Rees:16,Monarkha:11}, and the highest known 2D electronic mobilities \cite{Shirahama:95}. The recent studies \cite{Konstantinov:12,Chepelianskii:15} used a strong magnetic field causing vanishing diagonal conductivity, and resonant microwave radiation (MW). For a circular pool of surface electrons, currents induced at the center of a Corbino electrode geometry above the liquid exhibited nonlinear oscillations in the audiofrequency range. These were attributed to edge magnetoplasmons \cite{Konstantinov:13, Arai:12,Monarkha:16,Zadorozhko:18}, but there was also evidence \cite{Clemson:14,Monarkha:19} of an even lower-frequency modulation.

In this Letter, we study the electric currents induced in five electrodes above the liquid by the motion of electrons on the surface, under a strong magnetic field and microwave irradiation. Time-resolved multi-scale analysis methods are used to reconstruct the electrons' underlying oscillatory dynamics from the recorded signals.

\begin{figure}[t!]
\centering
\includegraphics[width=0.75\linewidth]{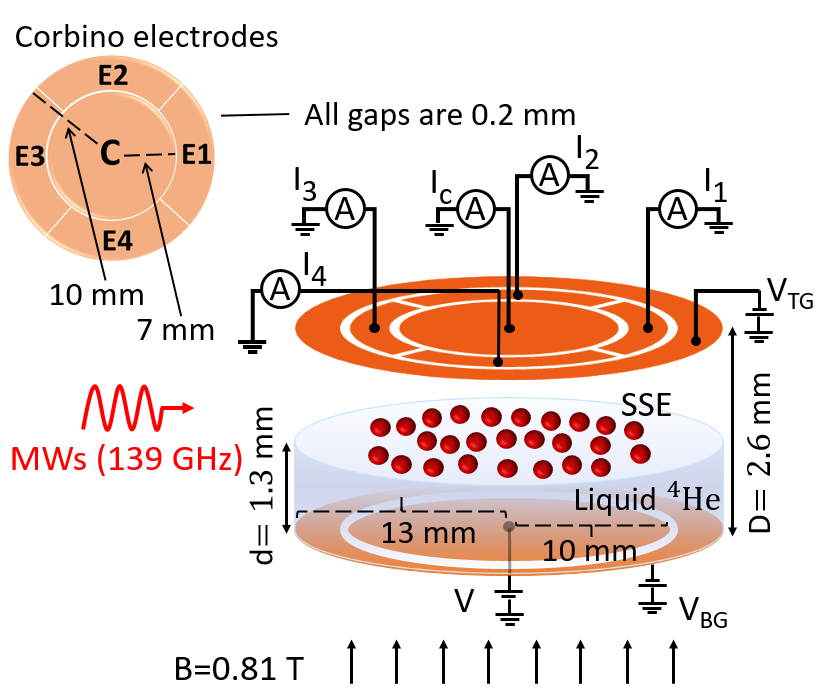}
\caption{\label{fig:E} Schematic (not to scale) of the experimental arrangement used for investigating the electron gas on the surface of superfluid helium. For detailed description, see text.}
\end{figure}

\begin{figure*}[t!]
\includegraphics[width=0.85\linewidth]{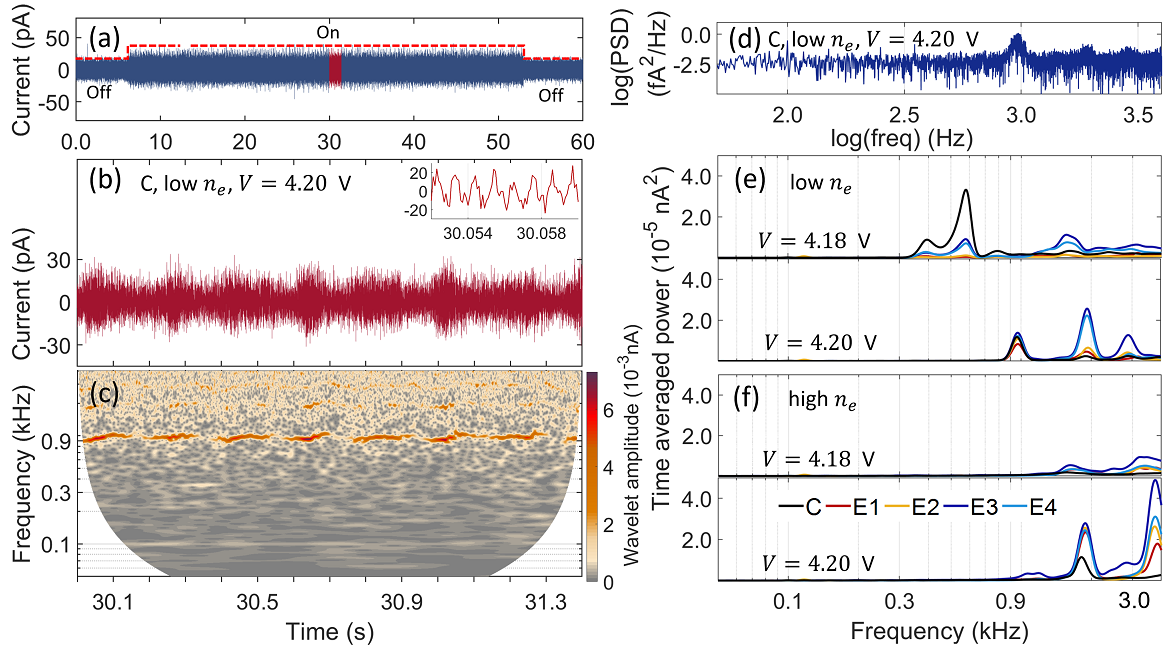}
\caption{\label{fig:current}(a) Signal on the center electrode C recorded for 60 seconds at a pressing voltage of $V\,=\,4.20$~V and electron density $n_e$\,=\,1.4$\times {10^{6}}$~\si{cm^{-2}}. The MW (139~GHz) is switched On/Off where indicated. (b) Enlarged reddened 1.4~s part of the signal in (a) used to obtain the characteristics of the oscillations, where the inset shows a further zoom to illustrate the pattern that emerges. Our conclusions are unaffected by the time interval chosen for this analysis. (c) Wavelet transform with frequency resolution 3~Hz. (d) Power spectral density of the oscillations in (b) showing the fundamental frequency and higher harmonics; its average gradient is near zero (see SM). (e), (f)  Time-averaged wavelet power of signals with MW On for each of the electrodes, as indicated: (e) is for the lower $n_e$\,=\,1.4$\times {10^{6}}$~\si{cm^{-2}} and two different pressing voltages; (f) is for the higher $n_e$\,=\,2.2$\times {10^{6}}$~\si{cm^{-2}} and the same two pressing voltages. The $y$-axis of (c) and the $x$-axes of (d-f) are logarithmic.}
\end{figure*}

Fig.~\ref{fig:E} illustrates the experimental arrangement. The copper cell is attached to the mixing chamber of a dilution refrigerator at $T$\,$\sim$\,0.3~K. The system of electrodes represents a parallel\,-\,plate capacitor with two horizontal circular plates of 20 mm diameter, separated by $D\,=\,2.6$~mm. The liquid ${}^{4}$He surface is set midway between the plates, so that the helium depth $d\,=\,1.3$~mm. The top plate consists of two concentric Corbino electrodes: a central disk C and an outer ring divided into four equal segments E1, E2, E3 and E4. The bottom disk-shaped electrode is positively biased to press the electrons against the helium surface. The top and bottom plates are each surrounded by a negatively-biased guard-ring ($ V_{\si {TG}}$\,=\,$V_{\si{BG}}$\,=\,$-$\,0.5~V) to confine the electrons within their pool. The helium level is controlled within $\approx$\,\SI{50} {\micro\metre} by monitoring the capacitance between the top and bottom plates. A pulse of electrons, thermionically emitted from a tungsten filament, accumulates above the liquid surface. A positive potential $V$ on the bottom electrode during charging controls the electron density $n_{e}$. The areal electron density $n_{e}$\,=\,$\epsilon_{\rm He} \epsilon_{0}{V}/ed$, where $\epsilon_{He}$ and $\epsilon_{0}$ are the helium dielectric constant and the vacuum permittivity, respectively, and $e$ is the electronic charge. Resonant microwaves propagating parallel to the helium surface with frequency  $\omega_{12}/2\pi$\,=\,139~GHz, and power 0~dBm, excite an intersubband transition between the ground and the first excited electronic surface states. The intersubband transition is tuned to resonance by adjustment of the holding electric field. A fixed magnetic field $B$\,=\,0.81~T perpendicular to the liquid surface leads to the zero resistance state \cite{Konstantinov:10}. The microwaves are activated for $\approx$ 47 seconds. Current signals $I_{c}, I_{1}, I_{2}, I_{3}$ and $I_{4}$ are measured simultaneously from C, E1, E2, E3 and E4, respectively. They are converted to voltage signals by current preamplifiers with a gain $10^{-9}$~A/V and a frequency bandwidth 10~kHz. The pressing voltage is set to a value $4.16<V<4.22$~V with an electron density of either $n_e=1.4\times10^{6}$ or $n_e=2.2\times10^{6}$~\si{cm^{-2}}. The signals are sampled for 60~s with a sampling frequency of 100~kHz. The microwaves are switched on 6~s after the start of recording and switched off 7~s before the end. 

\begin{figure*}[t!]
\includegraphics[width=0.85\linewidth]{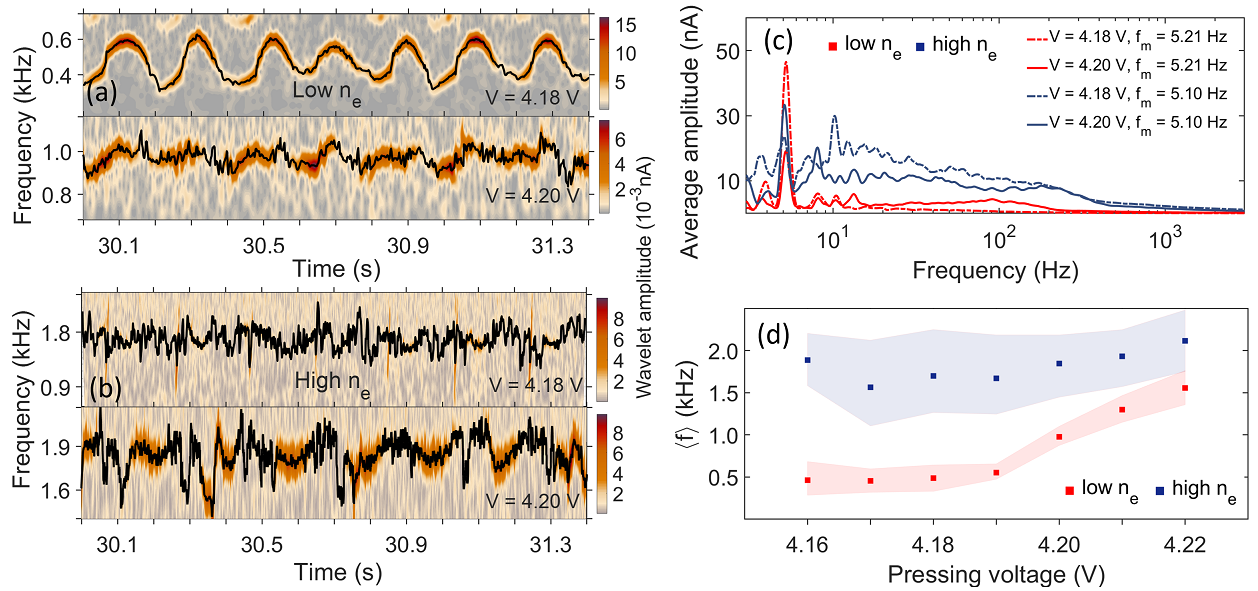}
\caption{\label{fig:Ex1}(a) Ridge extractions (black lines) for the signal on C, obtained from the time-frequency representation \cite{Iatsenko:16}, showing the instantaneous frequency variations of the main component, for the lower electron density $n_e\,=\,1.4 \times10^6$~\si{cm^{-2}} and two different pressing voltages $V$. Colour bars show the intensity of the oscillations. (b) As in (a) but for higher electron density $n_e\,=\,2.2\times10^6 $~\si{cm^{-2}}. (c) Average amplitudes of the ridge extractions in (a) and (b). The frequencies of the sharp peaks near 5~Hz are indicated.  (d) Mean oscillation frequency on C as a function of $V$ for the lower (red squares) and higher (blue squares) electron densities $n_e$, respectively. The shadows show the full frequency range of the oscillations.
}
\end{figure*}

The main results and analyses are summarised in Figs.\ \ref{fig:current}-\ref{fig:coh}; further detail is provided in the Supplemental Material (SM). Fig.\ \ref{fig:current}(a) shows a typical signal, with (b) an enlargement of the part that we analyse. Signals from all electrodes increase in amplitude under MW irradiation. For low $n_e$ and small $V$, the greatest increases are at electrodes C and E3 (closest to the MW input).

Signals were pre-processed by subtraction of a best-fit cubic polynomial \cite{Iatsenko:15b} to remove non-oscillatory trends. Wavelet analysis \cite{Kaiser:94,Iatsenko:15b,Clemson:16} using the complex Morlet wavelet was then applied to obtain time-frequency representations of the wavelet amplitude and phase, with logarithmic frequency resolution. The spectral properties of the time series $x(t)$ are described by the complex spectral function $W(\omega_k,t_n) = W_{k,n} = a_{k,n} + ib_{k,n}$ where subscripts $k$ and $n$ denote the data discreteness. Time is discretised by $t_{n}$, so the amplitude $|W_{k,n}|$ = $\sqrt{(a^{2}_{k,n} + b^{2}_{k,n})}$, and the phase $\theta_{k,n}$ = $\arctan{(b_{k,n}/a_{k,n})}$ for each time $t_{n}$ and frequency $\omega_{k}$. Thus, for two signals $x_{1}(t)$ and $x_{2}(t)$, the relative phase difference $\Delta \theta_{k,n}$ = $\theta_{2 k,n}$ - $\theta_{1 k,n}$.

Fig.\ \ref{fig:current}(c) shows the wavelet amplitude as a function of frequency and time, cf.\ the traditional calculation of power spectral density in (d), with linear frequency resolution,  averaged over time and thus losing the information in the frequency variations. Time-averaged wavelet powers for signals recorded on each of the electrodes at $V\,=\,4.18$ and $V\,=\,4.20$~V are shown in (e) for lower $n_e$ and in (f) for higher $n_e$.

It is evident from Fig.\ \ref{fig:current}(c) that the basic frequency is varying. Similar behaviour was observed earlier using a single electrode \cite{Konstantinov:13,Clemson:14, Monarkha:19}. The {\em instantaneous frequency} of each mode can be determined by ridge extraction \cite{Iatsenko:16}, which traces in time the loci of the highest amplitude/power. Typical results are shown in Figure.~\ref{fig:Ex1}(a) and (b).

The higher frequencies in the wavelet transform are attributable to harmonics \cite{Sheppard:11} (see SM). The more pronounced peak in the time-averaged spectral power for C at low $n_e$ and $V\,=\,4.18$~V in Fig.\ \ref{fig:current}(e) implies electron motion mostly involving C. For given values of $V$ and $n_e$, the frequency is found to be the same on all electrodes. At low electron density, the power is unevenly distributed among the electrodes. When $V\,=\,4.16\,-\,4.20$~V, most of the power is associated with the central electrode, but it becomes more uniformly distributed among the 4 outer electrodes for $V\,=\,4.21$ and 4.22~V. For the higher $n_e$, Fig.\ \ref{fig:current}(f), the spectral power is more uniformly distributed between the edge electrodes but it is consistently lower for the central electrode. This suggests that the highest frequencies (see Fig.\ \ref{fig:Ex1}(d) correspond to electrons moving predominantly below the edge electrodes, while the lower frequencies correspond to their moving mainly below the central electrode.

\begin{figure*}[t!]
\includegraphics[width=0.85\linewidth]{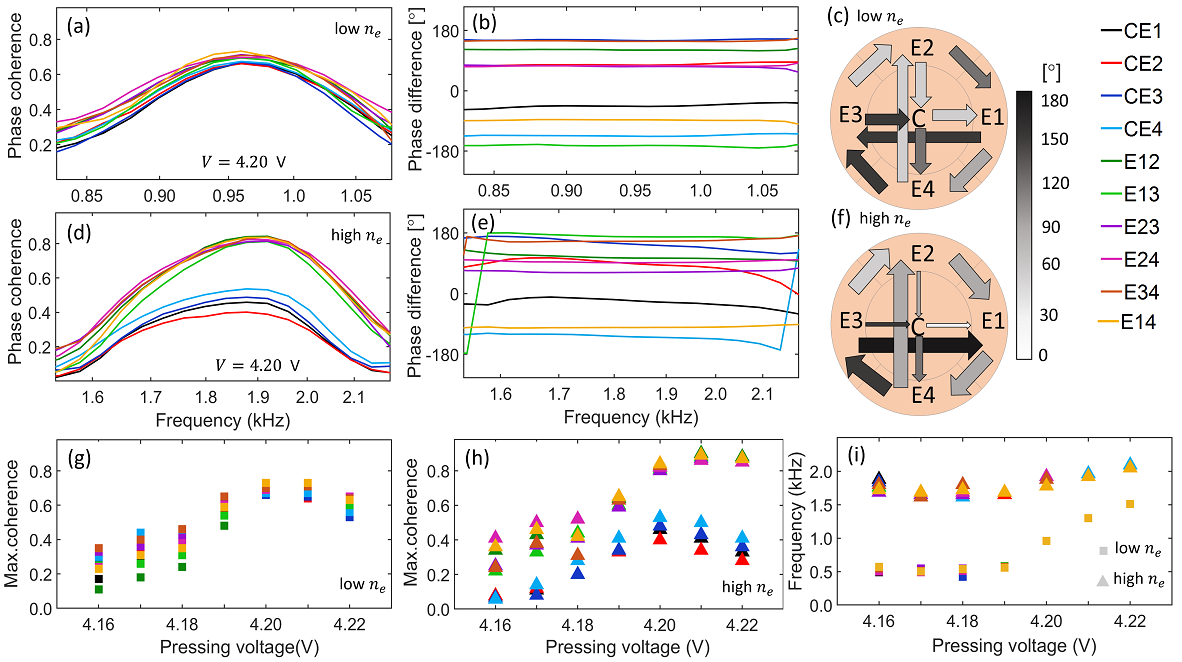}% Here is how to import EPS art
\caption{\label{fig:coh}
(a-b) Wavelet phase coherence and phase difference of pairs of electrodes for the lower electron density of $n_e\,=\,1.4 \times 10^6$~cm$^{-2}$ at a pressing voltage of $V\,=\,4.20$~V for a fixed depth of liquid helium. High phase coherence is observed in the range of 0.85\,-\,1.05~kHz with a constant phase difference.
(c) The circular schematic summarises the coherences and phases shifts between the currents. The thickness of the arrows indicates the magnitude of the coherence and the white/black shading the size of the phase shift.
(d-f) Same as (a-c) but with the higher $n_e\,=\,2.2 \times 10^{6}$~\si{cm^{-2}}. The lines are color-coded to indicate particular electrode pairs.
(g-h) Maximum coherence values for all pairs of electrodes obtained at different $V$ for both electron densities, showing that the resonance condition, i.e.\ similar coherence between all electrode pairs, is satisfied only at the lower $n_e$. (i) The change in frequency of the maximum coherence at each $n_e$ value as functions of $V$ for all electrode pairs.}
\end{figure*}

The low frequency modulation $\approx$\,5~Hz revealed by the ridge extractions in Fig.\ \ref{fig:Ex1}(a) is almost independent of $V$ and $n_e$. It may arise from the interaction of surface electrons with gravity waves, also known as ripplons \cite{Konstantinov:13,Monarkha:19}: for radially symmetric modes, the lowest resonant frequency is $f$\,$\approx 0.609 \sqrt gd/R$, where $g$ is the acceleration due to gravity, and $d$ and $R$ are respectively the height and radius of the circular liquid pool \cite{LandauFluid:87}. This yields 5.28~Hz, comparable with the observations.

A possible explanation of the kHz oscillations is that they arise from edge-magnetoplasmons (EMP) for which Volkov and Mikhailov \cite{volkov:91} calculate a frequency of 
\begin{equation}
\omega_{EMP}= 2\pi f_{EMP} =-\frac{q\sigma_{xy}}{2\pi \epsilon \epsilon_{0}}{} \Bigl(\ln{|q|w} +C_{f}\Bigr),
\label{ref:IEMP}
\end{equation}
where $\sigma_{yx}$ $\propto$  $n/{B}$, the constant $C_{f}$ depends on details of the density profile, $w$ is the width of the transition layer, and $q$ is the wavevector component along the boundary. This gives frequencies of $\approx$ 2.28 and 3.58~kHz for low and high electron densities, respectively, i.e.\ close to the experimental values. Figure.~\ref{fig:Ex1}(d) shows how the mean oscillation frequency for C changes with $V$ for different $n_e$. We note that the mean frequency starts to increase at $V \sim 4.19$~V when the electron density is low with only a minimal difference between the maxima and minima of the oscillations (red shadow), perhaps due to an accumulation of electrons under the edge electrodes \cite {Monarkha:19} (SM). For larger $n_e$, the mean frequency is higher and almost independent of $V$. %This suggests that the interaction of the electrons with the helium surface rises with decreasing $n_e$.

The wavelet phases can be used to calculate the phase difference between the signals on any chosen pair of electrodes. By averaging the sine and cosine components of the phase differences over time, we find the phase coherence function
\begin{equation}
C_ \theta(\omega_k)=\sqrt{\langle \cos\Delta \theta_k{}_n \rangle^2 + \langle \sin\Delta \theta_k{}_n\rangle^2},
\label{ref:eqn7}
\end{equation}
where the coefficients $\cos \Delta \theta_{k,n}$ and  $\sin \Delta \theta_{k,n}$ are given in \cite{Bandrivskyy:04a}. Its value ranges between 0 and 1. When $C_ \theta(\omega_k) \simeq 0$, the phase difference changes continuously so that there in no phase coherence, whilst $C_ \theta(\omega_k) \simeq $ 1 means that the phase difference remains constant corresponding to perfect phase coherence. The phase difference varies within  $\pm \pi$ and shows how one signal lags or leads another. We used the method of surrogates \cite{Schreiber:96,Lancaster:18a} to test the statistical significance of the computed coherence, employing 100 iterative amplitude-adjusted Fourier transform (IAAFT) surrogates. The statistically significant coherence is calculated by subtracting the 95th percentile of the surrogate numbers.

We have studied the mutual relationships of the current oscillations in the five electrodes by calculation of the wavelet phase coherence and phase differences between different electrode pairs. Results are shown for low and high $n_e$ in Figs.~\ref{fig:coh}(a-c) and (d-f), respectively, for $V=4.20$~V; for results with other values of $V$, see SM. At low $n_e$, significant coherence was found for all electrode permutations for $V=4.20$~V (Fig.~\ref{fig:coh}(a)). For high $n_e$ at $V=4.20$~V (Fig.~\ref{fig:coh}(d)), there is lower coherence between the central electrode and each of the edge electrodes, but high coherence when the edge electrodes are paired. From Fig.~\ref{fig:coh}(b) and (e), the phase difference is constant for both low and high electron density around the frequency where maximal phase coherence occurs. For high $n_e$ Fig.~\ref{fig:coh}(e) shows that some of the currents are almost in antiphase. 

The most plausible explanation of the phase differences between the signals from differently-positioned electrodes is that there is a complex clockwise circumferential electron flow, and there are also flows towards and away from C: the circular schematics (c) and (f) illustrate electron flow at the same $V=4.20$~V but at different $n_e$. It is clear that changing $n_e$ changes the flow pattern. The flow direction is independent of $V$ at the higher $n_e$, but dependent on it at lower values (see SM). The complexity of the flow may in part be attributable to interaction between the electrons and the gravity waves. Fig.~\ref{fig:coh}(g,h) show that, for all paired electrodes, the maximum phase coherence changes with $V$ for both $n_e$: (g) shows that, for low $n_e$, the resonance condition is satisfied at $V=4.20$~V, implying that the coherence is uniform over the electrodes. Fig.~\ref{fig:coh}(i) shows that the frequency of maximum coherence changes with $V$ at both electron densities, similarly to the behavior of the mean frequency of oscillation on each of the electrodes Fig.~\ref{fig:Ex1}(d).

In conclusion, we have confirmed that the main kHz signals are generated by inter-edge magnetoplasmons  \cite{volkov:91} modulated by interaction with surface ripplons at about 5 Hz \cite{Konstantinov:13,Monarkha:19}.  Note that, in attributing the $\sim 5$~Hz modulation to the effect of ripplons, we are effectively taking account of vertical motion, i.e. of the third dimension. Our use of segmented electrodes, coupled with time-frequency analysis, has illuminated the dynamics of the surface electrons. If the power spectral densities alone were calculated, the signals would look mostly like white noise. Adding the time dimension enables the varying frequency to be followed and brings new insight. We find that increasing $V$ can change the pattern of phase differences and coherence between the signals induced at differently positioned electrodes. Treating the time-resolved dynamics with logarithmic frequency resolution, opens up new possibilities for understanding these paradigmatic far-from-equilibrium phenomena, bringing together the quantum and classical processes involved. 

We acknowledge valuable discussions with Julian Newman, Paul Wileman, Sasha Proctor, Lawrence Sheppard, Yunus Abdulhameed, Mark Dykman and Yuriy Monarkha. The experimental data were obtained by the Quantum Condensed Phases Research Team, RIKEN CEMS, Japan. HS is supported by Jazan University, Saudi Arabia. KK is supported by the Ministry of Science and Technology, Taiwan, ROC, under Grant No. MOST 109-2112-M-009-021, and JSPS KAKENHI Grant Number JP17H01145. PVEMcC is supported by the Engineering and Physical Sciences Research Council (UK) under Grant No.\ EP/P022197/1.

\bibliography{ElectronHelium}

%\nocite{*}

\end{document}

\title{Visualization of Oscillatory Electron Dynamics on the Surface of Liquid Helium
SUPPLEMENTAL MATERIAL}

\author{Hala Siddiq}
\affiliation{Department of Physics, Lancaster University, Lancaster LA1 4YB, United Kingdom}
\affiliation{Jazan University, Faculty of Science, Department of Physics, Jazan, Saudi Arabia}
\author{Kostyantyn Nasyedkin}
\affiliation{Quantum Condensed Phases Research Team, RIKEN CEMS, Wako, Saitama 351 0198, Japan}
\affiliation{Neutron Scattering Division, Oak Ridge National Laboratory, Oak Ridge, TN 37831, USA}
\author{Kimitoshi Kono}
\affiliation{Quantum Condensed Phases Research Team, RIKEN CEMS, Wako, Saitama 351 0198, Japan}
\affiliation{International College of Semiconductor Technology, National Yang Ming Chiao Tung University, Hsinchu 300, Taiwan}
\author{Dmitry Zmeev}%
\affiliation{Department of Physics, Lancaster University, Lancaster LA1 4YB, United Kingdom}
\author{Peter V. E. McClintock}%
\affiliation{Department of Physics, Lancaster University, Lancaster LA1 4YB, United Kingdom}
\author{Yuri A. Pashkin}
\affiliation{Department of Physics, Lancaster University, Lancaster LA1 4YB, United Kingdom}
\author{Aneta Stefanovska }
\affiliation{Department of Physics, Lancaster University, Lancaster LA1 4YB, United Kingdom}
\date{\today}% It is always \today, today,
             %  but any date may be explicitly specified

\maketitle

\section{introduction}
The experimental system consists of a layer of liquid helium with a 2D sheet of electrons above its surface. The surface electrons (SEs) and liquid helium are sandwiched between electrodes forming a parallel-plate capacitor (see Fig 1 of main paper). An electric field $E_{\perp}$ is created perpendicular to the surface by applying a positive ``pressing voltage'' to the bottom electrode. The SE occupy quantized energy levels called subbands~\cite{monarkha2019magneto}, which are determined by competing forces in the system: an electric field force from $E_\perp$, an attractive image force from within the body of liquid helium and a repulsive surface barrier caused by the helium atoms~\cite{monarkha2013two}.
%The surface state of electrons on liquid helium subbands with energies $\epsilon _{n}$, where $n$\,=\,1, 2, …, create due to the electric field $E_\perp$, attractive image force inside the helium and the replusive surface barrier from atoms of helium.
%This means the motion of electrons is quantised in the direction perpendicular to the surface but free in the directions adjacent to the surface. %We effectively have a 2D electron gas levitating above a very clean surface of liquid helium.
In the experiment, transitions between the first and second subbands are excited by resonant microwave radiation with angular frequency $\omega$ such that  $\hbar\omega \approx \epsilon_2-\epsilon_1$ by tuning $\omega$ with $E_{\perp}$through the linear Stark shift. A strong magnetic field $B$ is applied perpendicular to the motion of electrons to fulfill the relation $ \omega/\omega_c \,=\,l+1/4$ in order to achieve the zero resistance regime.

%In order to understand the physics of a system of electrons moving on the surface of liquid helium and its interaction with microwaves (MWs) under a DC magnetic field, with changing the electron densities and pressing voltage.

\begin{figure*}[h!]
\includegraphics[width=5cm]{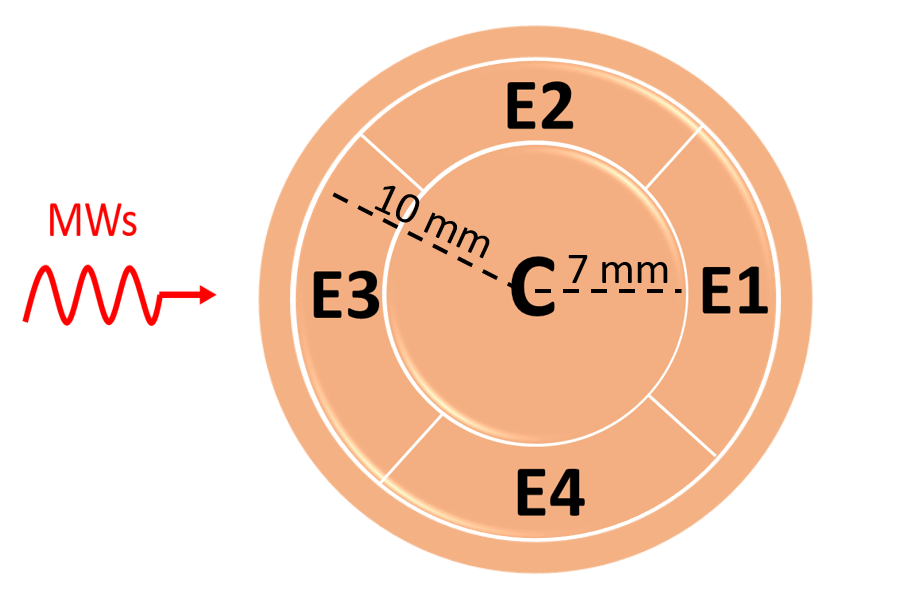}
\caption{\label{fig:1} Schematic view of the five top electrodes with their guard ring. The five electric current signals, $I_{c}$, $I_{1}$, $I_{2}$, $I_{3}$ and $I_{4}$, are measured simultaneously from the central electrode and the four segmented electrodes E1, E2, E3, E4, respectively. The gaps between the electrodes are 0.2~mm. The microwaves propagate parallel to the helium surface with a frequency $f$\,=$\,139/2\pi$~GHz, and 0~dBm power.}
\end{figure*}

The aim of the experiment is to understand the spontaneous oscillatory behaviour of the SEs and, in particular, how it changes as a function of electron density and pressing voltage. Nonlinear dynamics methods are used to explore the characteristic features of this system. The time evolution of electric currents induced in five electrodes above the liquid surface was measured. They are labelled C, E1, E2, E3 and E4 to represent the central electrode and the four edge electrodes as shown in Fig ~\ref{fig:1}. Both the top and bottom electrodes are surrounded by the guard rings that are negatively biased ($V_{\si{TG}}$ and $V_{\si{BG}}$\,=\,$-$\,0.5~V, respectively) to confine the electrons within their pool and form a sharp edge on the electron profile. The distance between the top and bottom electrodes is $D$\,=\,2.6~mm. The depth of liquid helium ${}^{4}$He $d$\,=\,1.3~mm and the magnetic field $B$\,=\,0.81~T are kept constant throughout the experiment. The pressing voltage ($V$) is varied from $V$\,=\,4.16 to 4.22~V; and the electron density is varied within the range $n_e\,=\, 1.4$\,to\,$2.2 \times 10^6$~\si{cm^{-2}}.
%The gap between the electrode pieces is 0.2\,mm. 

\begin{figure*}[h!]
\includegraphics[width=17cm]{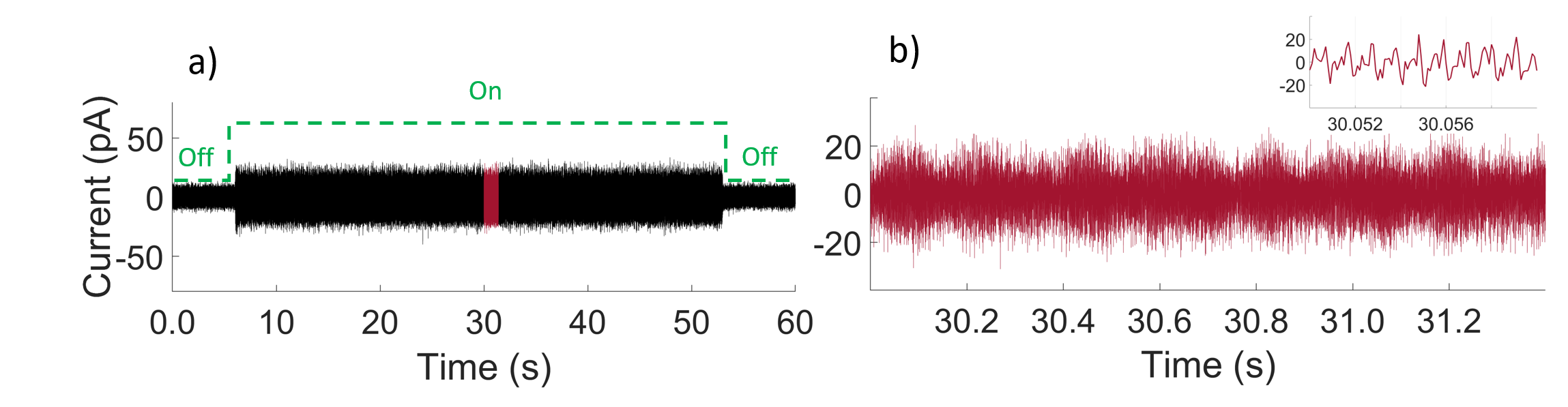}
\caption{\label{fig:t} (a) Example of the whole length of time series recorded for 60 seconds at a sampling frequency of 100~kHz resampled to 10~kHz. The MW (139~GHz) is switched On and Off where the On state is inside the green dashed lines. (b) The section of the signal in red, recorded for 1.4 seconds, is investigated to elucidate the characteristics of the oscillations. Inset shows current oscillations on a shorter time interval.}
\end{figure*}

%The experimental parameters are:
%\begin{list}
%\item$n_{e}$ - electron density,\\
% \item[] B -  magnetic field which is applied perpendicularly to the liquid surface\\
%\item[] d - depth of liquid helium\\
%\item[] V - pressing voltage \\
%\end{list}
For all of the measurements the MW circular frequency is 139~GHz. The parameters of the current preamplifiers and the oscilloscope are $10^{-9}$~A/V gain, 10~kHz preamplifier bandwidth and 100~kHz sampling frequency. Fig.~\ref{fig:1} shows the electrode arrangement used to measure the five current signals ($I_{c}$, $I_{1}$, $I_{2}$, $I_{3}$ and $I_{4}$), which were recorded for 60 sec from each of the five electrodes C, E1, E2, E3 and E4.  Fig.~\ref{fig:t} (a) shows an example of a recorded time trace to indicate the length of the data. Measurements were performed at T\,=\,0.3~K, in magnetic field $B$\,=\,0.81~T. Application of the MW resulted in an increase in amplitude of the measured signals, as seen in Fig. 2. Fig.~\ref{fig:t} (b) shows part of the signal in Fig. 1(a) with length 1.4 sec  that we used to study the characteristics of the oscillations, and the inset with an even shorter segment is provided to show the pattern of the signal.

Such systems have traditionally been treated in terms of quantum mechanics~\cite{kawakami2019image, monarkha2019magneto, zadorozhko2021motional}. They can, however, be entirely deterministic and we are currently analyzing this system within a classical framework. We apply time-resolved, nonautonomous, nonlinear dynamics methods to extract a wealth of information from the experimental data. This method provides temporal resolution and yields information about localized waves. In using it to reconstruct the dynamics of the SE we observe:

\begin{itemize}
    \item A variability in the frequency of the spontaneous kHz oscillations. The main reason is that the electron is moving in three\,-\,dimensional (3D) relative to the electrodes gravity waves moving the liquid surface vertically: the surface is not flat as conventionally assumed in quantum theoretical descriptions of the system).
    
    \item Motion of the electrons around the cell. The pattern and velocity of movement change with the pressing potential and electron density. %Where in quantum dynamics believe that the movement of electrons only by the changing the quantum level. I THINK THIS IS A FALSE ANTITHESIS
    
    %\item The pressing voltage and the electron density had a significant influence on the velocity of electron.\
    
    \item Resonance conditions in term of the phase coherence, which becomes constant between the five signals.\

\end{itemize}

\subsection{Power Spectral Density}
Systems that change their behaviour in a predictable way with time are usually described as deterministic. In the opposite limit, physical systems that seem at first sight to behave randomly are often treated as being stochastic, though this is not always appropriate. It may lead to a loss of potentially valuable information that can be recovered by use of time-resolved nonlinear dynamical methods \cite{clemson2016reconstructing,clemson2014discerning,iatsenko2015linear,Newman:2018}. The commonest method of visualizing the dynamical properties of a signal is the Fourier transform. It can be used to present time traces in the frequency domain. The time series are characterised by the total length of the signal, $T\,=\,N\Delta$, where $\Delta$ is the time interval between samplings and $N$ is the number of data points. The other important parameters are the sampling frequency $f_s$\,=\,1/$\Delta$ and the maximum observable frequency, equal to half the sampling frequency ($f_s$/2). The latter is called the Nyquist critical value, where the extracted frequency modes are limited to the range $f_k$\,=\,$\frac{k}{N\Delta}$, where $k$\,=\,0, 1, .., $\frac{N}{2}$. The discrete Fourier transform (DFT) \cite{press2007numerical} is given by
\begin{equation}
\label{eqn:1}
 G_{k} = \sum_{j=0}^{N-1}x_{j}e^{2\pi ijk/N}.
\end{equation}
The power spectral density (PSD) represents the power distribution over the frequency modes within the time series. The power content of the signal \cite{press2007numerical} is given by
\begin{align}
\label{eqn:eqlabel}
\begin{split}
    P(f_{0}) = \frac{1}{N^2} |G_0|^{2},
    \\
    P(f_k) = \frac{1}{N^2} \biggl[  |G_k|^{2} + |G_{N-k}|^2 \biggr],
    \\
    P(f_{N/2}) = \frac{1}{N^2} |G_{N/2}|^{2}.
    \end{split}
\end{align}
The power spectral density of a stochastic process has the form
\begin{equation}
\label{eqn:e}
    P(f) = \frac{\rm constant}{f^\alpha},    
\end{equation}
where $\alpha$ $\geq$ 0  and determines the functional dependence of the spectrum. Taking the logarithm of each side of Eq.(3), we obtain

\begin{equation}
\label{eqn:4}
    \log(P) = \log({\rm constant}/{f^\alpha}) = - \alpha{}\, \log(f) + \log({\rm constant}),
\end{equation}

\begin{figure*}[h!]
\includegraphics[width=17.2cm]{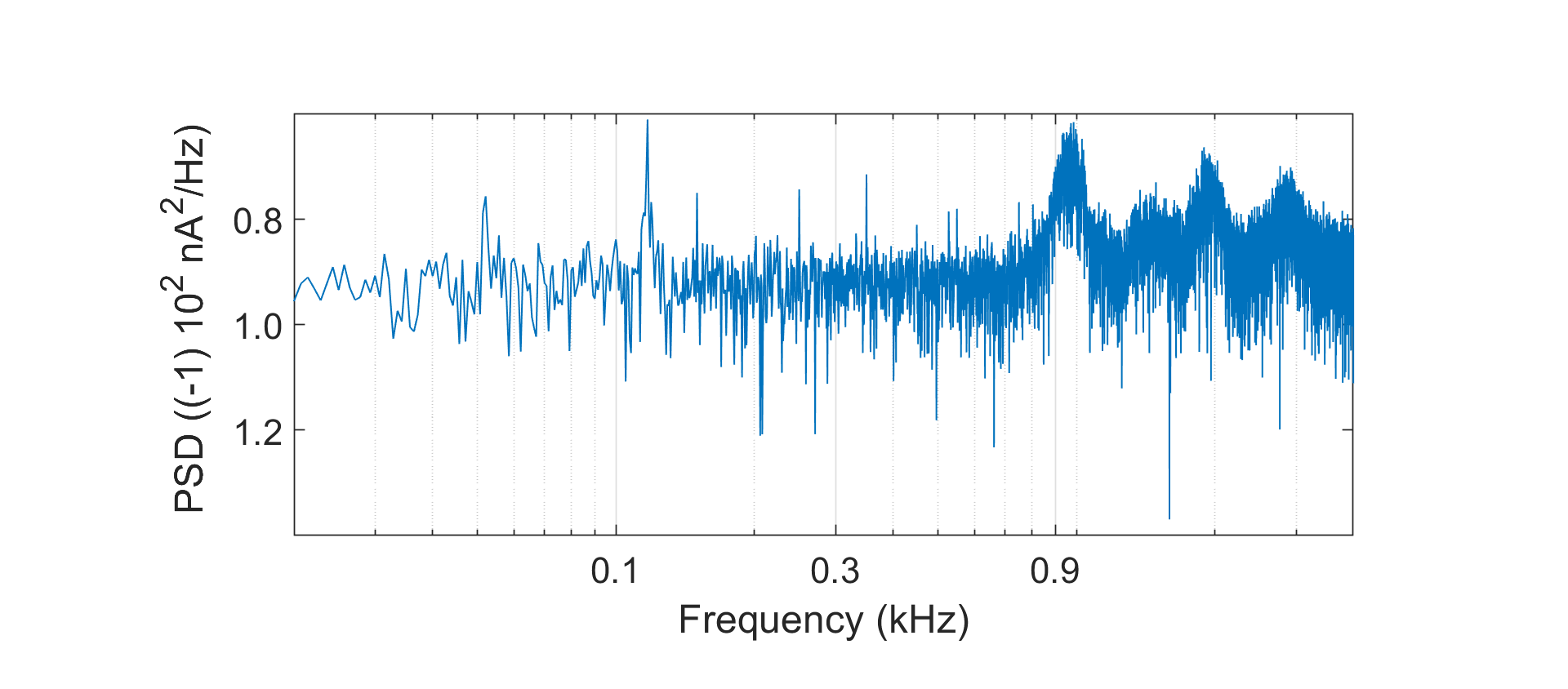}
\caption{\label{fig:psd} Power spectral density (PSD) of the time series presented in Fig.~\ref{fig:t} (b). The sampling
frequency is 10~kHz. The signal contains sharp peaks at frequencies 0.17 0.95, 1.94 and 2.9~kHz.
}
\end{figure*}
where $\alpha$ is a parameter that is computed from the gradient of the straight line graph. Fig.~\ref{fig:psd} shows the power spectral density of the time trace shown in Fig.~\ref{fig:t} b) by using equations \ref{eqn:1}, \ref{eqn:eqlabel} and \ref{eqn:4}. 
%The gradient is defined by 
%\begin{equation}
   %-\alpha = \frac{\sum_{k=0}^{N/2-1}(\log(f_k)-\mu_{\log(f)})(\log(P_k)-\mu_{\log(p)}) }{\sum_{k=0}^{N/2-1}(\log(f_k)-\mu_{\log(f)})^2}
%\end{equation}
%where $\mu_{\log(f)}$ and $\mu_{\log(p)}$ are the mean values \cite{lerch2012introduction}.

Our computations are for an $\alpha$ value that is~-1 $\le$~ $\alpha$~$\le$ 0.9, leading to a. %Therefore, the spectral density with this $\alpha$\,=\,1 s commonly referred to a system with $1/f$ or pink noise.
spectral density commonly referred to as ``1/f" or ``white" noise \cite{Ward:2007}. 
%Indeed, since our system is affected by external perturbations such as MW, $V_p$ and $B$. It exhibits a time-dependent dynamic characteristics of a nonautonomous system. %Indeed, such our system that affected  by external perturbations such (MW, $Vp$ and $B$) introducing explicit time-dependence (Nonautonomous).
%Moreover, since the dynamics of electron is measured from the initial condition and given inputs, it means that the current signals are non-stochastic (deterministic).
%The prior assumption of the signal being nonautonomous and deterministic, the traditional approach has been treated the data as stochastic. This results in the loss of valuable information, which we recover and introduce later.

%However, such this traditional approach (PSD) in treating the data as stochastic has resulted  in the loss of valuable information, that we will introduce later, where the time dependence is being regarding as noise.

%In fact, the data that is measured from movements of electrons above the surface of liquid helium under strong magnetic field and microwave irradiation require convenient methods for analysis. Here, we introduce these methods that we will discuss in details in the next in section. 

\section{Signal analysis}
\begin{figure*}[h!]
\includegraphics[width=17.2cm]{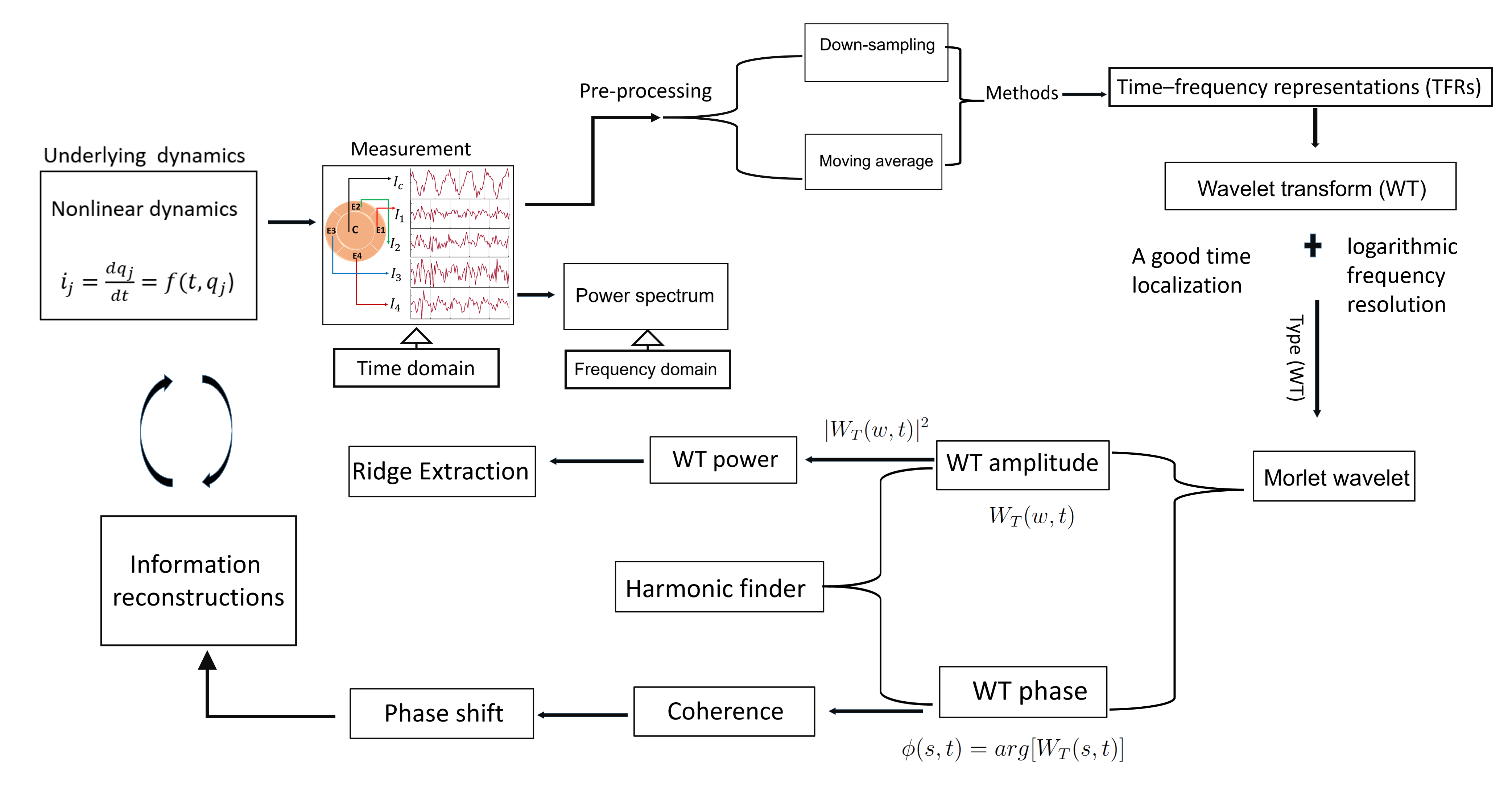}
\caption{\label{fig:2} A workflow chart of the methods discussed in this paper.}
\end{figure*}

Physical systems whose parameters vary in time are called ``dynamical systems''. 
%In physics, system whose parameters vary in time can be termed as a dynamical system. 
%Dynamical systems are commonly given in the form of\begin{equation}\frac{dx}{dt} = f(x),\end{equation}
Electrons above liquid helium in the presence of a magnetic field directed normally and exposed to microwave radiation is a good example of such a system. There exists an external mechanism (such as the microwave-electron interaction) which allows the system to exchange energy with its surroundings, resulting in an inherent time-variability and high nonlinearity. 
%in which allow to the system to exchange mass and energy with the surroundings and therefore, this leads to an inherent time-variability and represents highly nonlinear.
We need suitable analysis methods to acquire insight into the behavior of the system generating the data. Thus, we aim to use nonlinear dynamical methods to extract information about time-varying oscillatory modes and their mutual interactions. 
Figure~\ref{fig:2} provides an overview of the methods that were used to study the oscillatory dynamics of the recorded signals. 

\subsection{Pre-processing}
Pre-processing of the time series is required to provide optimal conditions for analysis.
\paragraph{\textbf {Down-sampling}}
\noindent Down-sampling is a process of reducing the initial sampling rate of the data. The sampling frequency is 100~kHz, which gives a higher resolution than is needed, so all the signals are down-sampled to 10~kHz. 
\paragraph {\textbf{De-trending}}
\noindent Prior to the analysis of the signals, all data were de-trended by using a moving average (default in Matlab commands). Non-oscillatory trends appear as low-frequency variations of the signals, and the purpose of detrending is to remove their possible interference with low-frequency oscillatory modes within the range of interest.

%Down-sampling is a process of reducing the sampling rate of the signal. The signal has two important features that are the time interval $L$=N$\Delta$t,
%and the sampling frequency $f_s$ =1/$\Delta$t, the
%rate at which sample are calculated.The Nyquist - Shannon sampling frequency is $f_s$/2 the
%maximum observable frequency that is equal to half the sampling frequency. Where the lowest observable frequency is given by 1=L

\subsection{Time-frequency representations}
A nonautonomous system is affected by external perturbations, producing explicit time-dependences in the system. Methods of analysis that assume a stationary frequency distribution (Fourier transform) are insufficient to compute all relevant information from the data. %Also, the useful signal is treated as noise where the frequency is time-averaged.\
Instead, windowing methods such as windowed Fourier transform (WFT) and the wavelet transform (WT) may be used to extract spectral information.

The windowed Fourier transform was developed to overcome limitations of the Fourier transform when used to analyze the nonstationary signals, and to create time-frequency representations of the signals. The WFT works with a time series $x(u)$ of length $T$, by sliding a rectangular function of length {$\tau$}\,$\textless$\,$T$ ( window ) over the signal. Within the window, the signal can be treated as stationary and the Fourier Transform (FT) can usefully be applied. However, the WFT method still has limitations because of a trade-off between time localisation and frequency resolution, due to the uncertainty principle, which states that we cannot determine the precise frequency and the exact time simultaneously. Therefore, the size of the window is adjusted depending on whether the user needs good frequency resolution or a good time localisation. By using a large window we can obtain good frequency resolution but poor time localisation: the frequency resolution is proportional to the length of the window. Similarly, the low-frequency information cannot be extracted when using a narrow window.

The wavelet transform (WT) was developed to overcome these problems. It makes use of an adaptive window that yields good resolution in both time and frequency, with logarithmic frequency resolution. It is calculated by moving a wavelet function along the signal and it can be tuned according to the frequency ranges that we want to investigate \cite{clemson2014discerning}. The wavelet transform is obtained from \cite{clemson2016reconstructing},
\begin{equation}
W_T(s,t)=\frac{1}{\sqrt s} \int_{-\frac{L}{2}}^{\frac{L}{2}}\varPsi(s, u-t) f(u) du,
\label{eq: wavelet CWT}
\end{equation}
where $\varPsi(s,t)$ is the mother wavelet (providing the basis of the continuous WT method), $s$ is a scaling parameter used to change the wavelet frequency distribution and shift its time as needed. The frequency scale in the WT is continuous so that, for any arbitrary frequency, the wavelet components can be calculated. The mother wavelet that we choose is the Morlet wavelet, given by
\begin{equation}
\varPsi(s,t)= \frac{1}{\sqrt[4]{\pi}} (e^{\frac{2 \pi if_rt}{s}}-e^{\frac{-2 \pi {f_r}^2}{2}}) e^{-\frac{t^2}{2s^2}}.
\label{eq:morlet wavelet }
\end{equation}
The parameter $f_r$ is the frequency resolution, which determines the trade-off between time localisation and frequency resolution in the wavelet coefficients calculated from the signal under study. Higher values of $f_r$ give lower time resolution, but have good frequency resolution. However, at very small frequencies the wavelet becomes meaningless. 

Coefficients from performing a WT with the Morlet wavelet can be complex-valued. This property can be used to determine the instantaneous amplitude and phase for each frequency and time \cite{Kaiser:94} .

\subsection{Wavelet power}

The wavelet power spectrum can be obtained by calculating the integral of the square of the modulus of the wavelet transform over frequency \cite{clemson2014discerning,clemson2016reconstructing}:
\begin{equation}
P_W(w,t)= \int_{w-\frac{dw}{2}}^{w+ \frac{dw}{2}}|W_T(w,t)|^2 dw.
\label{eq:wavelet power}
\end{equation}
It gives a vector representing the power of the whole data set. It can be plotted as a function of frequency to show the power distribution between different spectral components and to identify the frequency range of the main oscillatory components in the signal under investigation.

\subsection{Harmonic detection}\label{sec:H}
\noindent  We applied the wavelet transform (WT) to the data in order to trace the time variability of the fundamental oscillation frequencies, and to obtain time-dependent phase information. However, we may observe other frequency modes appearing in the WT, that are integer multiples of the fundamental frequency. Where We expect high frequency harmonics in the data due to the nonlinearity of the measured signal, we need a method to detect relationships between the oscillations and identify the harmonics of any component in the signal. The method provided by Sheppard et al \cite{Sheppard:11} can determine whether the oscillation is a harmonic of the main frequency of the signal or an independent oscillation of  higher-frequency. This method of finding harmonics is based on mutual information, by analogy with the Shannon entropy and surrogate testing. The mutual information (M) provides a measure of the missing entropy from a conditional distribution while considering information about another variable. The Shannon entropy introduces a measure of the unpredictability. The entropy is high if the distribution is uniform, and it is low if the distribution is sharply peaked. 
%The basic assumption for algorithm is that higher-correlations can be destroyed by randomization of Fourier phases in time while preserving the linear correlations \cite{raeth2009surrogates}.

\textbf{Procedure for identifying harmonics}:
First, we need to extract the phases from wavelet transform at each point in time for each frequency in the signal. The phases are then discretized into 24 bins. These phases are obtained to calculate the mutual information, which is calculated for every possible pair of phases in the signal. 
\begin{figure}[h!]
\includegraphics[width=17.2cm]{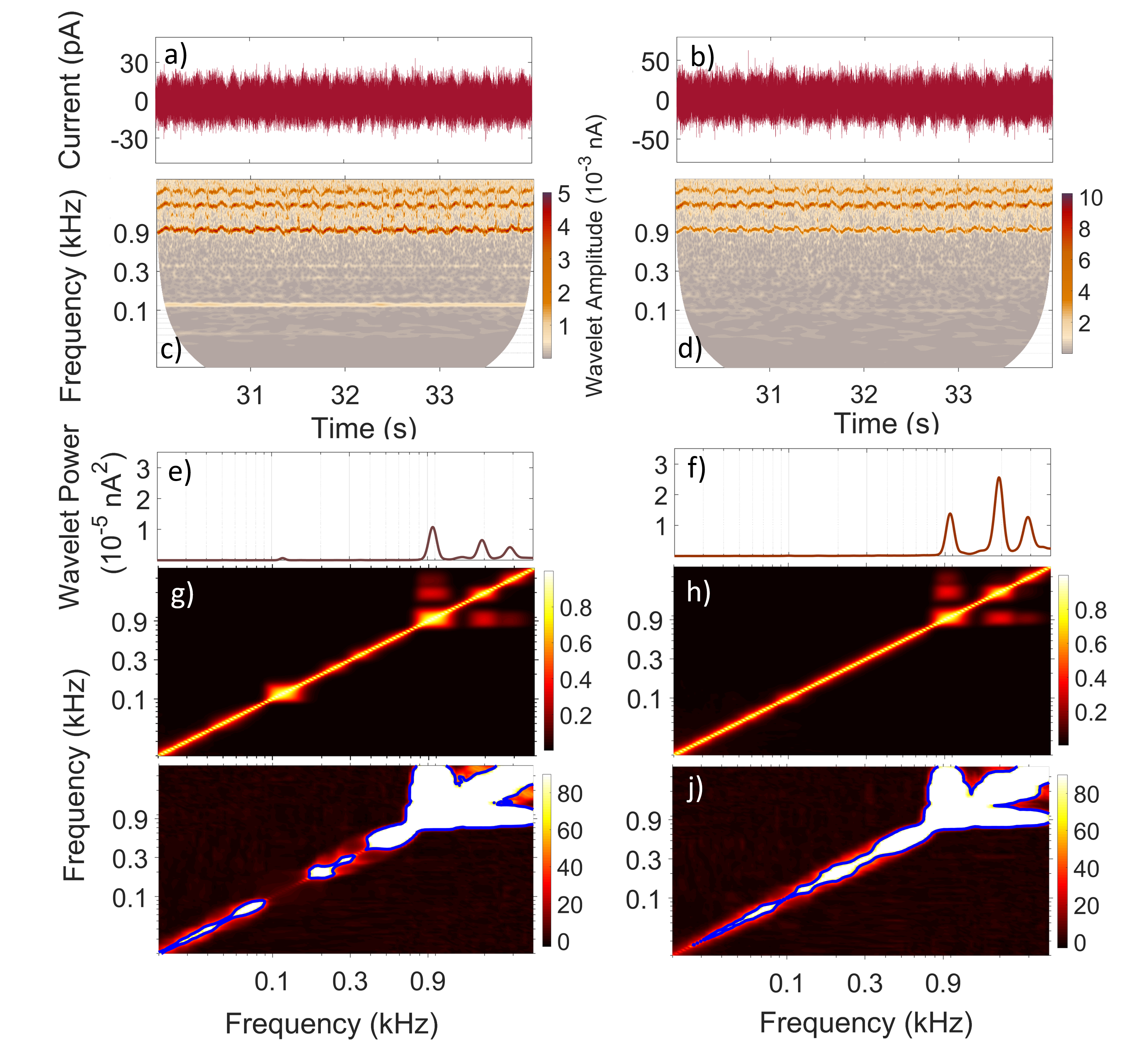}
\caption{\label{fig:HAR} (a-b) Current signal for electrodes (2,3) (4 s sections) recorded for 60 seconds in total at a sampling frequency of 100~kHz resampled to 10~kHz, pressing voltage 4.20~V, B\,=\,0.81~T and $n_e\,=\,1.4 \times 10 ^{6}$~\si{cm^{-2}}. (c-d) Wavelet transforms with frequency resolution 4~Hz and (e-f) time-averaged wavelet power of the signals in (a-b) showing the fundamental frequency and higher harmonic oscillations. The $z$-axes in panels (c-d) represents the wavelet amplitude; the y-axis of (b) and the x-axis of (e-f) are on logarithmic scales. (g-h) The mutual information M. (i-j) shows the actual mutual information values relative to the surrogate distribution, which we will refer to as an M-plot. The threshold of standard deviations above the mean of the 100 surrogate distribution is marked in blue. High harmonics appears in the representation due to nonlinearity. The central frequency used in calculations is 1~Hz.}
\end{figure}

The discrete entropy of the 24 phases is given by
\begin{equation}
H(\phi_1) = - \sum_{\phi_1=1}^{24} p(\phi_1){}\log_2(\phi_1).
\label{eq: entropy}
\end{equation}
 The mean entropy of the conditional distribution is calculated using the equation
\begin{equation}
H(\phi_1||\phi_2) = - \sum_{\phi_1=1}^{24}p(\phi_2) \sum_{\phi_1=1}^{24} p(\phi_1||\phi_2) {}\log_2 p(\phi_1||\phi_2). 
\label{eq: conditional entropy}
\end{equation}
The mutual information of the data is calculated from the difference between equation \ref{eq: entropy} and equation \ref{eq: conditional entropy},
\begin{equation}
M(\phi_1, \phi_2)= H(\phi_1)- H(\phi_1||\phi_2). 
\label{eq: mutual information}
\end{equation}

 The two frequencies $\omega_1$ and $\omega_2$ of the phases approach each other as the mutual information for $\phi(\omega_1,t)$ and $\phi(\omega_2,t)$ approaches unity. Since the mutual information is biased by the correlation of the phase signal caused by binning, it is necessary to perform surrogate-testing \cite{Lancaster:18a}. %As the mutual information are more biased by the correlation that provide in the phase signal that caused by bining, so the surrogate testing is needed.%Then, statistical tests (surrogates) require to check the significant cross-correlation between each pairs
 Surrogates are signals that are designed to preserve all of the properties of the original signals, except the property relating to the hypothesis that is being tested. 
 %This testing required to test for significance in harmonic detections are the amplitude adjusted Fourier transform
%(AAFT) and the iterative amplitude adjusted Fourier transform (IAAFT). The
%null hypothesis for both is that the phases in the time series are independent for all
%frequencies, which means that surrogates can be generated by the randomization
%of the time-phase information
 Commonly-used methods for generating surrogates are the amplitude-adjusted Fourier transform (AAFT) and the iterative amplitude-adjusted Fourier transform (IAAFT). These methods conserve the amplitude distribution in real space and reproduce the power spectrum (PS) of the original signals. The basic assumption of (AAFT) and (IAAFT) is that higher-order correlations can be destroyed by randomization of Fourier phases in time, while preserving the linear correlations \cite{raeth2009surrogates}. A local maximum in the mutual information calculated for a pair of phases from the original data is deemed to indicate a harmonic if it occurs some number of standard deviations above the mean value of the surrogate's mutual information.

Fig.~\ref{fig:HAR}(a) and (b) show 4-second sections of the current signals from the 60 second measurements by electrodes E2 and E3. The measurements are taken simultaneously with a sampling frequency of 100~kHz, resampled to 10~kHz, and with a pressing voltage of 4.20~V, $B\,=\,0.81$~T, and with $n_e$\,=\,1.4$\times10^6$~\si{cm^{-2}}. 
 %Figure \ref{fig:HAR} (a,b) show the section of current signals (4 sec) are taken of whole current signal of recorded simultaneously for 60 second at sampling frequency of 100\,kHz resampled to 10\,kHz at pressing voltage 4.20\,V, B\,=\,0.81\,T and $n_e$\,=\,1.4\times $10^6$ \si{cm^{-2}}.
The time frequency representation (WT) and the time-averaged wavelet power are presented in Fig.~\ref{fig:HAR} (c,d) and Fig.~\ref{fig:HAR} (e,f) for electrodes E2 and E3, respectively, with frequency resolution frequency~3~Hz. The E2 and E3 signals oscillate at the same frequency, the mean  frequency of the main mode of each signal being around 951~Hz where the second and third frequencies are 1884 and 2886~Hz, respectively, as shown in Fig.~\ref{fig:HAR} (c). %For electrode 3, the fundamental frequency is around 970\,Hz  and the frequency of the second and third harmonics are 1884 and 2886\,Hz, respectively, as shown in Fig.\ref{fig:HAR} (d)
It appears from the identical frequencies that E2 and the E3 are coupled. We have calculated the coherence of the two signals, and found that they are coherent in the frequency range of the oscillatory modes: see also below.
 
To determine whether the second and third harmonics of the signals from E2 and E3 are independent modes, or higher harmonics of the fundamental mode, the harmonic finder method is used (see above). Figures~\ref{fig:HAR} (g,i) for E2, (h,j) for E3 show the mutual information plot relative to the mean and standard deviation of the surrogates. A threshold of standard deviations above the mean of the 100 surrogate distribution is marked in blue. High harmonics appear in the representation due to nonlinearity in both data. It was found that the second and third frequencies for each electrode (E2,E3) are higher harmonics of the main mode, which results from a strong nonlinearity that is generated from the coupling of the surface liquid helium and the moving of the electrons.

%%%%%%%%%%%%%%%%%%%%%%%%%%%%%%%%%%%%%%%%%%%%%%%%%%%%%%%%%%%%%%%%55
\subsection{Ridge Curve Extraction}
\begin{figure*}[h!]
\includegraphics[width=17.2cm]{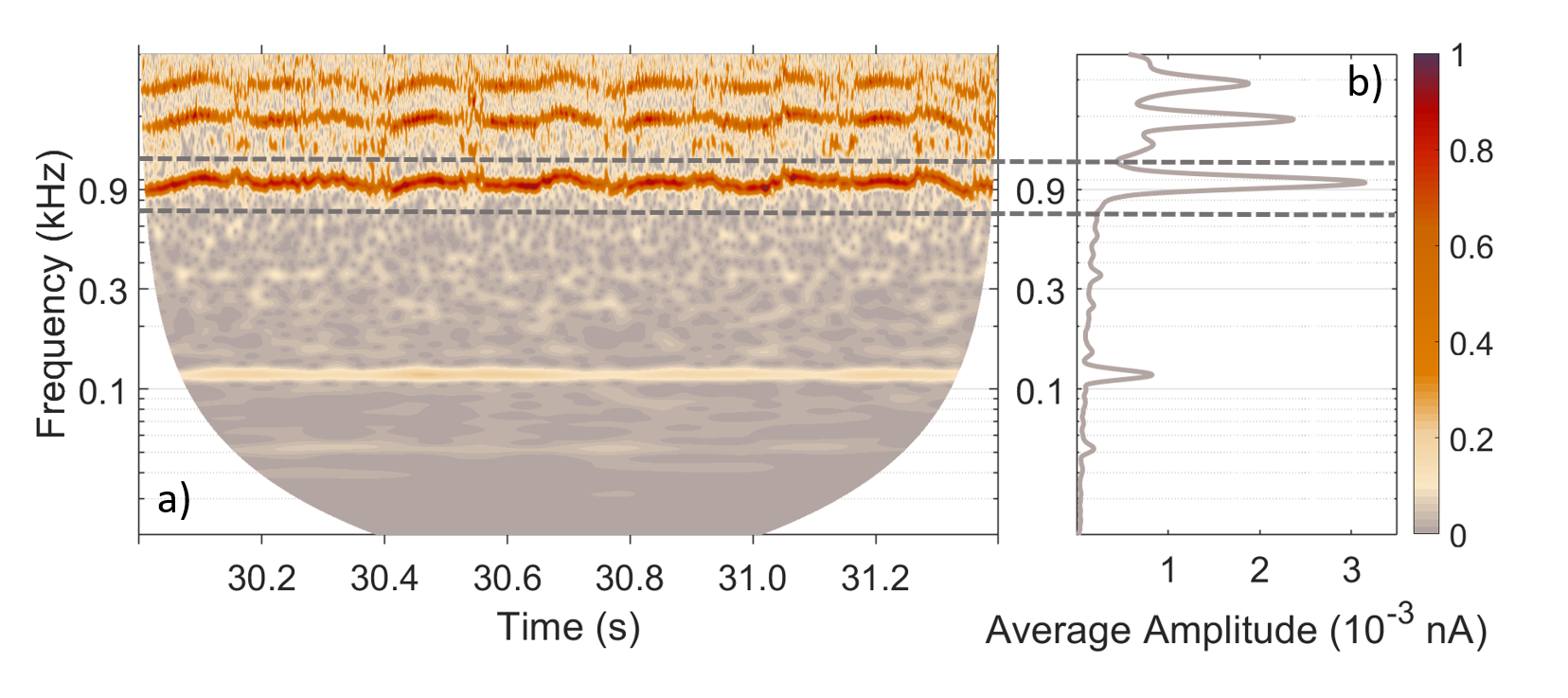}
\caption{\label{fig:ridg} Time-frequency representation (TFR) and average amplitude plot of the set signal presented in Fig.~\ref{fig:t}. Oscillatory components are shown as amplitude peaks in the TFR and average amplitude plot. The frequency band chosen for ridge extraction include the entirety of the oscillatory components (limits shown by the grey dashed lines).}
\end{figure*}

Ridge curve extraction is a method for extracting the oscillatory modes from a signal \cite{iatsenko2016extraction}. The time\,-\,frequency representations of the signals are computed by use of the WT. The oscillatory components  appear as amplitude peaks in the time-frequency representations (TFRs) and are called ``ridge curves". We can determine the oscillatory components by using both the time-frequency representation and the time-averaged plot. The ridge extraction can be defined by selecting the frequency band which includes the entirety of the oscillatory components within the signal. The frequency band should include the whole width of the oscillatory component, including the peak and the surrounding blur due to the uncertainty principle as shown in Fig.~\ref{fig:ridg}. We need to extract only one oscillatory component that appears in the frequency band. 

 \subsection{Phase coherence and phase difference.}{\label{sec:1}}
\begin{figure*}[h!]
\includegraphics[width=17cm]{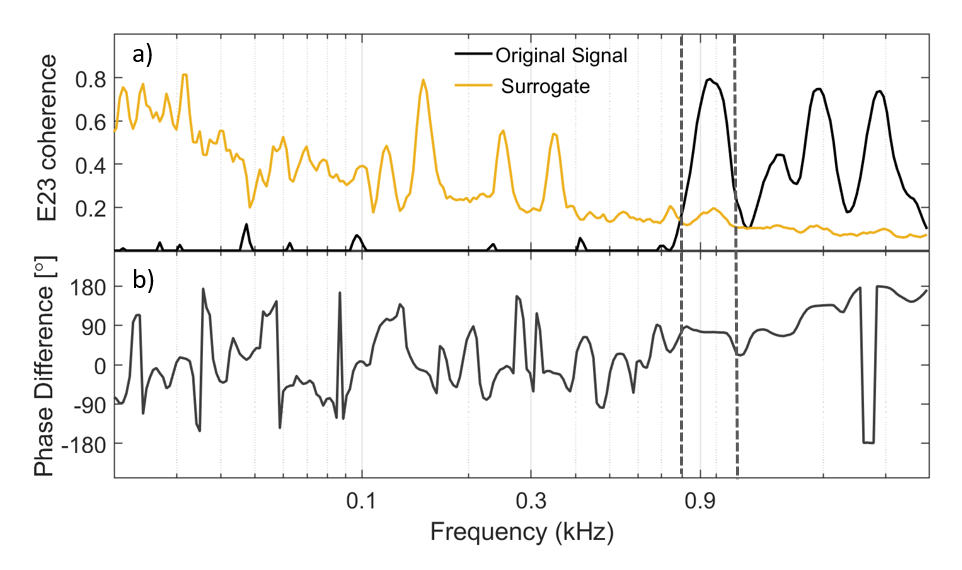}
\caption{\label{fig:cE}(a-b) Wavelet phase coherence and phase difference of signals from the two of electrodes (E1,E2). The main frequency component is chosen as the significant phase coherence and the corresponding phase difference is shown by the black dashed line. Coherence below the 95th percentile of the surrogates (yellow curve) is not considered significant.}
\end{figure*}

When we observe oscillations at the same frequency in two different time series and find that the difference between their instantaneous phases $\phi_{1 k,n}$, $\phi_{2 k,n}$ is constant, then we can say that the oscillations are coherent at that frequency. The wavelet phase coherence (WPC) between the two signals $x_1(t)$ and $x_2(t)$ is obtained through their respective wavelet transforms as defined by equation (6). $W_{si}$ is the wavelet transform of signal $x_i$. The WPC of the two signals \cite{Bandrivskyy:04a} are given as %%{x_1;x_2}(s,t)$ as$

\begin{equation}
WPC_{x_1;x_2} (f)=\frac{1}{L}\int_{o}^{L} e^{iarg[W_{x_1}(s,t)W^*_{x_2}(s,t)]}dt|\label{eq: phas}, 
\end{equation}
where the scale $s$ is related to the frequency $f$ as $s\,=\,1/(2\pi f)$. The phase coherence function $C_{\theta}(fk)$ is obtained by computing and averaging over time the sine and cosine components of the phase differences for the whole signal, effectively defining the time-averaged WPC as
\begin{equation}
C_ \theta(\omega_k)=\sqrt{\langle \cos\Delta \theta_k{}_n \rangle^2 + \langle \sin\Delta \theta_k{}_n\rangle^2}.
\label{ref:eqn7}
\end{equation}
The phase coherence function $C_{\theta}(fk)$ as defined in equation (\ref{ref:eqn7}) is exactly the discrete version of the phase coherence formula in equation (\ref{eq: phas}). In addition, we can also calculate the phase difference $\Delta\theta_{kn}$ between two signals according to
\begin{equation}
  \Delta\theta_{kn}=\theta_{2k,n}-\theta_{1k,n}.
\end{equation}
The value of $\Delta\theta_{kn}$ lies within $180^{\circ}$ and gives information about the phase of one oscillator relative to the other. 

Connections may be found between two signals or components due to the interactions between them, or common external influences affecting them. 
%There may have possible connection between two signals or two components found due the interaction between the processes of these signals or there are external influence on these processes.
The calculated coherence between processes is not reliable unless we can repeat the calculation while the interaction or common influence is absent. One way to test for significance coherence uses surrogate signals \cite{kugiumtzis2002surrogate}. The type of the surrogates that we use with our signals is the iterative amplitude adjusted Fourier transform (IAAFT) that we introduced in (\ref{sec:H}). The null hypothesis here is that the phase between the signals is  independent for all frequencies. It means that surrogates can be produced by randomization of the time-phase information \cite{clemson2016reconstructing}.

In our study, the wavelet phase coherence and the corresponding phase difference were calculated between pairs of signals, for example: electrode 1 (E1) and electrode 2 (E2) signals as shown in Fig~\ref{fig:cE}. The phase coherence at each frequency was considered significant, if the original signal is higher than 95th percentile of 100 surrogates in the frequency range of interest. Fig~\ref{fig:cE} shows that the phase coherence exists in the range 826\,-\,111~Hz where the two peaks above this range are considered to be high harmonics. The phase difference (constant) was only taken into consideration at the frequencies where significant coherence was observed.

%The two phases will extract $\theta_{1 k,n}$ and $\theta_{2 k,n}$ with $\Delta \theta_{k,n}$ = $\theta_{2 k,n}$ - $\theta_{1 k,n}$ being the relative phase difference. The phase coherence function  $C_ \theta(\omega_k)$ is calculated  by averaging in time the components of the sine and cosine of the phase differences for the signal. The phase coherence is given by .............................................................................................................................................................................................................................

%We study oscillatory dynamics of currents recorded from five electrodes placed on the top of a mixing chamber. The currents result from movements of electrons above the surface of liquid helium under strong magnetic field and microwave irradiation. Methods, that optimize Heisenberg uncertainty principle in terms of time localization and frequency resolution, have been applied to extract information on basic oscillatory modes and their frequency modulation.  Instantaneous frequencies of the modes have been extracted to obtain phase coherences and phase shifts between the five electrode positions. 

\newpage

\section{Results and Discussion}
\subsection{Absence and presence of electrons}
Before introducing the main results, we present a preliminary survey of the results. We consider measurements made both in the absence and presence of surface electrons, for different values of pressing voltage (V), magnetic field (B), and microwave radiation (MW). %Before we introduce the main measurements of our studying. We present some measurements in two cases of absence and presence the electron with fixing and varying parameters such as pressing voltage ($V$) and magnetic field ($B$) and the microwave radiation (MW), in order to see when the oscillations will be observe.

\begin{figure*}[h!]
\includegraphics[width=17.2cm]{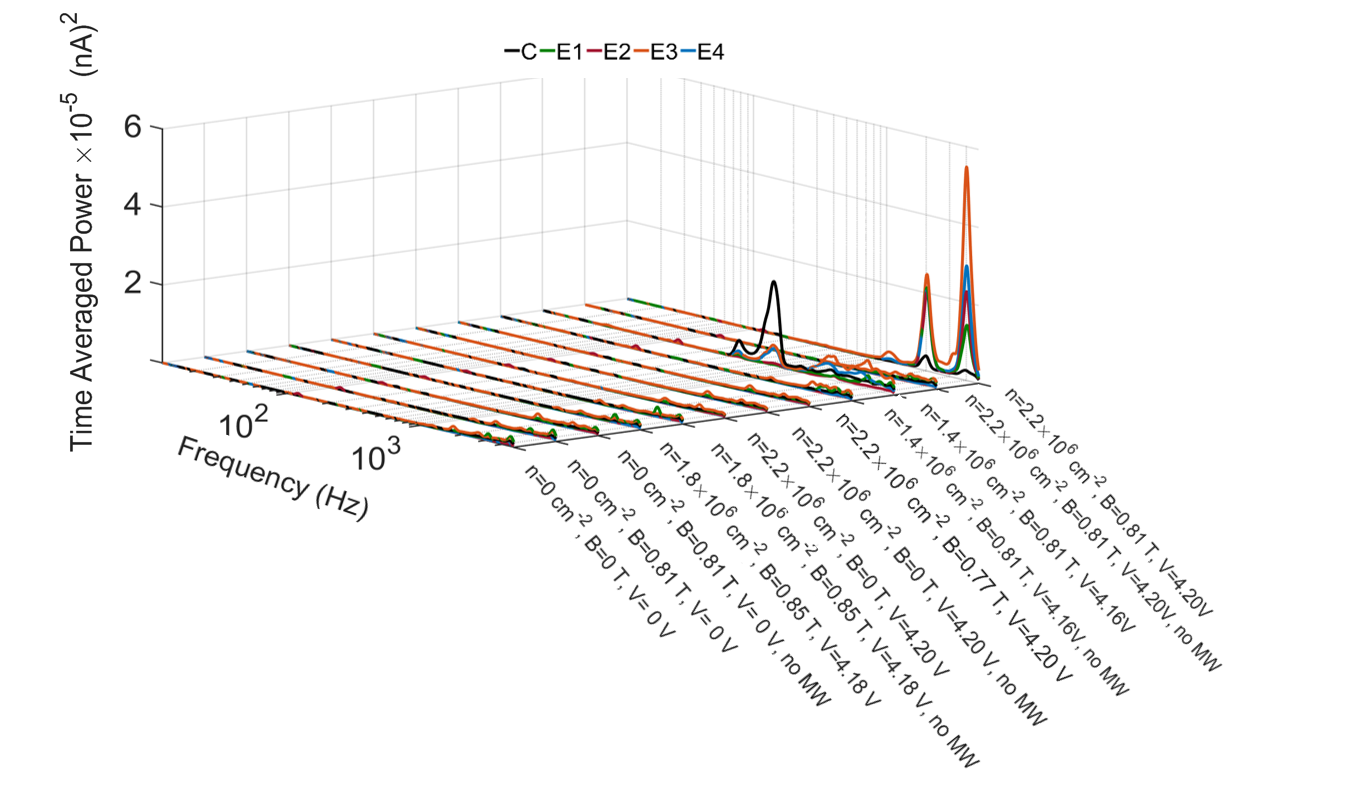}
\caption{\label{fig:d} Time-averaged power as a function of frequency, calculated from the wavelet transforms of signals measured for various experimental parameters, keeping the electron density at 0~$\si{cm}^{-2}$, $n_e$\,=\,1.4 $\times 10^{6}$~$\si{cm}^{-2}$, $n_e$\,=\,1.8 $\times 10^{6}$~$\si{cm}^{-2}$ and $n_e$\,=\,2.2 $\times 10^{6}$~$\si{cm}^{-2}$, throughout. The WTs were obtained for frequency resolution~3~Hz in the frequency range 0.2\,-\,4~kHz, with frequency plotted on a logarithmic scale. For each set of experimental values, all electrode measurements are plotted allowing for comparison. (i). In case $n\,=\,0~\si{cm}^{-2}$, while keeping the pressing voltage $V\,=\,0$~V at $B\,=\,0$~T in the presence of MW and at $B\,=\,0.81$~T with MW and without MW. (ii). In case $n_e$\,=\,1.8 $\times 10^{6}$~$\si{cm}^{-2}$ at fixed the value of pressing voltage $V\,=\,4.18$~V, $B\,=\,0.85$~T and with/without MW. (iii). In case $n_e$\,=\,2.2 $\times 10^{6}$~$\si{cm}^{-2}$ at fixed the value of pressing voltage $V\,=\,4.20$~V, the first set with no magnetic field $B\\,=\,0$~T with/without MW, the second set with $B\,=\,0.77$~T with MW. (iv) In case $n_e$\,=\,1.4 $\times 10^{6}$~$\si{cm}^{-2}$ at fixed the value of pressing voltage $V\,=\,4.16$~V and $B\,=\,0.81$~T with/without MW. (v) In case $n_e$\,=\,2.2 $\times 10^{6}$~$\si{cm}^{-2}$ at fixed the value of pressing voltage $V\,=\,4.20$~V and $B\,=\,0.81$~T with/without MW.}
\end{figure*}

Figure \ref{fig:d} of the time-averaged power of frequencies summarizes the results of the experiment that was performed with a cell containing liquid helium but with absence and presence of electrons. It shows one expects signals with such a high frequency and ill-defined frequency value (notice the smeared nature of the power over a large frequency-range) to be a result of inherent noise in the system in case (i),(ii), (iii) and (v) (when there is no MW applied). Therefore, in these cases there is no evidence for coherent behaviour in the oscillations in the frequency range (0.2\,-\,4~kHz). We classify signals producing these measurements as noise inherent to the system. 

The small power peaks measured for frequencies 0.1~kHz also exhibit the feature of existing for a range of different experimental parameters. Such an oscillation is certainly not due to electrons responding to applied fields/MWs and must be an artefact exclusive to electrode 2 (main power).

In case (iv) at low electron density, the frequencies of the oscillations are low, meaning that most electrons are at the centre of the chamber, especially before resonance values (shown later). 
%are low resulting in that most of the power of the electrons are at the centre of the chamber especially before the resonances values (that we will show in detail later on).
The electrons are concentrated at edges of the cell when the frequencies are higher (we will show through the text). Indeed, the oscillations occur when switching on the MW and applying a magnetic field of 0.81~T. In case (v), when MW are applied and at high electron density, the electrons appear to be at the edge of the cell, based on the readout from the five electrode system. %In case (v), with applying the microwave, at high electron density, the electrons seem to be at the edge of the cell based on the five electrodes system.

From cases (iv) and (v), the electron density and the pressing voltages can change the frequency of the oscillations. This indicates that the pressing voltage influences how the electrons orbit the cell.

\subsection{Time-frequency analysis}
\begin{figure*}[h!]
\includegraphics[width=17.2cm]{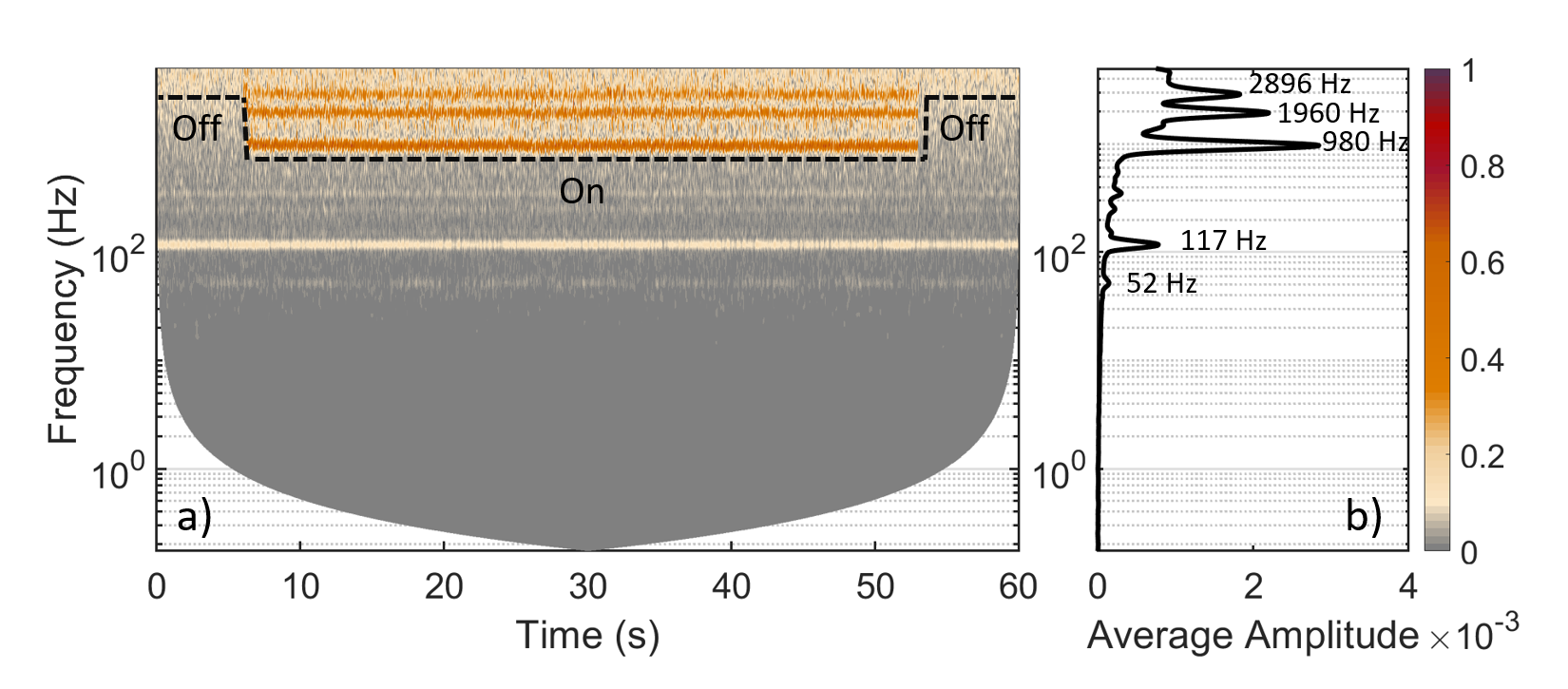}
\caption{\label{fig:Wo}(a)Time-frequency representation (TFR) and (b) average amplitude plot of the whole signal presented in Fig.~\ref{fig:t}. %Oscillatory components are shown as amplitude peaks in the TFR and average amplitude plot.
}
\end{figure*}

We show the time - frequency analysis for the whole signal presented in Fig \ref{fig:t} a), where the microwaves (MW) switched (Off - On - Off). Fig \ref{fig:Wo} a) indicates the evolution of a couple of frequency components, or modes, after MW irradiation. Note that the peak that is manifested in fig ~\ref{fig:Wo} b) at a frequency of 117~ Hz is the second harmonic of the main oscillation.

The sets of measurements provided below are for two different electron densities $n_e$:

\begin{itemize}

\item [A.] For $n_e$\,=\,1.4\,$\times 10^{6}$\,$\si{cm}^{-2}$ in Figures~\ref{fig:onesix}--\ref{fig:22} (lower electron density)

\item [B.] For $n_e$\,=\,2.2\,$\times 10^{6}$\,$\si{cm}^{-2}$ in Figures~\ref{fig:216}--\ref{fig:222} (higher electron density)

\end{itemize}

\noindent In each case, results are given for pressing voltages in the range $4.16 \leq V \leq 4.22$~V, with increments of 0.1~V. For all measurements, the magnetic field and the depth of liquid were fixed at {$B\,=\,0.81$~T} and {$d=1.3$~mm}, respectively. For each set of conditions, we present current time series similar to those shown in Fig.~\ref{fig:t} together with the corresponding WT and the time-averaged power in the middle 1.4\,sec of the signal. The group of figures presented for each set of conditions relates to the current recorded from each of the individual electrodes C, E1, E2, E3, and E4. The WTs are calculated with a frequency resolution of 3~Hz in the frequency range 0.011\,-\,4~kHz for the part of the time series 30\,-\,31.4~s. Where any of these parameters differed when calculating the WT, this is specified.

%%%%%%%%%%%%%%%%

\clearpage 

\centerline{\bf A. LOW ELECTRON DENSITY}
\vspace{0.3cm}

\underbar{\bf Pressing voltage $V\,=\,4.16$~V}

\begin{figure}[b!]
\centering
\includegraphics[scale=0.86]{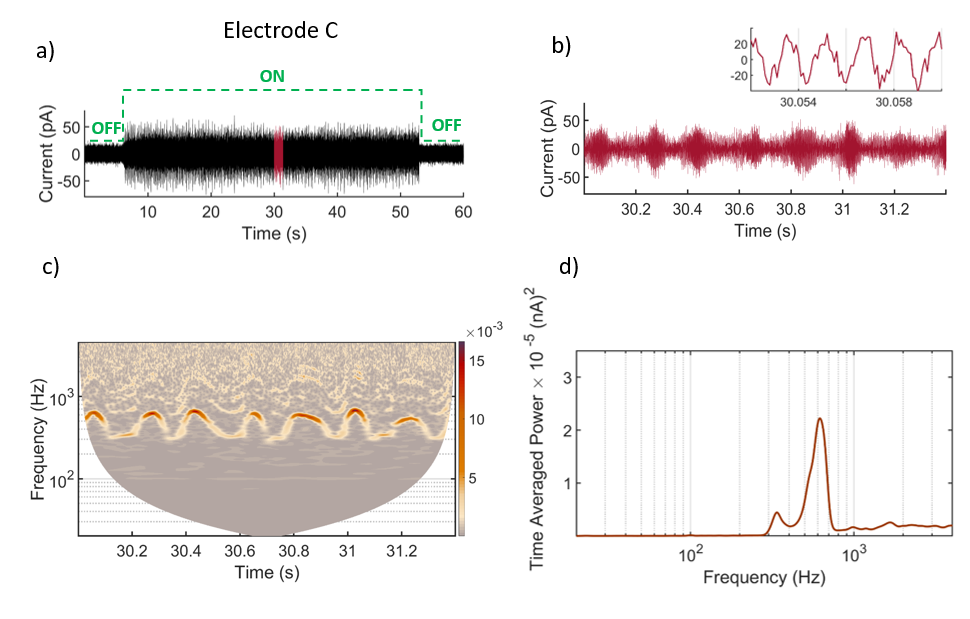}
\includegraphics[scale=0.86]{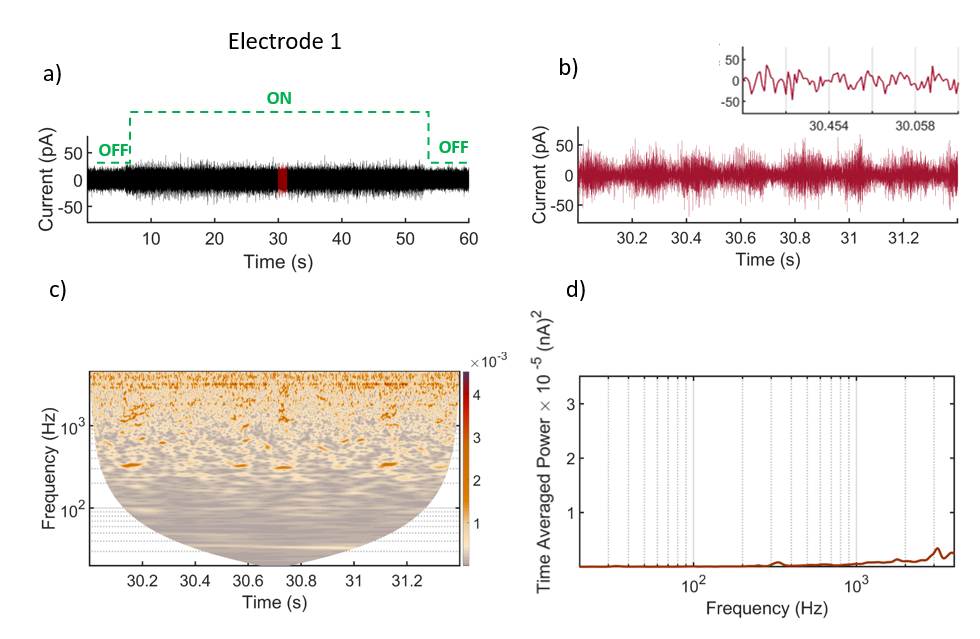}
\caption{\label{fig:onesix} Signals from electrodes C and E1, and their spectral analyses. (a) Signals were recorded for 60 seconds at a pressing voltage of $V\,=\,4.16$~V and electron density $n_e$\,=\,1.4$\times {10^{6}}$~\si{cm^{-2}}. The MW (139~GHz) is switched On/Off where indicated. (b) Enlarged reddened 1.4~s part of (a) used for analysis, where the inset shows a further zoom. (c) Wavelet transform with frequency resolution~3~Hz. (d) Time-averaged wavelet power of signals in (b) showing the fundamental frequency and higher harmonics. The $y$-axis of (c) and the $x$-axes of (d) are logarithmic}.
\end{figure}

\addtocounter{figure}{-1}
\clearpage 

\begin{figure}[b!]
\centering
\includegraphics[scale=0.9]{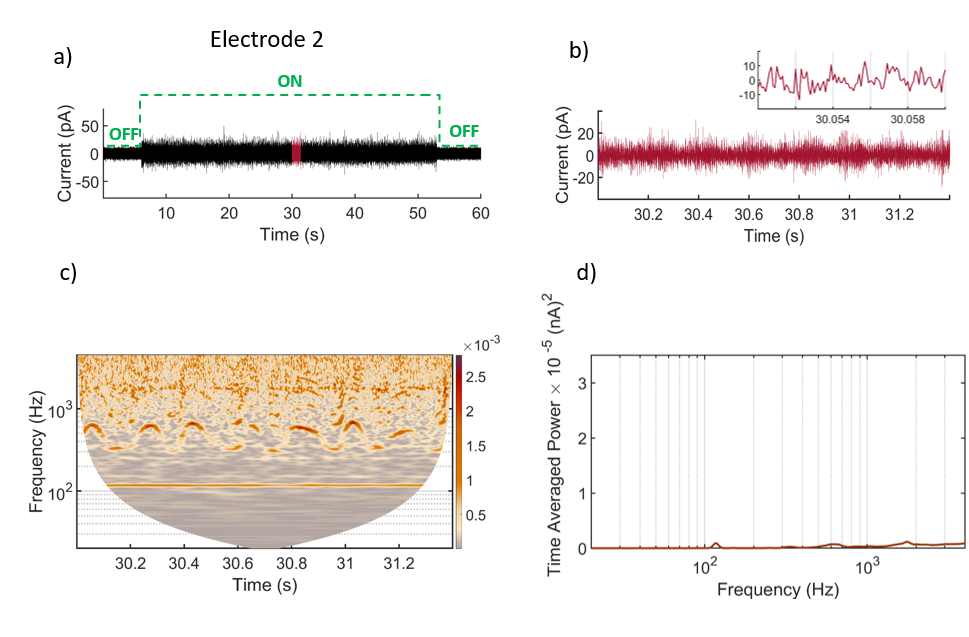}
\includegraphics[scale=0.9]{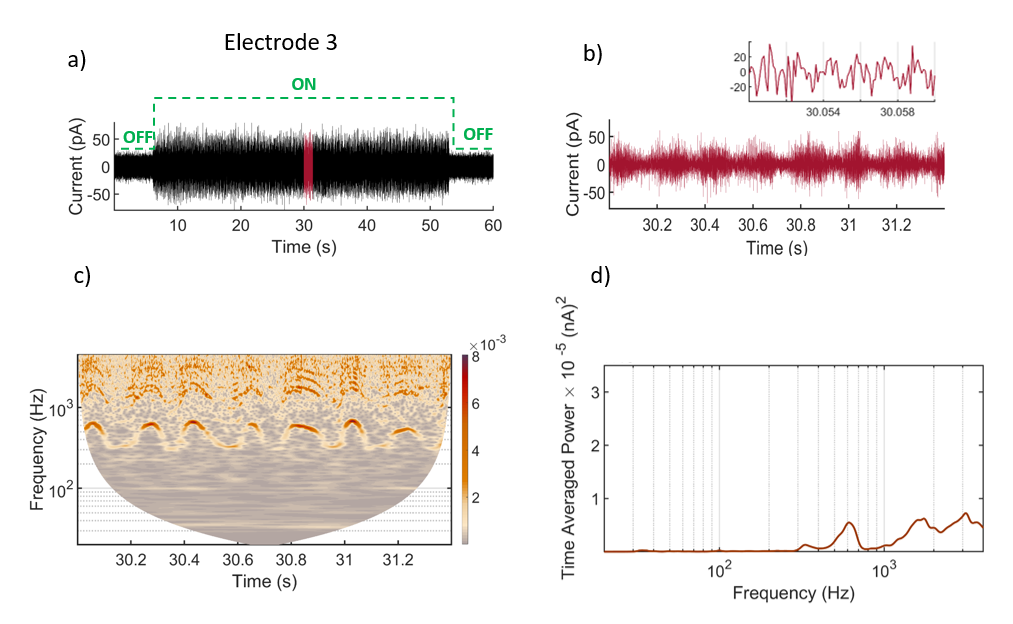}
\caption{(continued) Signals from electrodes E2 and E3, and their spectral analyses. (a) Signals were recorded for 60 seconds at a pressing voltage of $V\,=\,4.16$~V and electron density $n_e$\,=\,1.4$\times {10^{6}}$~\si{cm^{-2}}. The MW (139~GHz) is switched On/Off where indicated. (b) Enlarged reddened 1.4\,s part of (a) used for analysis, where the inset shows a further zoom. (c) Wavelet transform with frequency resolution~3~Hz. (d) Time-averaged wavelet power of signals in (b) showing the fundamental frequency and higher harmonics. The $y$-axis of (c) and the $x$-axes of (d) are logarithmic.}
\end{figure}  

\addtocounter{figure}{-1}
\clearpage 

\begin{figure*}
%\centering
\includegraphics[scale=0.9]{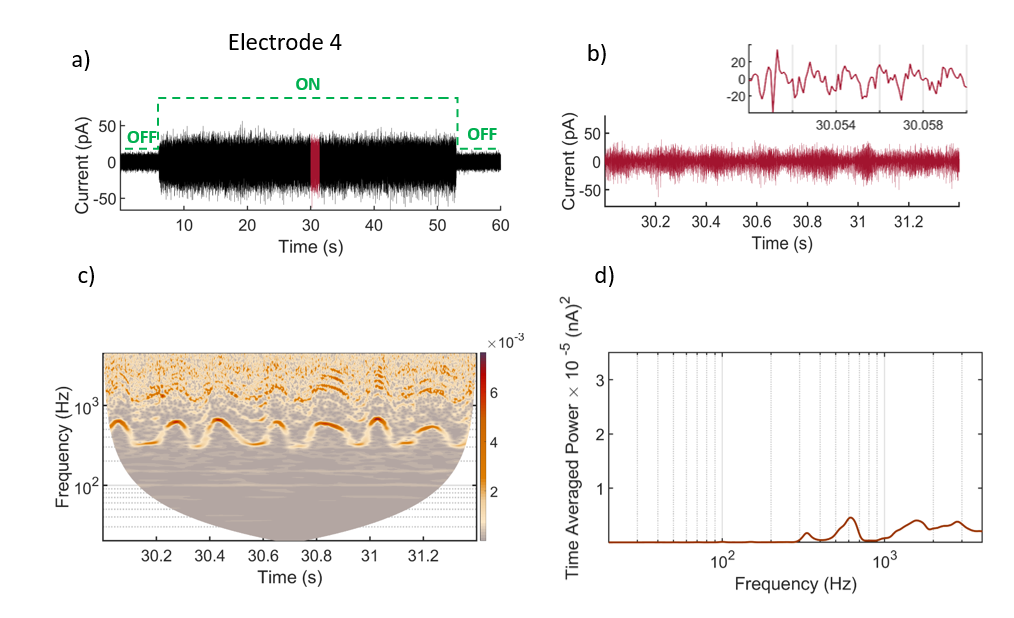}[c]
\caption{(continued) Signal from electrode E4, and its spectral analysis. (a) The signal was recorded for 60 seconds at a pressing voltage of $V\,=\,4.16$~V and electron density $n_e$\,=\,1.4$\times {10^{6}}$~\si{cm^{-2}}. The MW (139~GHz) is switched On/Off where indicated. (b) Enlarged reddened 1.4~s part of (a) used for analysis, where the inset shows a further zoom. (c) Wavelet transform with frequency resolution~3~Hz. (d) Time-averaged wavelet power of signals in (b) showing the fundamental frequency and higher harmonics. The $y$-axis of (c) and the $x$-axes of (d) are logarithmic.}
\label{fig:16}
\end{figure*}

\clearpage
\underbar{\bf Pressing voltage $V\,=\,4.17$~V}

\begin{figure}[b!]
\centering
\includegraphics[scale=0.87]{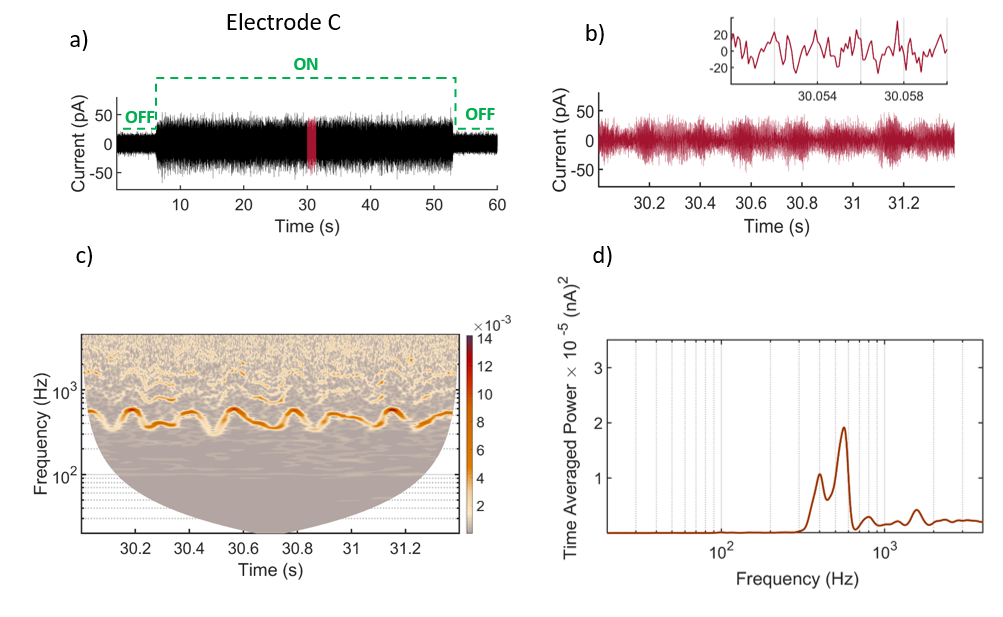}
\includegraphics[scale=0.87]{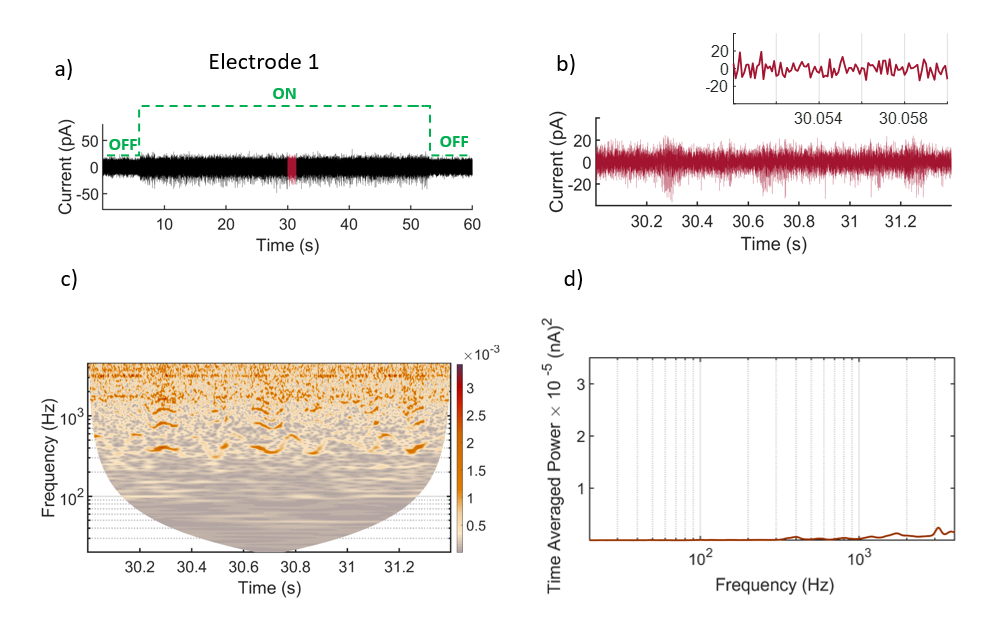}
\caption{ Signals from electrodes C and E1, and their spectral analyses. (a) Signals were recorded for 60 seconds at a pressing voltage of $V\,=\,4.17$~V and electron density $n_e$\,=\,1.4$\times {10^{6}}$~\si{cm^{-2}}. The MW (139~GHz) is switched On/Off where indicated. (b) Enlarged reddened 1.4~s part of (a) used for analysis, where the inset shows a further zoom. (c) Wavelet transform with frequency resolution~3~Hz. (d) Time-averaged wavelet power of signals in (b) showing the fundamental frequency and higher harmonics. The $y$-axis of (c) and the $x$-axes of (d) are logarithmic.}
\end{figure}

\addtocounter{figure}{-1}
\clearpage 

\begin{figure}[b!]
\centering
\includegraphics[scale=0.9]{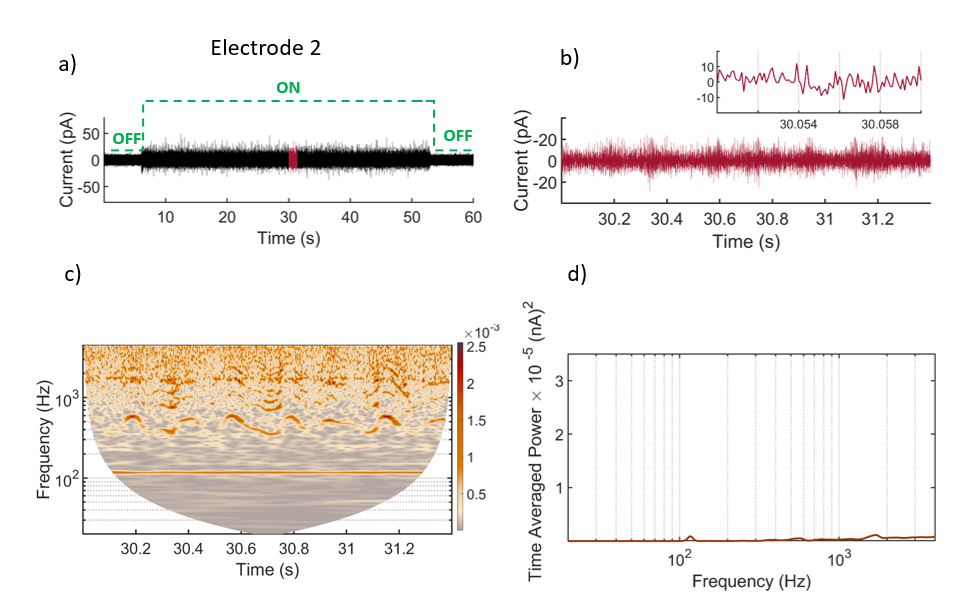}
\includegraphics[scale=0.9]{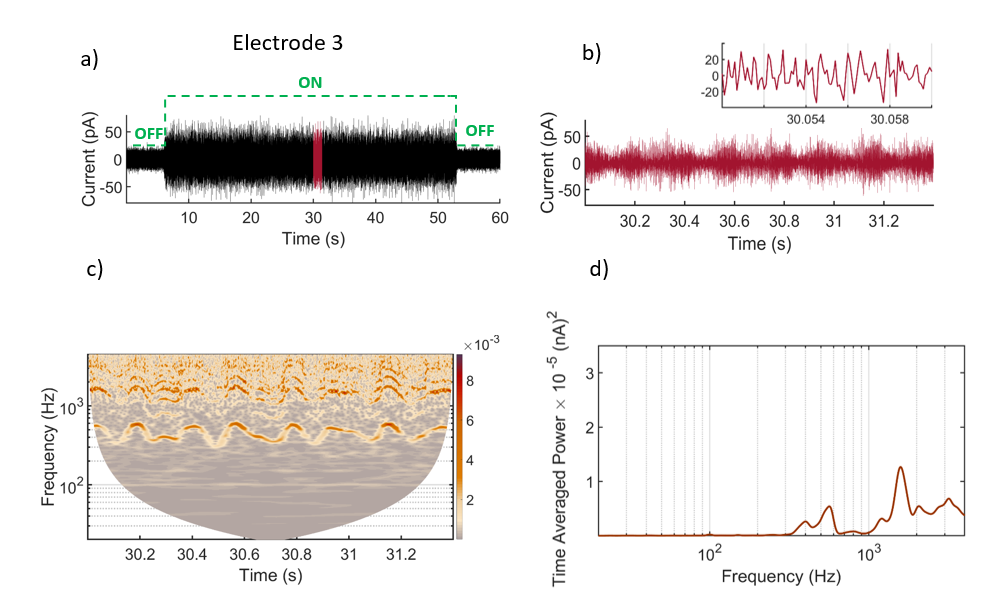}
\caption{(continued) Signals from electrodes E2 and E3, and their spectral analyses. (a) Signals were recorded for 60 seconds at a pressing voltage of $V\,=\,4.17$~V and electron density $n_e$\,=\,1.4$\times {10^{6}}$~\si{cm^{-2}}. The MW (139~GHz) is switched On/Off where indicated. (b) Enlarged reddened 1.4~s part of (a) used for analysis, where the inset shows a further zoom. (c) Wavelet transform with frequency resolution~3~Hz. (d) Time-averaged wavelet power of signals in (b) showing the fundamental frequency and higher harmonics. The $y$-axis of (c) and the $x$-axes of (d) are logarithmic.}
\end{figure}  

\addtocounter{figure}{-1}
\clearpage 

\begin{figure*}
%\centering
\includegraphics[scale=0.9]{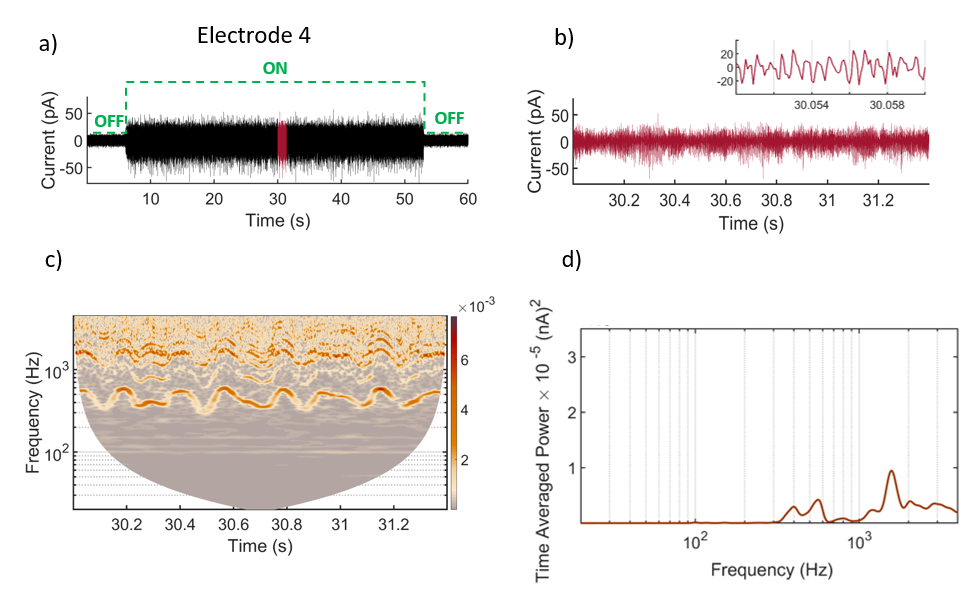}
\caption{(continued) Signal from electrode E4, and its spectral analysis. (a) The signal was recorded for 60 seconds at a pressing voltage of $V\,=\,4.17$~V and electron density $n_e$\,=\,1.4$\times {10^{6}}$~\si{cm^{-2}}. The MW (139~GHz) is switched On/Off where indicated. (b) Enlarged reddened 1.4~s part of (a) used for analysis, where the inset shows a further zoom. (c) Wavelet transform with frequency resolution~3~Hz. (d) Time-averaged wavelet power of signals in (b) showing the fundamental frequency and higher harmonics. The $y$-axis of (c) and the $x$-axes of (d) are logarithmic.}
\label{fig:16}
\end{figure*}

\clearpage
\underbar{\bf Pressing voltage $V\,=\,4.18$~V}

\begin{figure}[b!]
\centering
\includegraphics[scale=0.87]{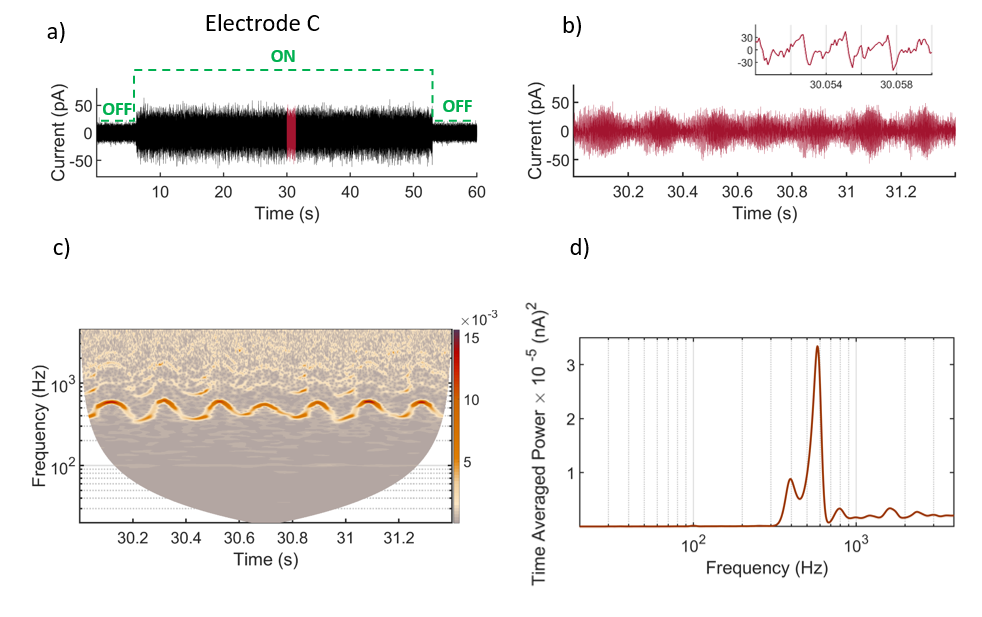}
\includegraphics[scale=0.87]{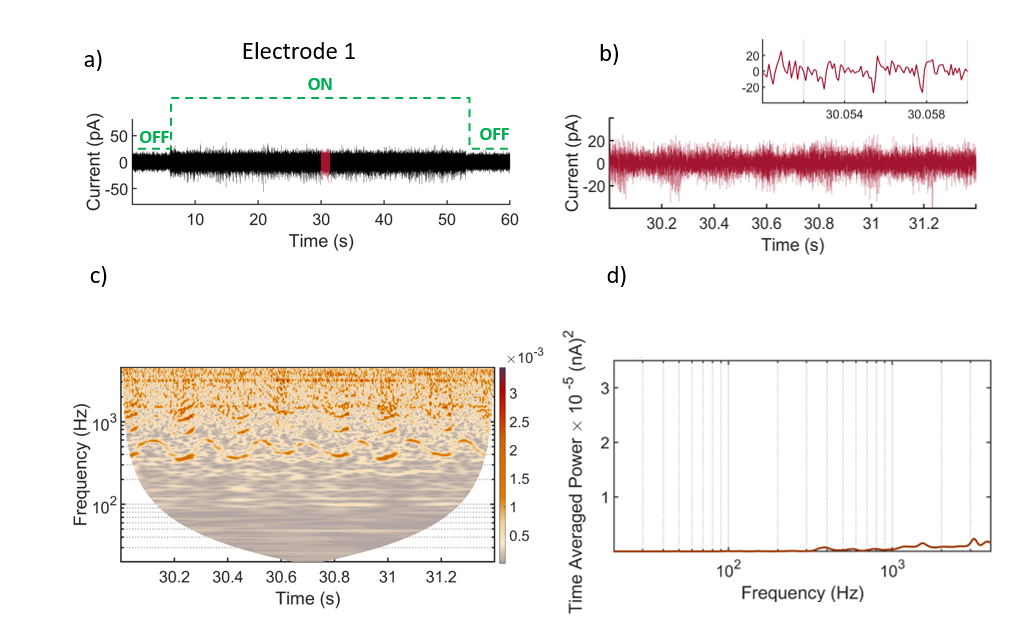}
\caption{ Signals from electrodes C and E1, and their spectral analyses. (a) Signals were recorded for 60 seconds at a pressing voltage of $V\,=\,4.18$~V and electron density $n_e$\,=\,1.4$\times {10^{6}}$~\si{cm^{-2}}. The MW (139~GHz) is switched On/Off where indicated. (b) Enlarged reddened 1.4~s part of (a) used for analysis, where the inset shows a further zoom. (c) Wavelet transform with frequency resolution~3~Hz. (d) Time-averaged wavelet power of signals in (b) showing the fundamental frequency and higher harmonics. The $y$-axis of (c) and the $x$-axes of (d) are logarithmic.}
\end{figure}

\addtocounter{figure}{-1}
\clearpage 

\begin{figure}[b!]
\centering
\includegraphics[scale=0.9]{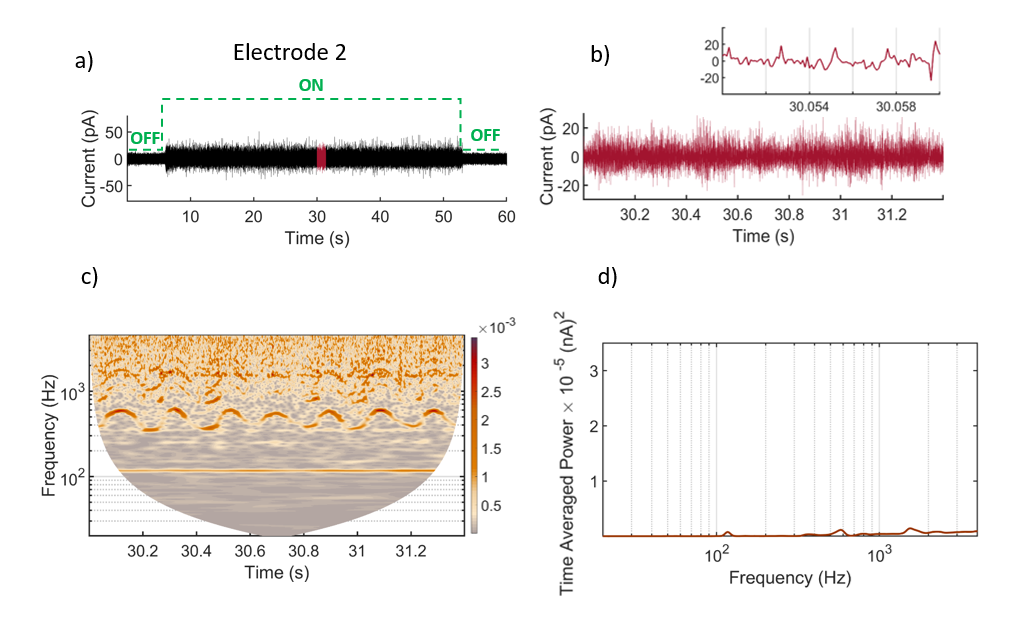}
\includegraphics[scale=0.9]{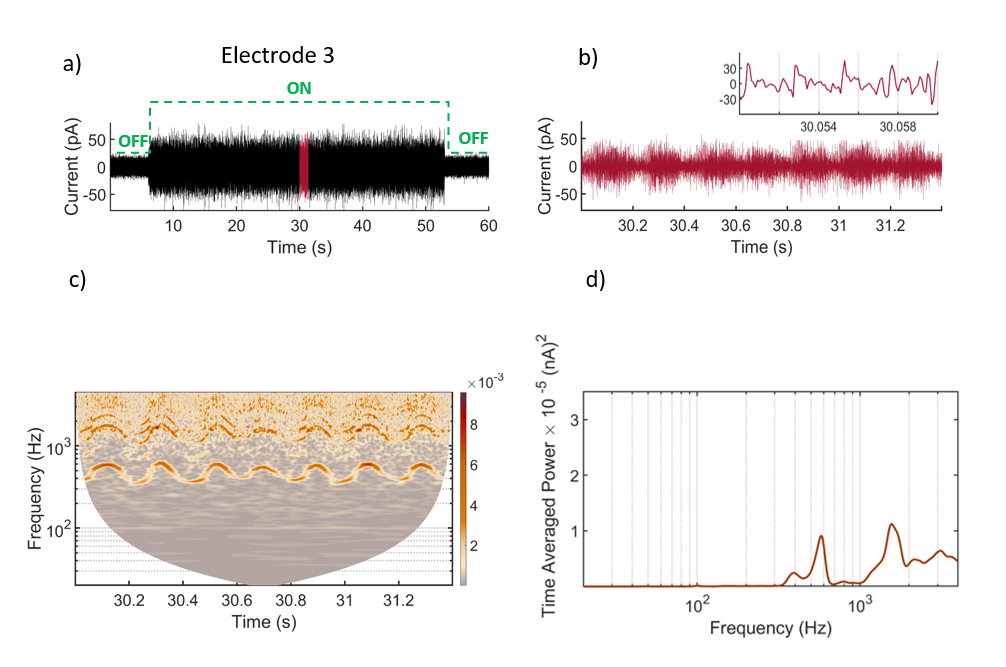}
\caption{(continued) Signals from electrodes E2 and E3, and their spectral analyses. (a) Signals were recorded for 60 seconds at a pressing voltage of $V\,=\,4.18$~V and electron density $n_e$\,=\,1.4$\times {10^{6}}$~\si{cm^{-2}}. The MW (139~GHz) is switched On/Off where indicated. (b) Enlarged reddened 1.4~s part of (a) used for analysis, where the inset shows a further zoom. (c) Wavelet transform with frequency resolution~3~Hz. (d) Time-averaged wavelet power of signals in (b) showing the fundamental frequency and higher harmonics. The $y$-axis of (c) and the $x$-axes of (d) are logarithmic.}
\end{figure}  

\addtocounter{figure}{-1}
\clearpage 

\begin{figure*}
%\centering
\includegraphics[scale=0.9]{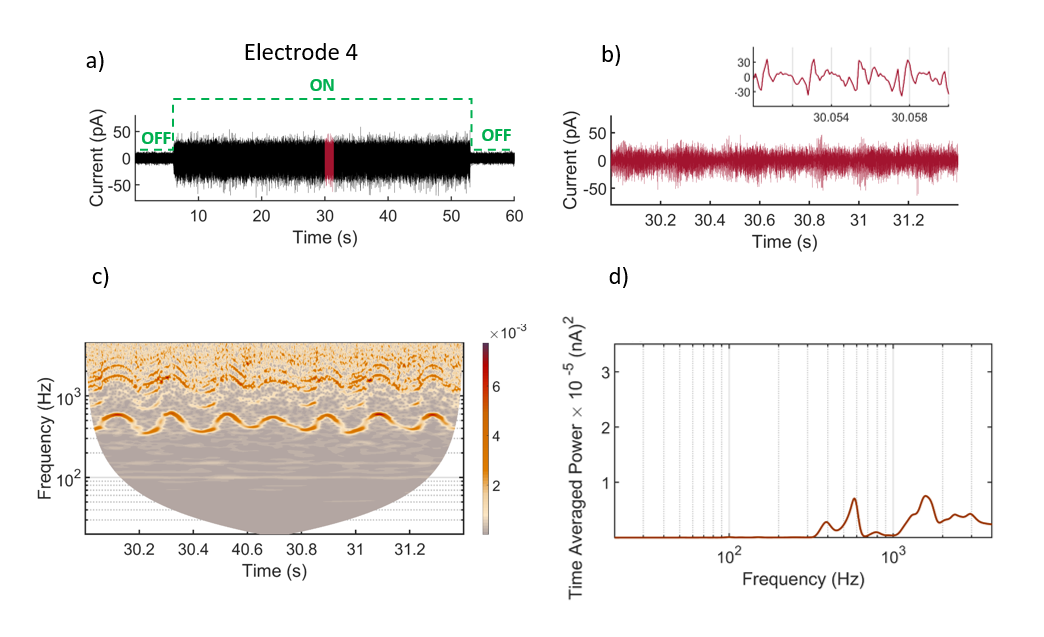}[c]
\caption{(continued) Signal from electrode E4, and its spectral analysis. (a) The signal was recorded for 60 seconds at a pressing voltage of $V\,=\,4.18$~V and electron density $n_e$\,=\,1.4$\times {10^{6}}$~\si{cm^{-2}}. The MW (139~GHz) is switched On/Off where indicated. (b) Enlarged reddened 1.4~s part of (a) used for analysis, where the inset shows a further zoom. (c) Wavelet transform with frequency resolution~3~Hz. (d) Time-averaged wavelet power of signals in (b) showing the fundamental frequency and higher harmonics. The $y$-axis of (c) and the $x$-axes of (d) are logarithmic.}
\label{fig:16}
\end{figure*}

\clearpage
\underbar{\bf Pressing voltage $V\,=\,4.19$~V}

\begin{figure}[b!]
\centering
\includegraphics[scale=0.87]{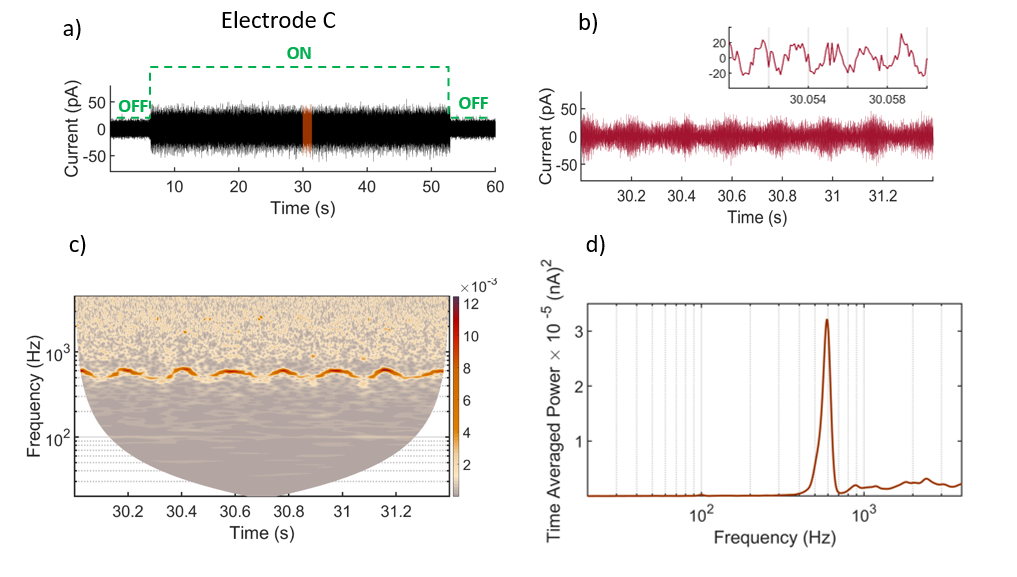}
\includegraphics[scale=0.87]{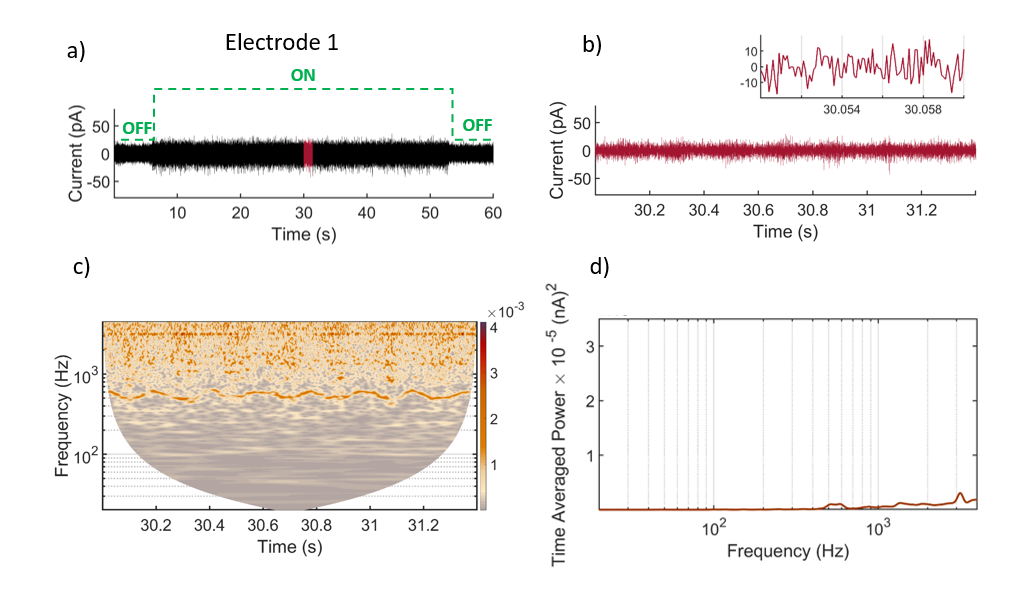}
\caption{ Signals from electrodes C and E1, and their spectral analyses. (a) Signals were recorded for 60 seconds at a pressing voltage of $V\,=\,4.19$~V and electron density $n_e$\,=\,1.4$\times {10^{6}}$~\si{cm^{-2}}. The MW (139~GHz) is switched On/Off where indicated. (b) Enlarged reddened 1.4~s part of (a) used for analysis, where the inset shows a further zoom. (c) Wavelet transform with frequency resolution~3~Hz. (d) Time-averaged wavelet power of signals in (b) showing the fundamental frequency and higher harmonics. The $y$-axis of (c) and the $x$-axes of (d) are logarithmic.}
\end{figure}

\addtocounter{figure}{-1}
\clearpage 

\begin{figure}[b!]
\centering
\includegraphics[scale=0.9]{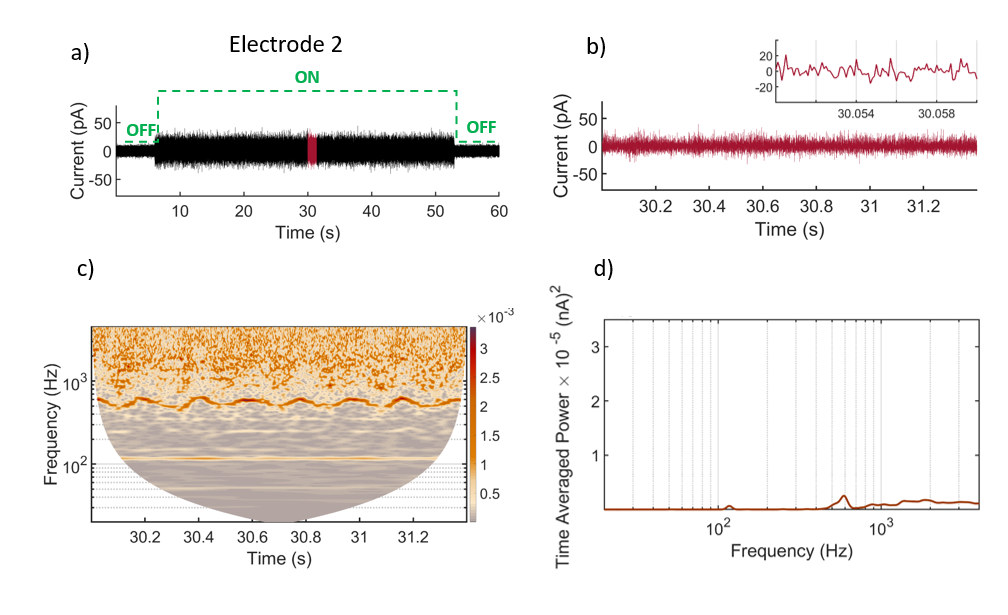}
\includegraphics[scale=0.9]{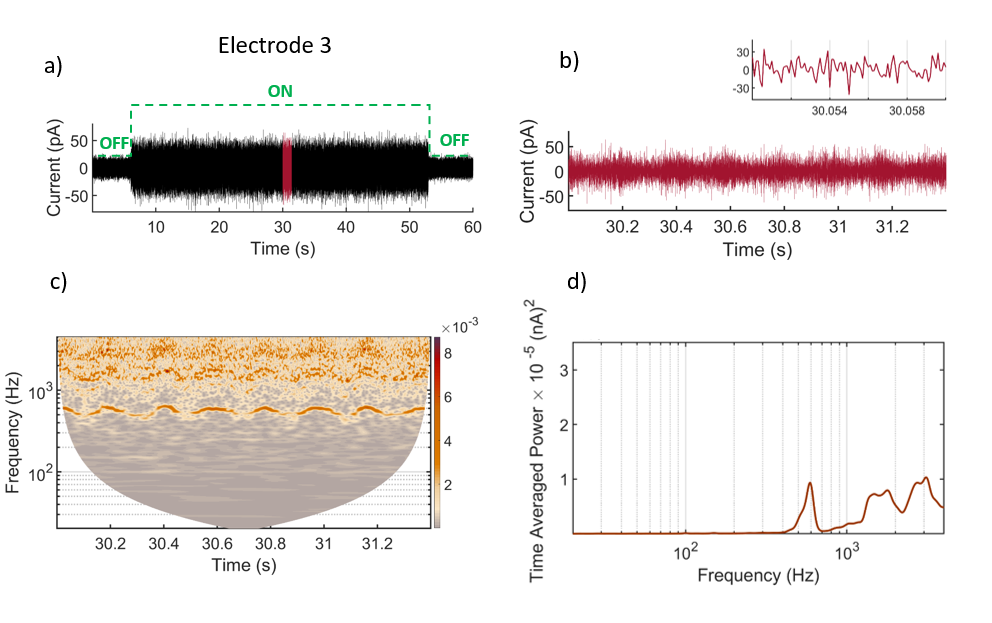}
\caption{(continued) Signals from electrodes E2 and E3, and their spectral analyses. (a) Signals were recorded for 60 seconds at a pressing voltage of $V\,=\,4.19$~V and electron density $n_e$\,=\,1.4$\times {10^{6}}$~\si{cm^{-2}}. The MW (139~GHz) is switched On/Off where indicated. (b) Enlarged reddened 1.4~s part of (a) used for analysis, where the inset shows a further zoom. (c) Wavelet transform with frequency resolution~3~Hz. (d) Time-averaged wavelet power of signals in (b) showing the fundamental frequency and higher harmonics. The $y$-axis of (c) and the $x$-axes of (d) are logarithmic.}
\end{figure}  

\addtocounter{figure}{-1}
\clearpage 

\begin{figure*}
%\centering
\includegraphics[scale=0.9]{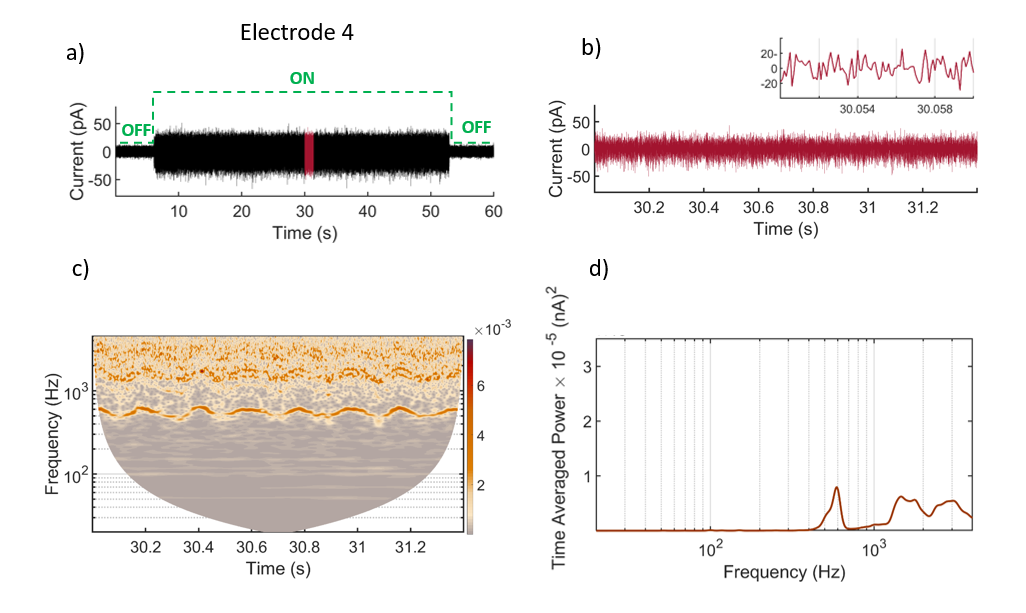}[c]
\caption{(continued) Signal from electrode E4, and its spectral analysis. (a) The signal was recorded for 60 seconds at a pressing voltage of $V\,=\,4.19$~V and electron density $n_e$\,=\,1.4$\times {10^{6}}$~\si{cm^{-2}}. The MW (139~GHz) is switched On/Off where indicated. (b) Enlarged reddened 1.4~s part of (a) used for analysis, where the inset shows a further zoom. (c) Wavelet transform with frequency resolution~3~Hz. (d) Time-averaged wavelet power of signals in (b) showing the fundamental frequency and higher harmonics. The $y$-axis of (c) and the $x$-axes of (d) are logarithmic.}
\label{fig:16}
\end{figure*}

\clearpage
\underbar{\bf Pressing voltage $V\,=\,4.20$~V}

\begin{figure}[b!]
\centering
\includegraphics[scale=0.87]{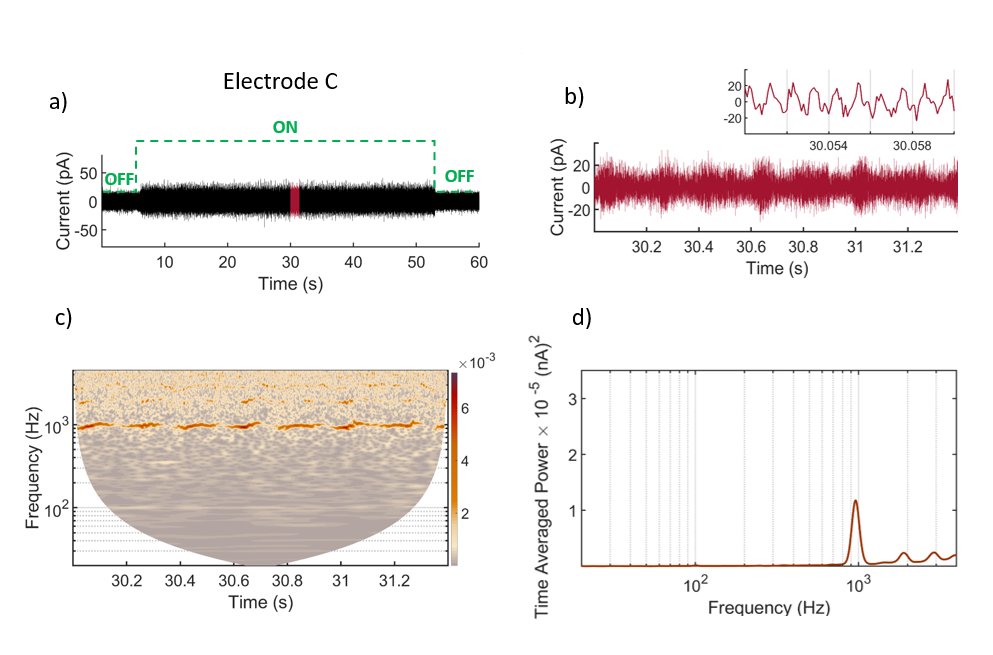}
\includegraphics[scale=0.87]{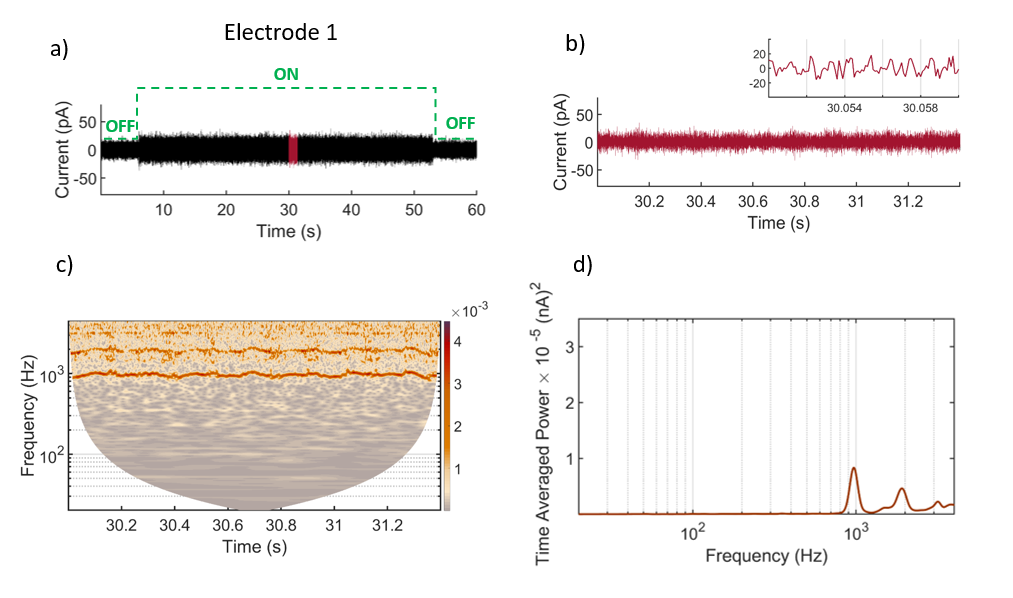}
\caption{ Signals from electrodes C and E1, and their spectral analyses. (a) Signals were recorded for 60 seconds at a pressing voltage of $V\,=\,4.20$~V and electron density $n_e$\,=\,1.4$\times {10^{6}}$~\si{cm^{-2}}. The MW (139~GHz) is switched On/Off where indicated. (b) Enlarged reddened 1.4~s part of (a) used for analysis, where the inset shows a further zoom. (c) Wavelet transform with frequency resolution~3~Hz. (d) Time-averaged wavelet power of signals in (b) showing the fundamental frequency and higher harmonics. The $y$-axis of (c) and the $x$-axes of (d) are logarithmic.}
\end{figure}

\addtocounter{figure}{-1}
\clearpage 

\begin{figure}[b!]
\centering
\includegraphics[scale=0.9]{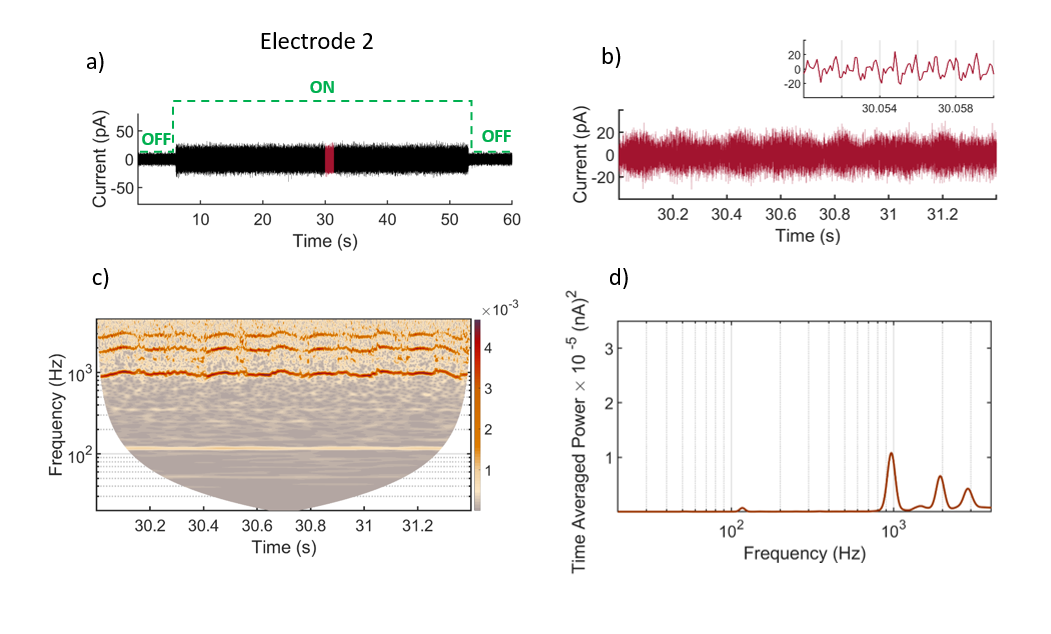}
\includegraphics[scale=0.9]{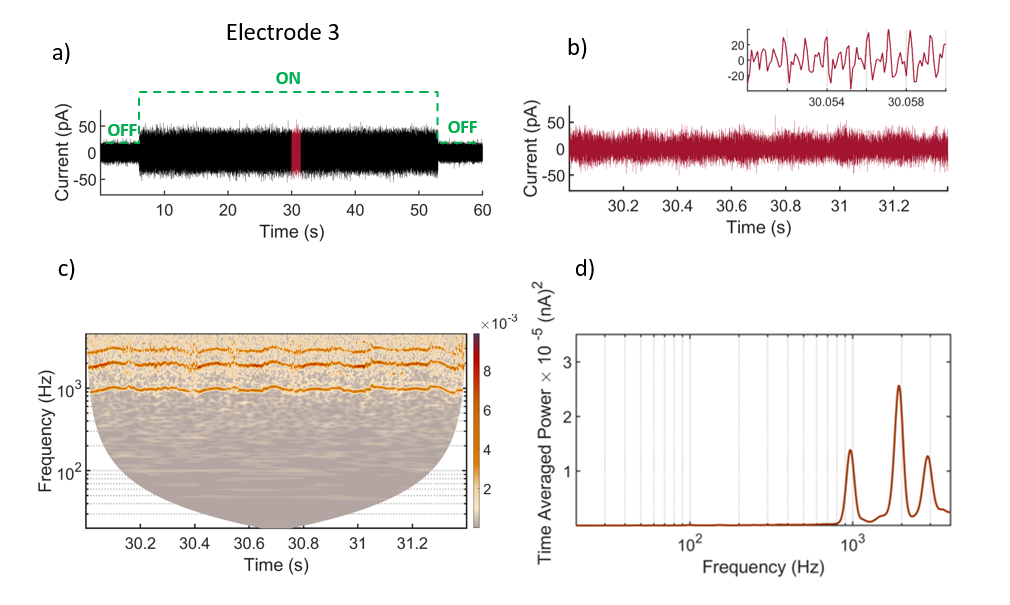}
\caption{(continued) Signals from electrodes E2 and E3, and their spectral analyses. (a) Signals were recorded for 60 seconds at a pressing voltage of $V\,=\,4.20$~V and electron density $n_e$\,=\,1.4$\times {10^{6}}$~\si{cm^{-2}}. The MW (139~GHz) is switched On/Off where indicated. (b) Enlarged reddened 1.4~s part of (a) used for analysis, where the inset shows a further zoom. (c) Wavelet transform with frequency resolution~3~Hz. (d) Time-averaged wavelet power of signals in (b) showing the fundamental frequency and higher harmonics. The $y$-axis of (c) and the $x$-axes of (d) are logarithmic.}
\end{figure}  

\addtocounter{figure}{-1}
\clearpage 

\begin{figure*}
%\centering
\includegraphics[scale=0.9]{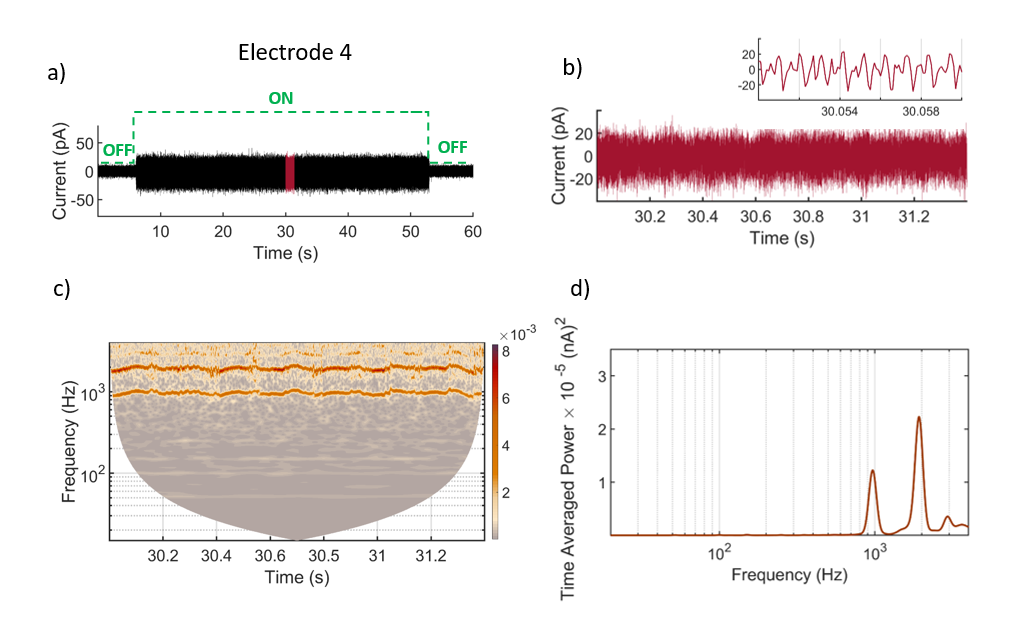}[c]
\caption{(continued) Signal from electrode E4, and its spectral analysis. (a) The signal was recorded for 60 seconds at a pressing voltage of $V\,=\, 4.20$~V and electron density $n_e$\,=\,1.4$\times {10^{6}}$~\si{cm^{-2}}. The MW (139~GHz) is switched On/Off where indicated. (b) Enlarged reddened 1.4~s part of (a) used for analysis, where the inset shows a further zoom. (c) Wavelet transform with frequency resolution~3~Hz. (d) Time-averaged wavelet power of signals in (b) showing the fundamental frequency and higher harmonics. The $y$-axis of (c) and the $x$-axes of (d) are logarithmic.}
\label{fig:16}
\end{figure*}

\clearpage
\underbar{\bf Pressing voltage $V\,=\,4.21$~V}

\begin{figure}[b!]
\centering
\includegraphics[scale=0.87]{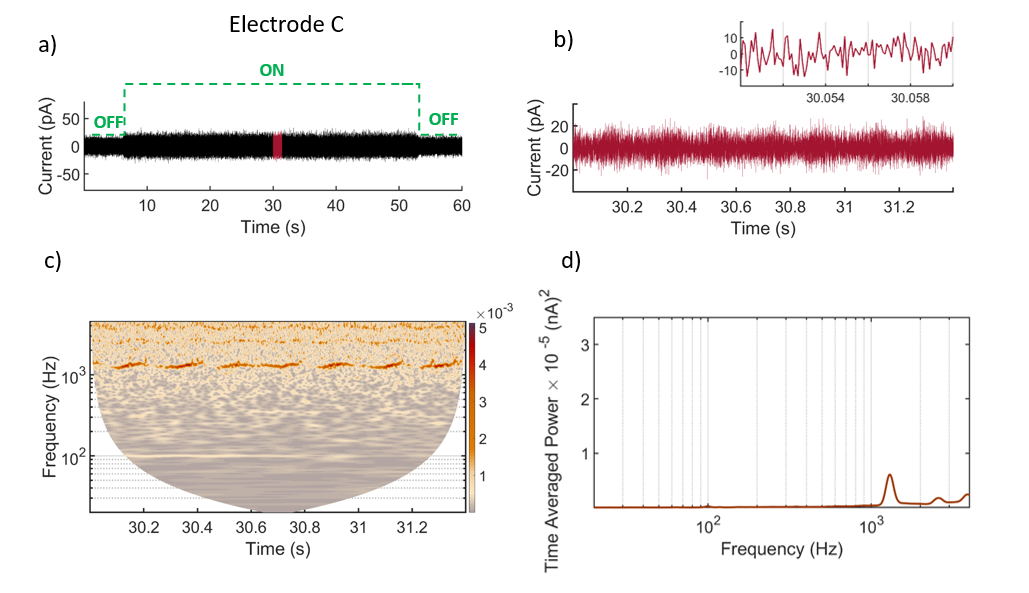}
\includegraphics[scale=0.87]{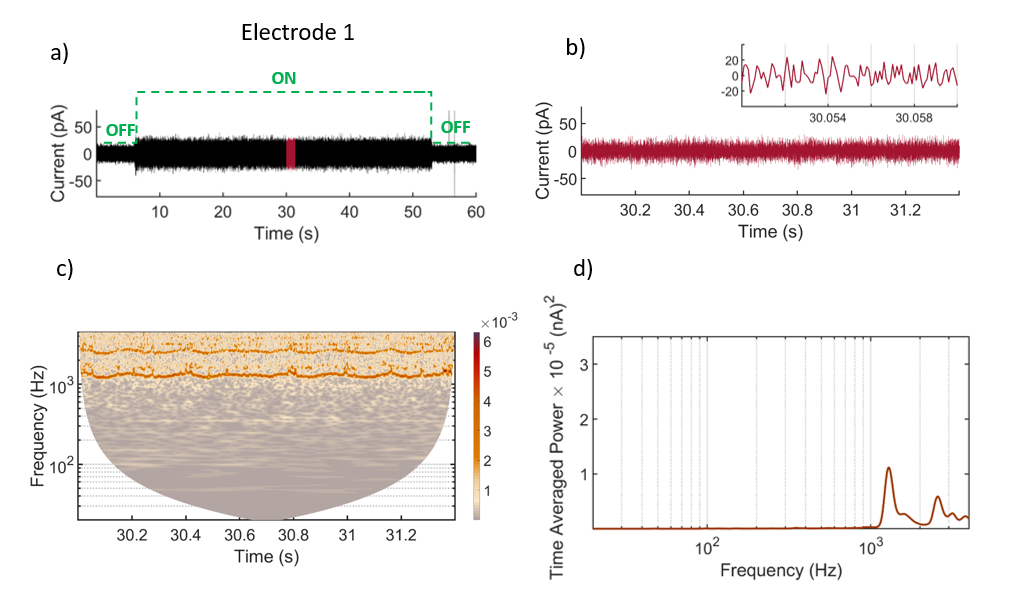}
\caption{ Signals from electrodes C and E1, and their spectral analyses. (a) Signals were recorded for 60 seconds at a pressing voltage of $V\,=\,4.21$~V and electron density $n_e$\,=\,1.4$\times {10^{6}}$~\si{cm^{-2}}. The MW (139~GHz) is switched On/Off where indicated. (b) Enlarged reddened 1.4~s part of (a) used for analysis, where the inset shows a further zoom. (c) Wavelet transform with frequency resolution~3~Hz. (d) Time-averaged wavelet power of signals in (b) showing the fundamental frequency and higher harmonics. The $y$-axis of (c) and the $x$-axes of (d) are logarithmic.}
\end{figure}

\addtocounter{figure}{-1}
\clearpage 

\begin{figure}[b!]
\centering
\includegraphics[scale=0.9]{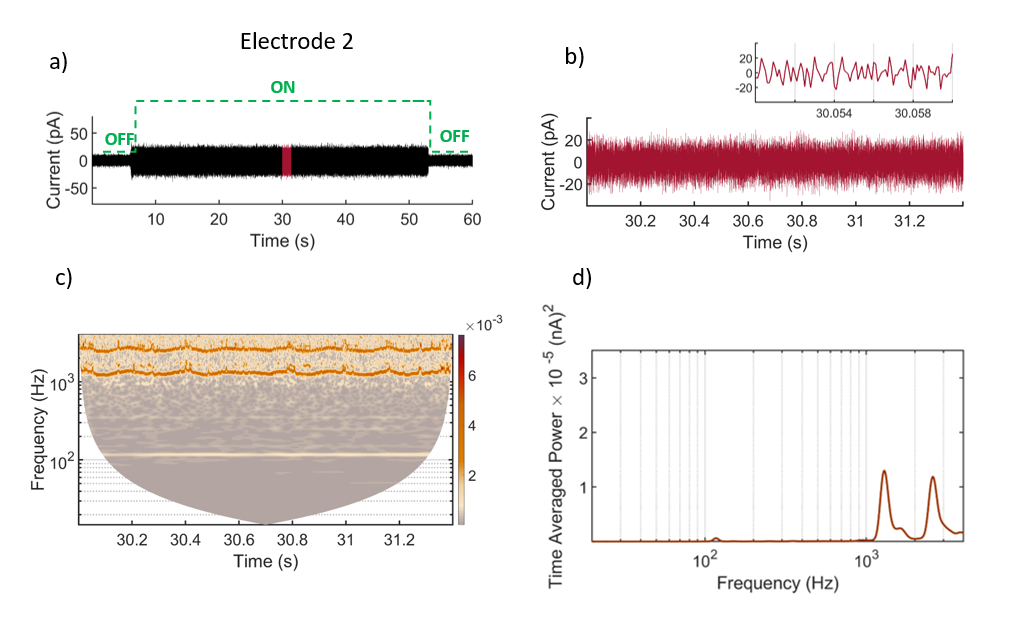}
\includegraphics[scale=0.9]{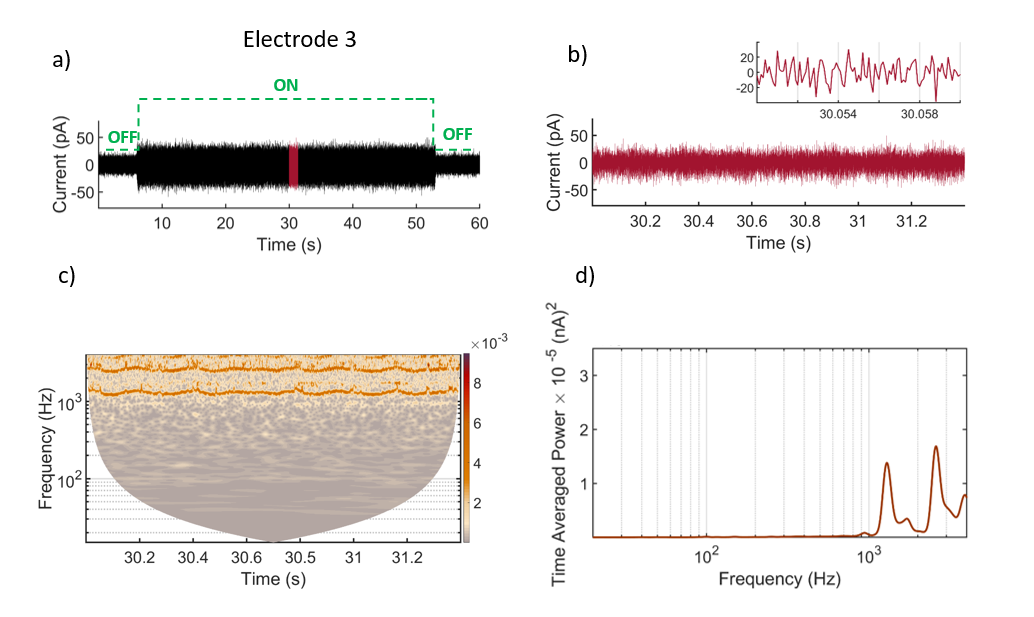}
\caption{(continued) Signals from electrodes E2 and E3, and their spectral analyses. (a) Signals were recorded for 60 seconds at a pressing voltage of $V\,=\,4.21$~V and electron density $n_e$\,=\,1.4$\times {10^{6}}$~\si{cm^{-2}}. The MW (139~GHz) is switched On/Off where indicated. (b) Enlarged reddened 1.4~s part of (a) used for analysis, where the inset shows a further zoom. (c) Wavelet transform with frequency resolution~3~Hz. (d) Time-averaged wavelet power of signals in (b) showing the fundamental frequency and higher harmonics. The $y$-axis of (c) and the $x$-axes of (d) are logarithmic.}
\end{figure}  

\addtocounter{figure}{-1}
\clearpage 

\begin{figure*}
%\centering
\includegraphics[scale=0.9]{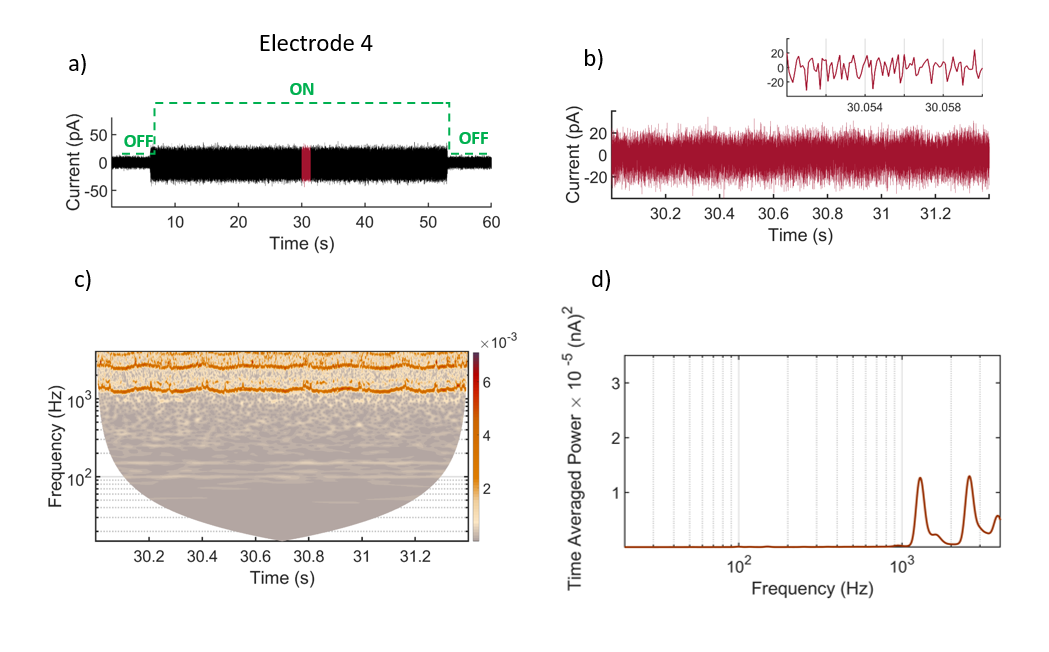}[c]
\caption{(continued) Signal from electrode E4, and its spectral analysis. (a) The signal was recorded for 60 seconds at a pressing voltage of $V\,=\,4.21$~V and electron density $n_e$\,=\,1.4$\times {10^{6}}$~\si{cm^{-2}}. The MW (139~GHz) is switched On/Off where indicated. (b) Enlarged reddened 1.4\,s part of (a) used for analysis, where the inset shows a further zoom. (c) Wavelet transform with frequency resolution~3~Hz. (d) Time-averaged wavelet power of signals in (b) showing the fundamental frequency and higher harmonics. The $y$-axis of (c) and the $x$-axes of (d) are logarithmic.}
\label{fig:16}
\end{figure*}

\clearpage
\underbar{\bf Pressing voltage $V\,=\,4.22$~V}

\begin{figure}[b!]
\centering
\includegraphics[scale=0.87]{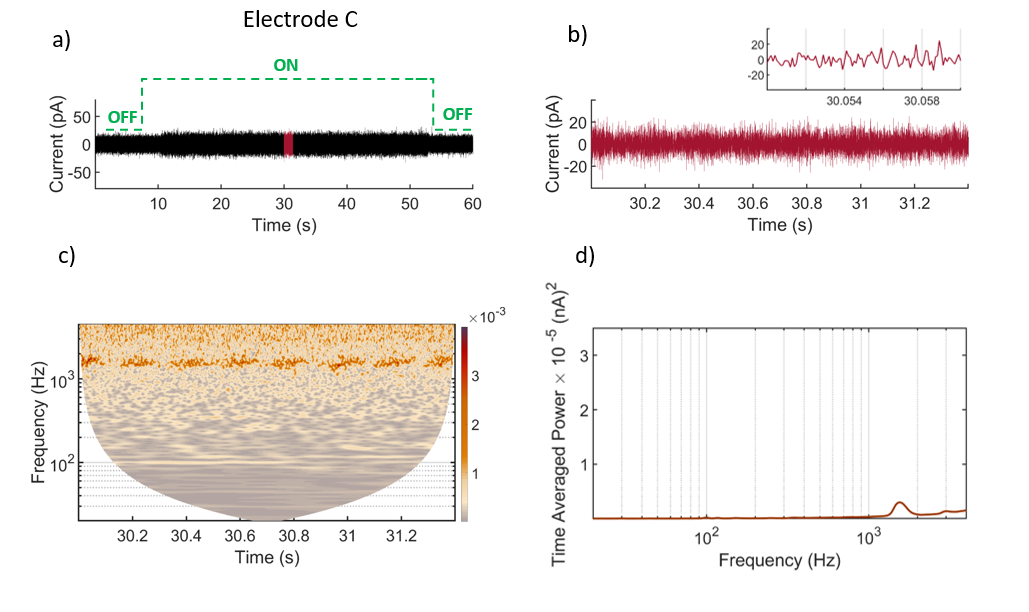}
\includegraphics[scale=0.87]{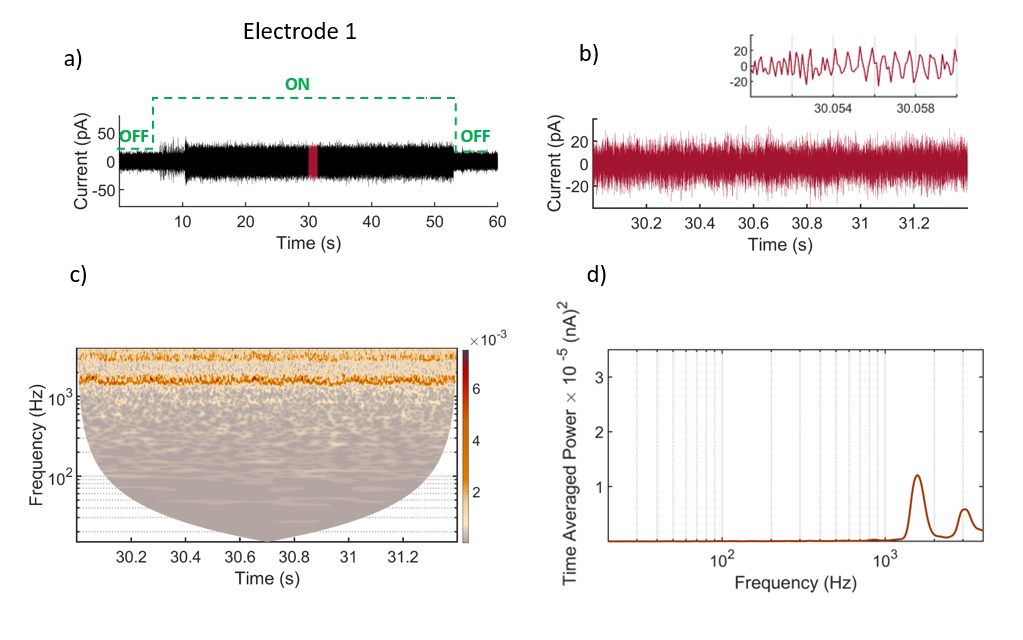}
\caption{ Signals from electrodes C and E1, and their spectral analyses. (a) Signals were recorded for 60 seconds at a pressing voltage of $V\,=\,4.22$~V and electron density $n_e$\,=\,1.4$\times {10^{6}}$~\si{cm^{-2}}. The MW (139~GHz) is switched On/Off where indicated. (b) Enlarged reddened 1.4~s part of (a) used for analysis, where the inset shows a further zoom. (c) Wavelet transform with frequency resolution~3~Hz. (d) Time-averaged wavelet power of signals in (b) showing the fundamental frequency and higher harmonics. The $y$-axis of (c) and the $x$-axes of (d) are logarithmic.}
\end{figure}

\addtocounter{figure}{-1}
\clearpage 

\begin{figure}[b!]
\centering
\includegraphics[scale=0.9]{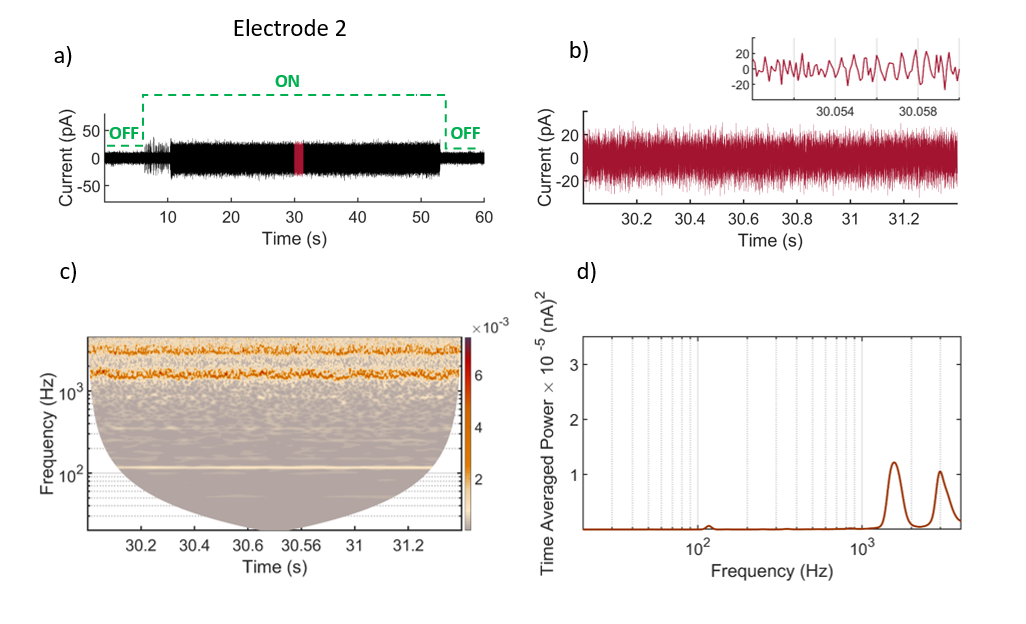}
\includegraphics[scale=0.9]{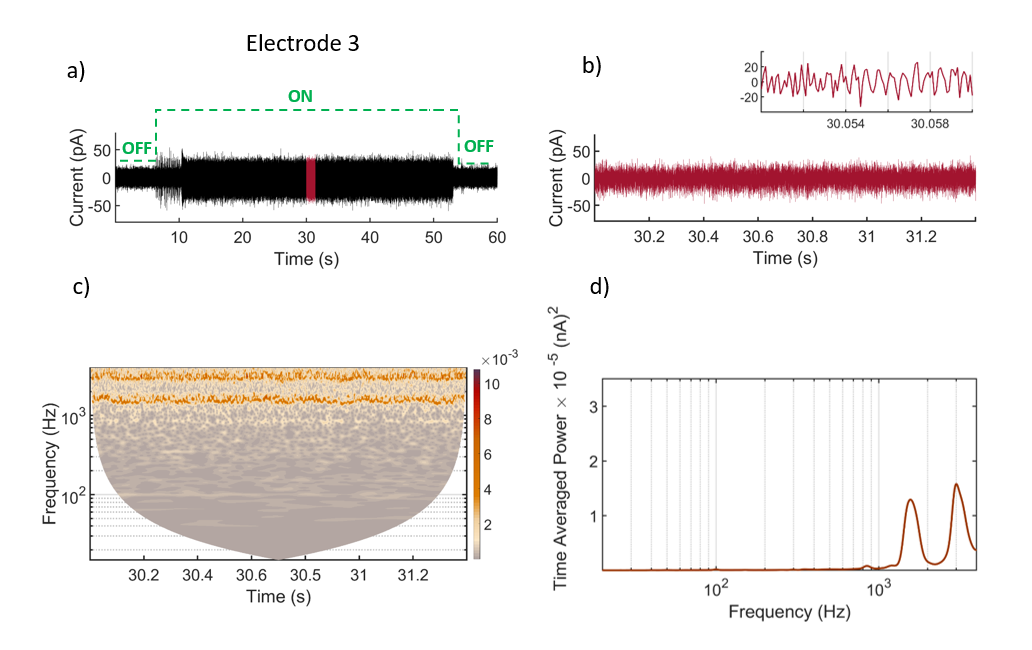}
\caption{(continued) Signals from electrodes E2 and E3, and their spectral analyses. (a) Signals were recorded for 60 seconds at a pressing voltage of $V\,=\,4.22$~V and electron density $n_e$\,=\,1.4$\times {10^{6}}$~\si{cm^{-2}}. The MW (139~GHz) is switched On/Off where indicated. (b) Enlarged reddened 1.4~s part of (a) used for analysis, where the inset shows a further zoom. (c) Wavelet transform with frequency resolution~3~Hz. (d) Time-averaged wavelet power of signals in (b) showing the fundamental frequency and higher harmonics. The $y$-axis of (c) and the $x$-axes of (d) are logarithmic.}
\end{figure}  

\addtocounter{figure}{-1}
\clearpage 

\begin{figure*}
%\centering
\includegraphics[scale=0.9]{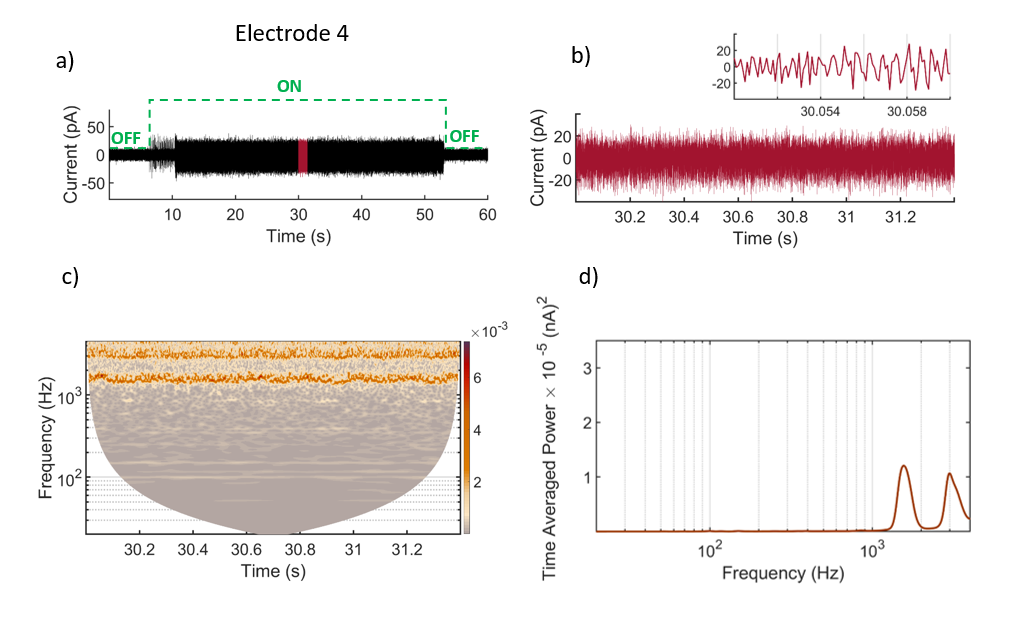}[c]
\caption{(continued) Signal from electrode E4, and its spectral analysis. (a) The signal was recorded for 60 seconds at a pressing voltage of $V\,=\,4.22$~V and electron density $n_e$\,=\,1.4$\times {10^{6}}$~\si{cm^{-2}}. The MW (139~GHz) is switched On/Off where indicated. (b) Enlarged reddened 1.4~s part of (a) used for analysis, where the inset shows a further zoom. (c) Wavelet transform with frequency resolution~3~Hz. (d) Time-averaged wavelet power of signals in (b) showing the fundamental frequency and higher harmonics. The $y$-axis of (c) and the $x$-axes of (d) are logarithmic.}
\label{fig:22}
\end{figure*}

\clearpage
%%%%%%%%%%%%%%%%%%%%%%%%%%%%%

\centerline{\bf B. HIGH ELECTRON DENSITY}
\vspace{0.3cm}

\underbar{\bf Pressing voltage $V\,=\,4.16$~V}

\begin{figure}[b!]
\centering
\includegraphics[scale=0.80]{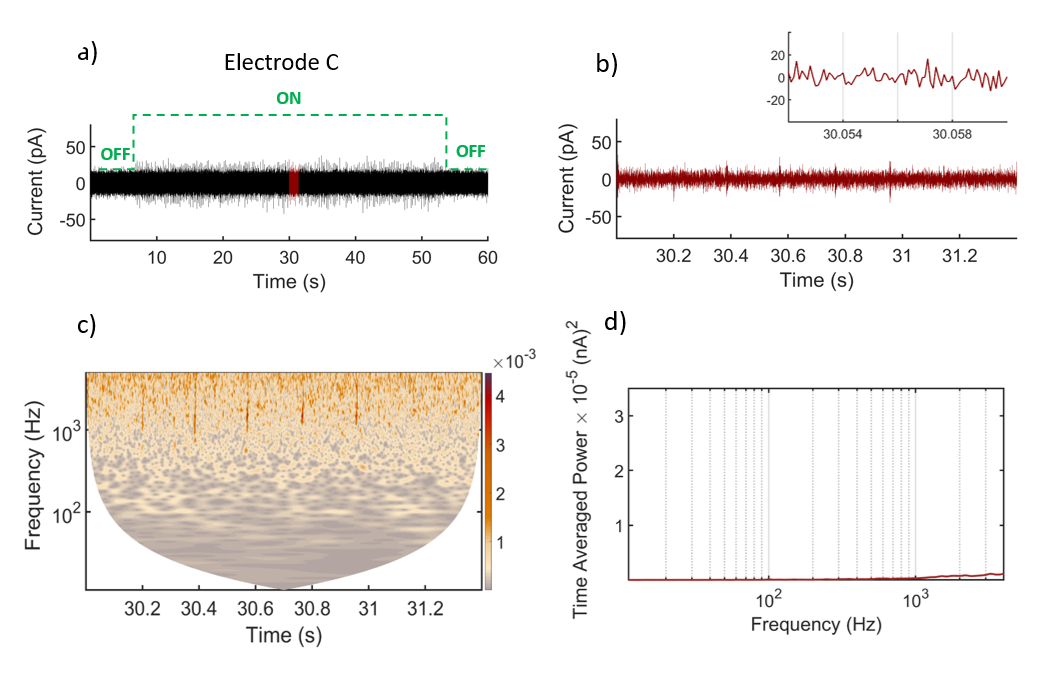}
\includegraphics[scale=0.80]{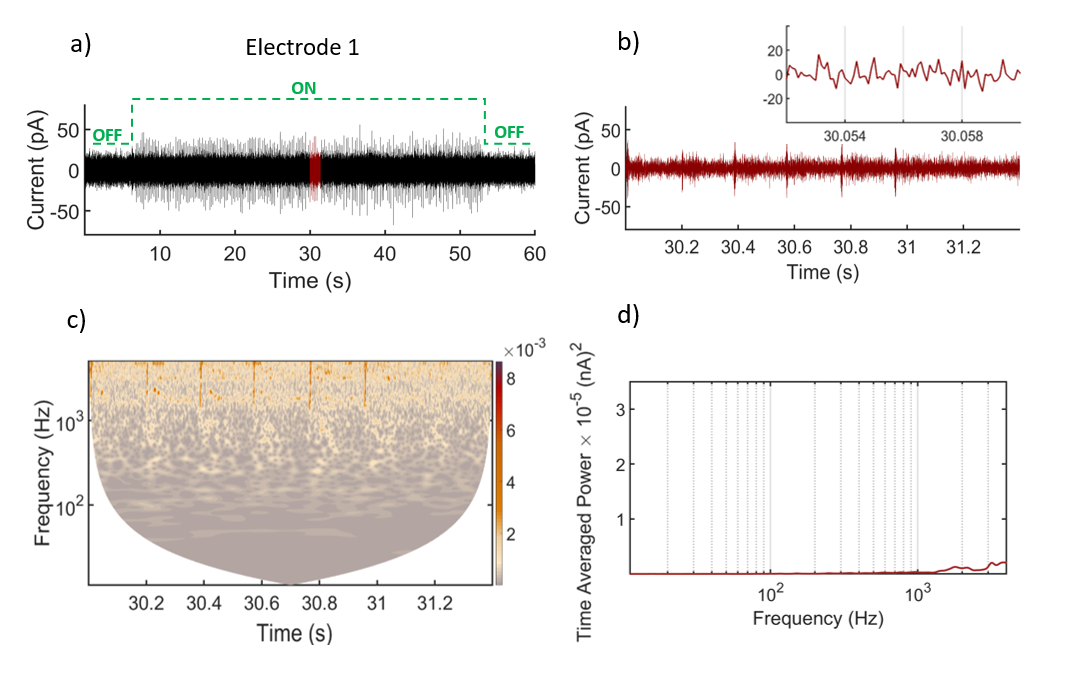}
\caption{ Signals from electrodes C and E1, and their spectral analyses. (a) Signals were recorded for 60 seconds at a pressing voltage of $V\,=\,4.16$\,V and electron density $n_e=2.2\times {10^{6}}$~\si{cm^{-2}}. The MW (139~GHz) is switched On/Off where indicated. (b) Enlarged reddened 1.4~s part of (a) used for analysis, where the inset shows a further zoom. (c) Wavelet transform with frequency resolution~3~Hz. (d) Time-averaged wavelet power of signals in (b) showing the fundamental frequency and higher harmonics. The $y$-axis of (c) and the $x$-axes of (d) are logarithmic.}
\label{fig:216}
\end{figure}

%\addtocounter{figure}{-1}
\clearpage 

\begin{figure}[b!]
%\centering
\includegraphics[scale=0.84]{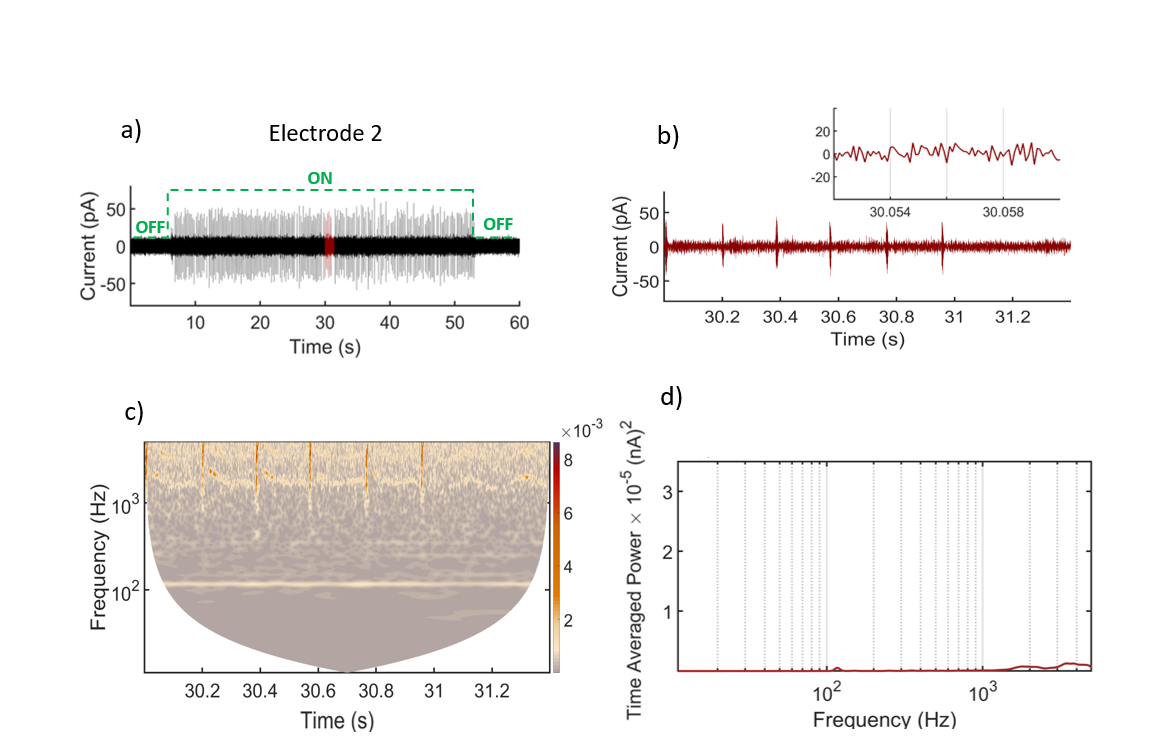}
\includegraphics[scale=0.85]{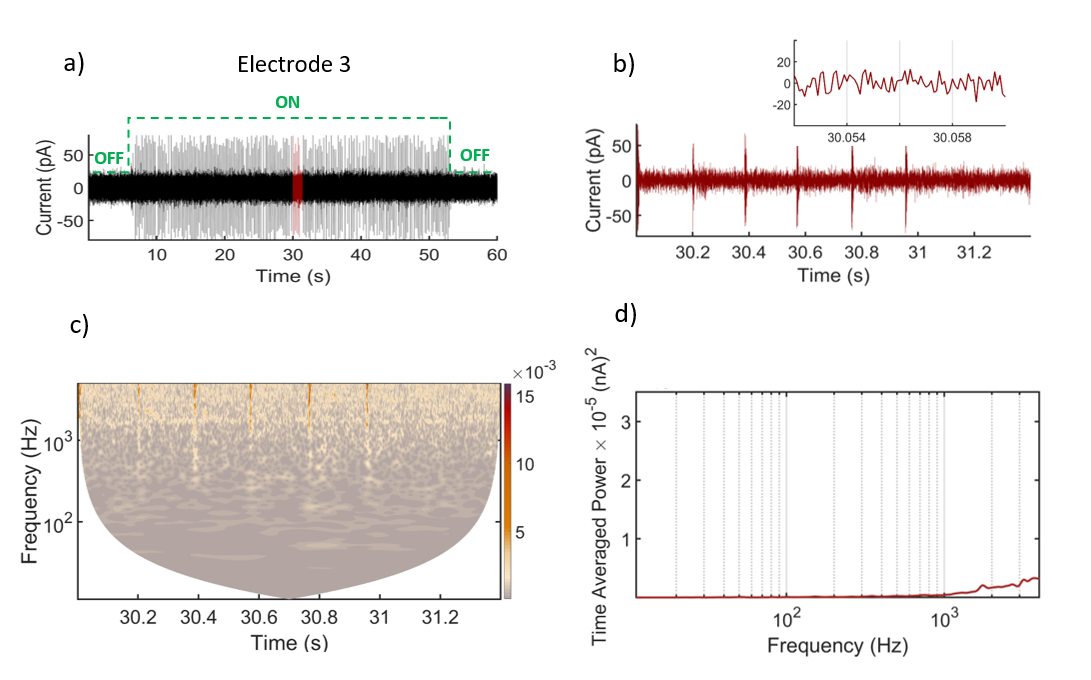}
\caption{(continued) Signals from electrodes E2 and E3, and their spectral analyses. (a) Signals were recorded for 60 seconds at a pressing voltage of $V$\,=\,4.16~V and electron density $n_e$\,=\,2.2$\times {10^{6}}$~$\si{cm^{-2}}$. The MW (139~GHz) is switched On/Off where indicated. (b) Enlarged reddened 1.4~s part of (a) used for analysis, where the inset shows a further zoom. (c) Wavelet transform with frequency resolution~3~Hz. (d) Time-averaged wavelet power of signals in (b) showing the fundamental frequency and higher harmonics. The $y$-axis of (c) and the $x$-axes of (d) are logarithmic.}
\end{figure}  

%\addtocounter{figure}{-1}
\clearpage 

\begin{figure*}
%\centering
\includegraphics[scale=0.87]{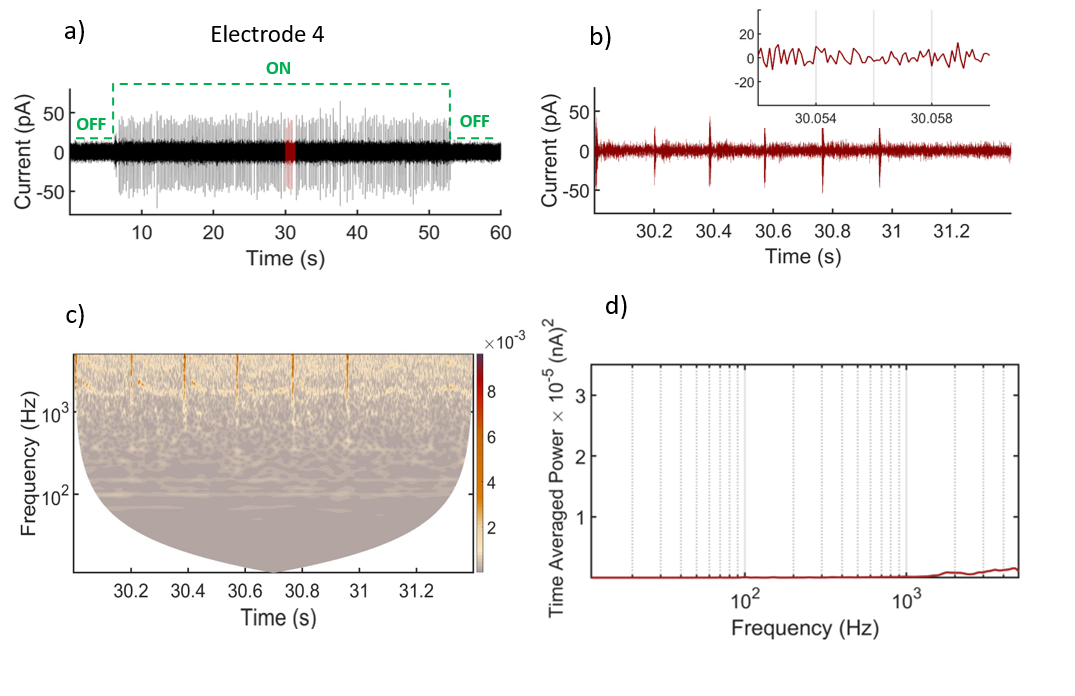}[c]
\caption{(continued) Signal from electrode E4, and its spectral analysis. (a) The signal was recorded for 60 seconds at a pressing voltage of $V\,=\,4.16$~V and electron density $n_e\,=\,2.2\times {10^{6}}$~\si{cm^{-2}}. The MW (139~GHz) is switched On/Off where indicated. (b) Enlarged reddened 1.4~s part of (a) used for analysis, where the inset shows a further zoom. (c) Wavelet transform with frequency resolution~3~Hz. (d) Time-averaged wavelet power of signals in (b) showing the fundamental frequency and higher harmonics. The $y$-axis of (c) and the $x$-axes of (d) are logarithmic.}
\label{fig:16}
\end{figure*}

\clearpage
\underbar{\bf Pressing voltage $V\,=\,4.17$~V}

\begin{figure}[b!]
\centering
\includegraphics[scale=0.85]{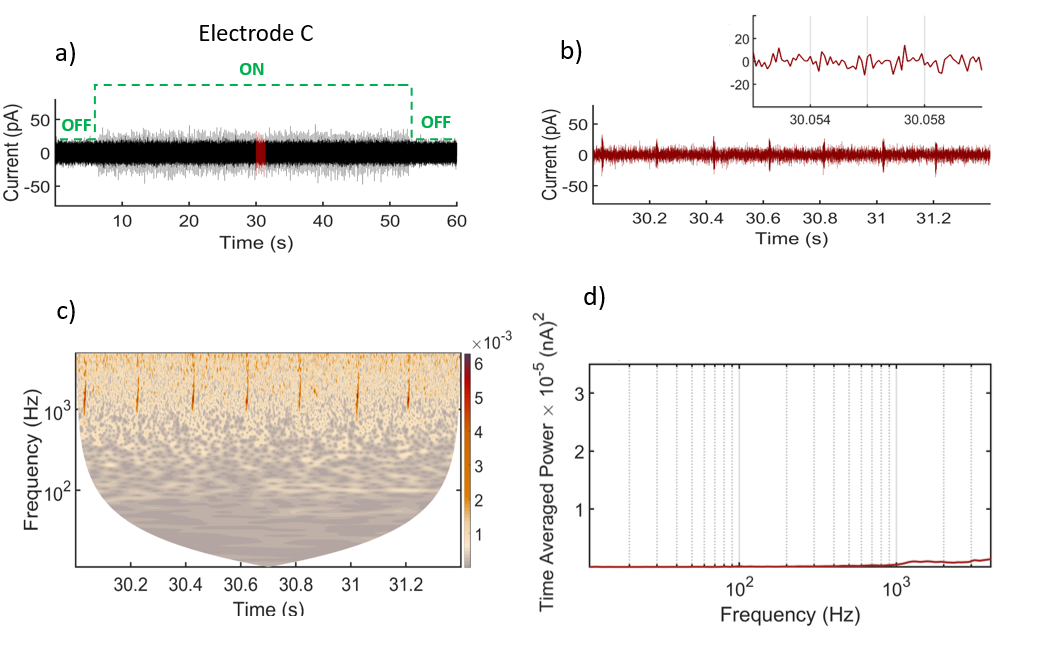}
\includegraphics[scale=0.85]{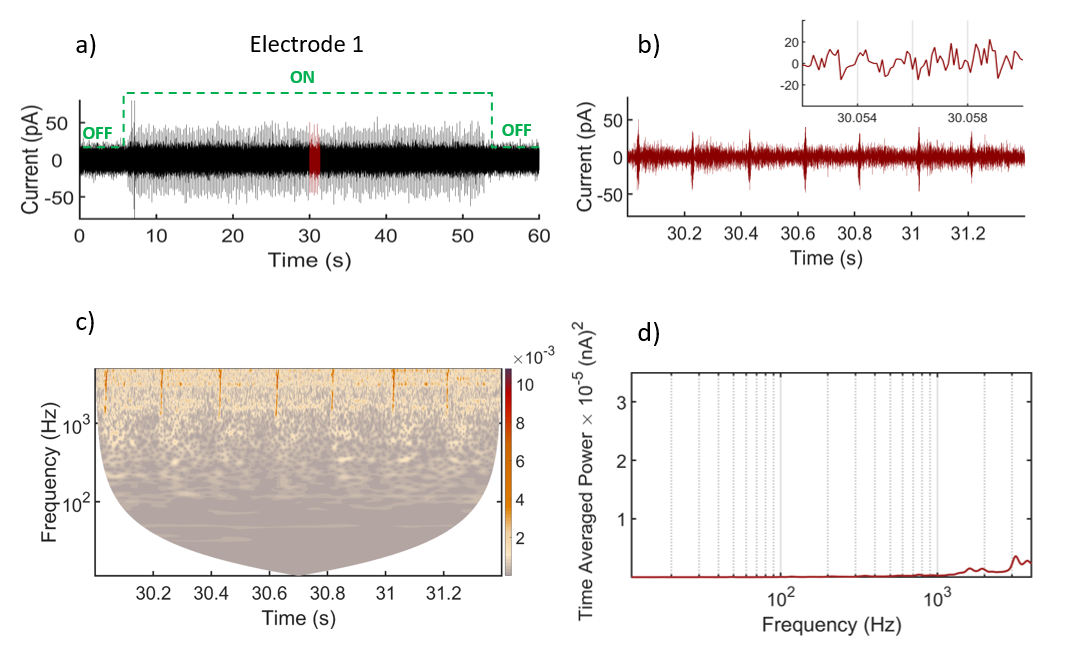}
\caption{ Signals from electrodes C and E1, and their spectral analyses. (a) Signals were recorded for 60 seconds at a pressing voltage of $V\,=\,4.17$~V and electron density $n_e\,=\,2.2\times {10^{6}}$~\si{cm^{-2}}. The MW (139~GHz) is switched On/Off where indicated. (b) Enlarged reddened 1.4~s part of (a) used for analysis, where the inset shows a further zoom. (c) Wavelet transform with frequency resolution~3~Hz. (d) Time-averaged wavelet power of signals in (b) showing the fundamental frequency and higher harmonics. The $y$-axis of (c) and the $x$-axes of (d) are logarithmic.}
\end{figure}

\addtocounter{figure}{-1}
\clearpage 

\begin{figure}[b!]
\centering
\includegraphics[scale=0.87]{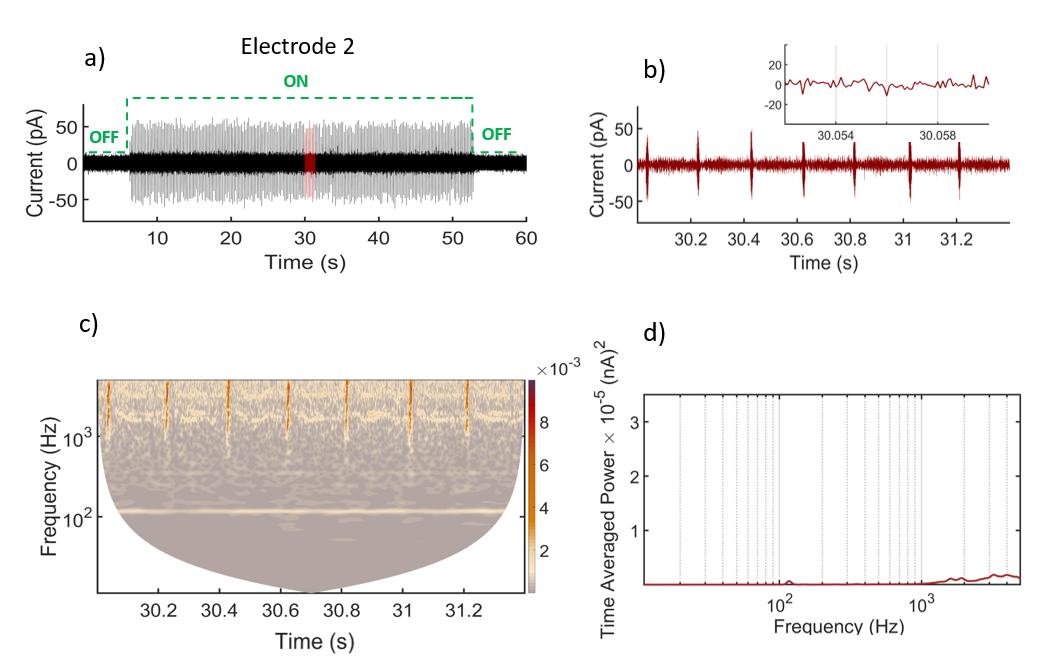}
\includegraphics[scale=0.87]{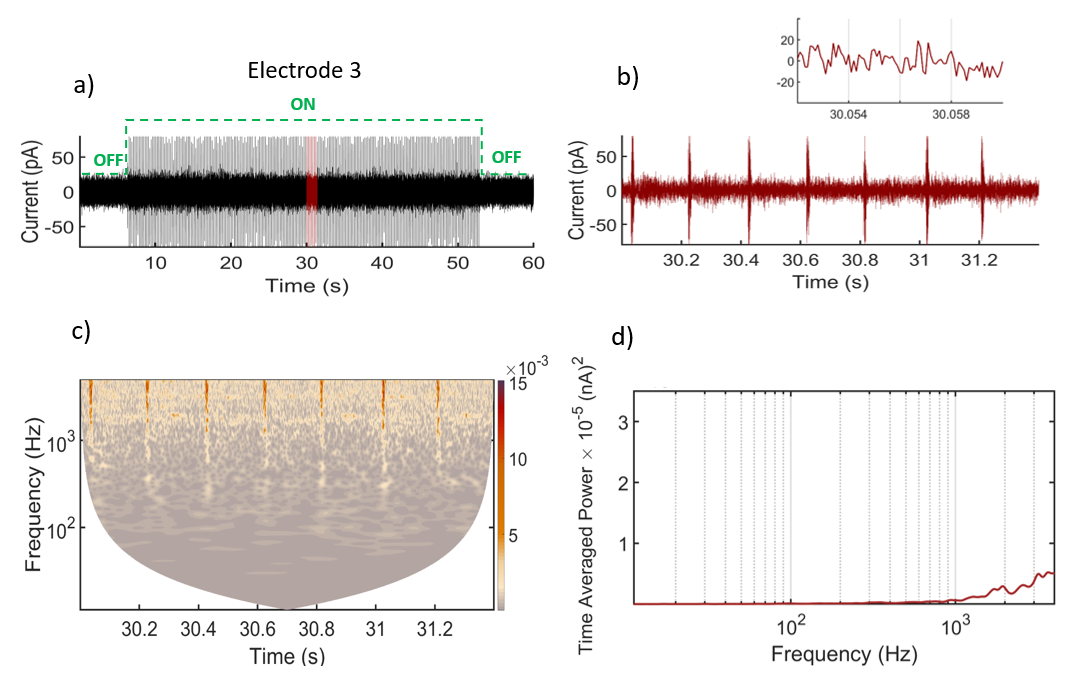}
\caption{(continued) Signals from electrodes E2 and E3, and their spectral analyses. (a) Signals were recorded for 60 seconds at a pressing voltage of $V = 4.17$\,V and electron density $n_e=2.2\times {10^{6}}$\,\si{cm^{-2}}. The MW (139\,GHz) is switched On/Off where indicated. (b) Enlarged reddened 1.4\,s part of (a) used for analysis, where the inset shows a further zoom. (c) Wavelet transform with frequency resolution $f_{0}$\,=\,3\,Hz. (d) Time-averaged wavelet power of signals in (b) showing the fundamental frequency and higher harmonics. The $y$-axis of (c) and the $x$-axes of (d) are logarithmic.}
\end{figure}  

\addtocounter{figure}{-1}
\clearpage 

\begin{figure*}
%\centering
\includegraphics[scale=0.9]{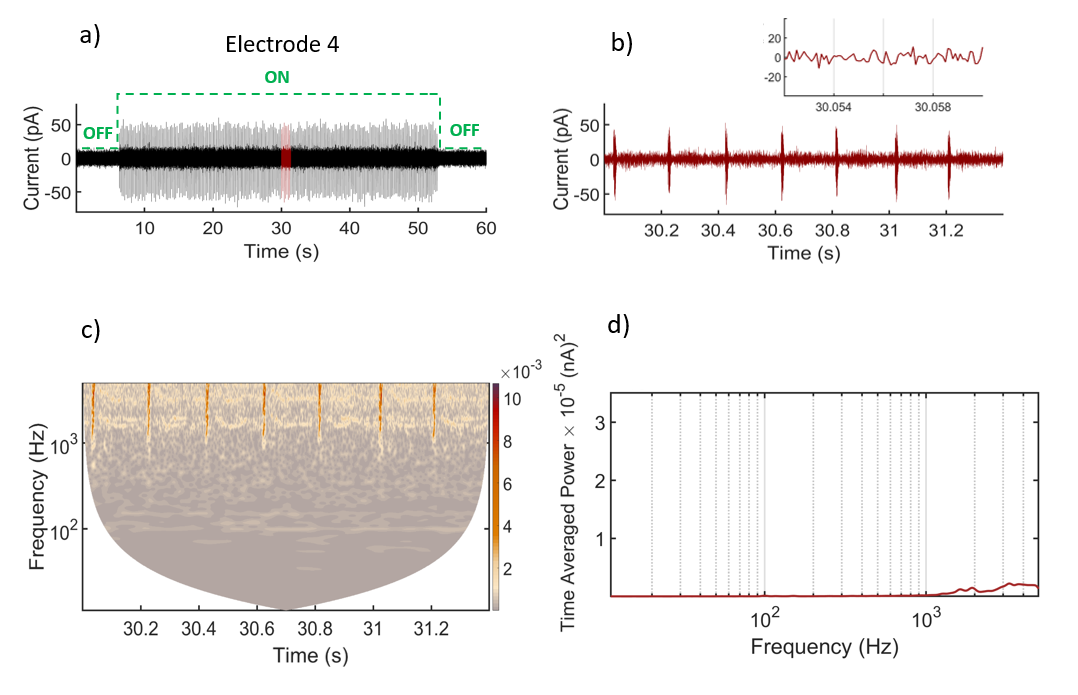}[c]
\caption{(continued) Signal from electrode E4, and its spectral analysis. (a) The signal was recorded for 60 seconds at a pressing voltage of $V\,=\,4.17$~V and electron density $n_e\,=\,2.2\times {10^{6}}$~\si{cm^{-2}}. The MW (139~GHz) is switched On/Off where indicated. (b) Enlarged reddened 1.4~s part of (a) used for analysis, where the inset shows a further zoom. (c) Wavelet transform with frequency resolution~3~Hz. (d) Time-averaged wavelet power of signals in (b) showing the fundamental frequency and higher harmonics. The $y$-axis of (c) and the $x$-axes of (d) are logarithmic.}
\label{fig:16}
\end{figure*}

\clearpage
\underbar{\bf Pressing voltage $V\,=\,4.18$~V}

\begin{figure}[b!]
\centering
\includegraphics[scale=0.80]{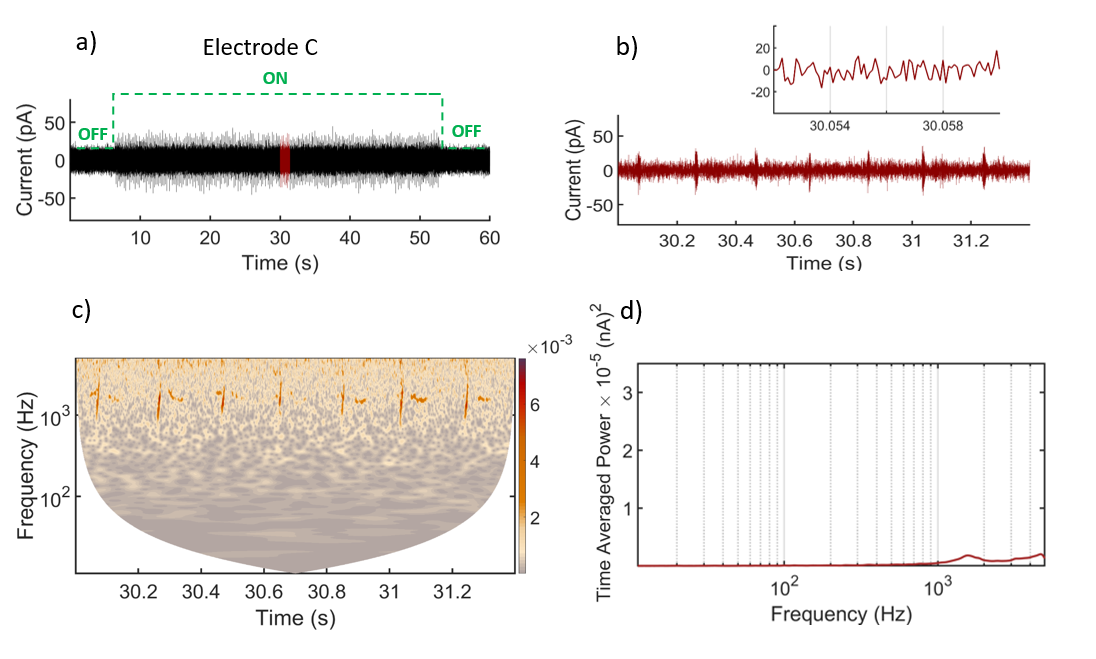}
\includegraphics[scale=0.80]{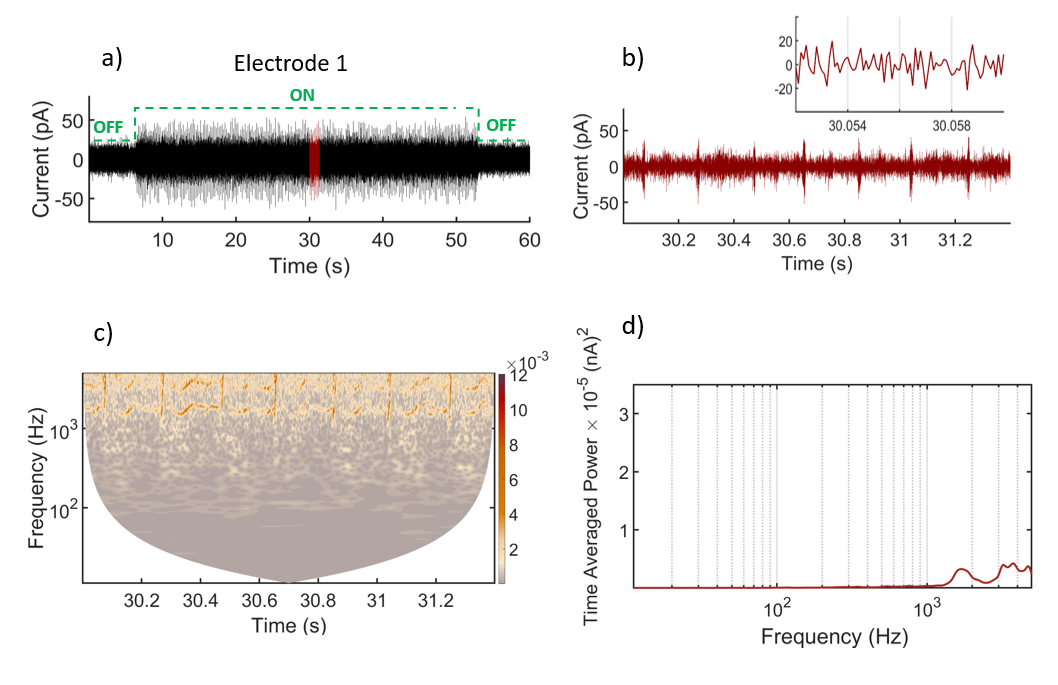}
\caption{ Signals from electrodes C and E1, and their spectral analyses. (a) Signals were recorded for 60 seconds at a pressing voltage of $V\,=\,4.18$~V and electron density $n_e\,=\,2.2\times {10^{6}}$~\si{cm^{-2}}. The MW (139~GHz) is switched On/Off where indicated. (b) Enlarged reddened 1.4~s part of (a) used for analysis, where the inset shows a further zoom. (c) Wavelet transform with frequency resolution~3~Hz. (d) Time-averaged wavelet power of signals in (b) showing the fundamental frequency and higher harmonics. The $y$-axis of (c) and the $x$-axes of (d) are logarithmic.}
\end{figure}

\addtocounter{figure}{-1}
\clearpage 

\begin{figure}[b!]
\centering
\includegraphics[scale=0.85]{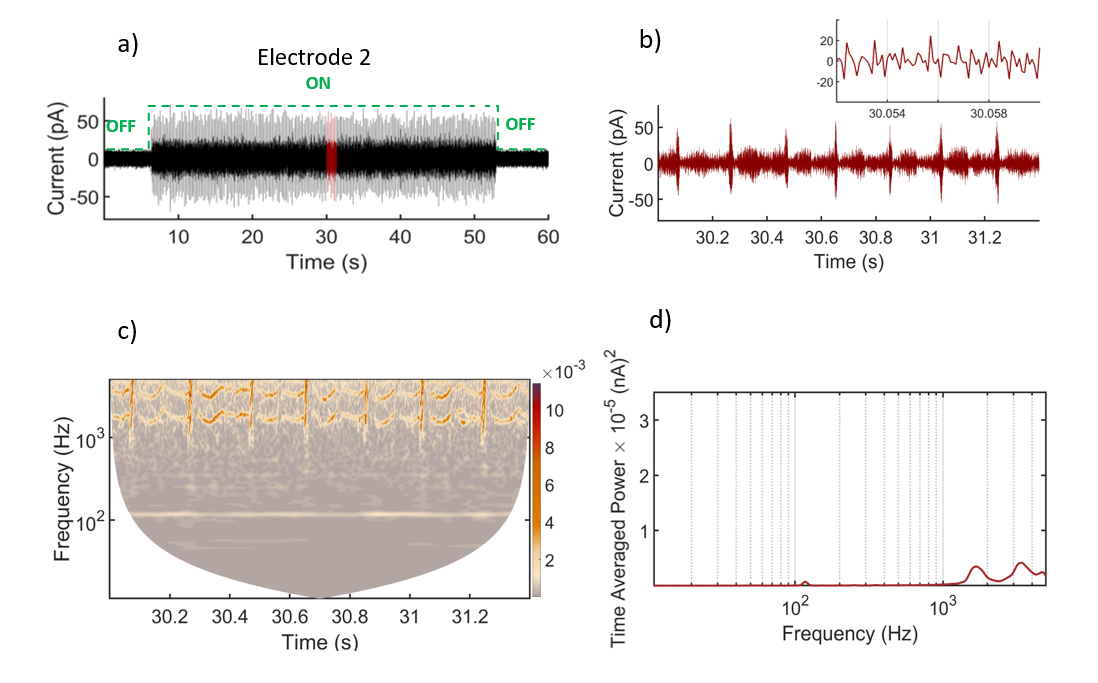}
\includegraphics[scale=0.85]{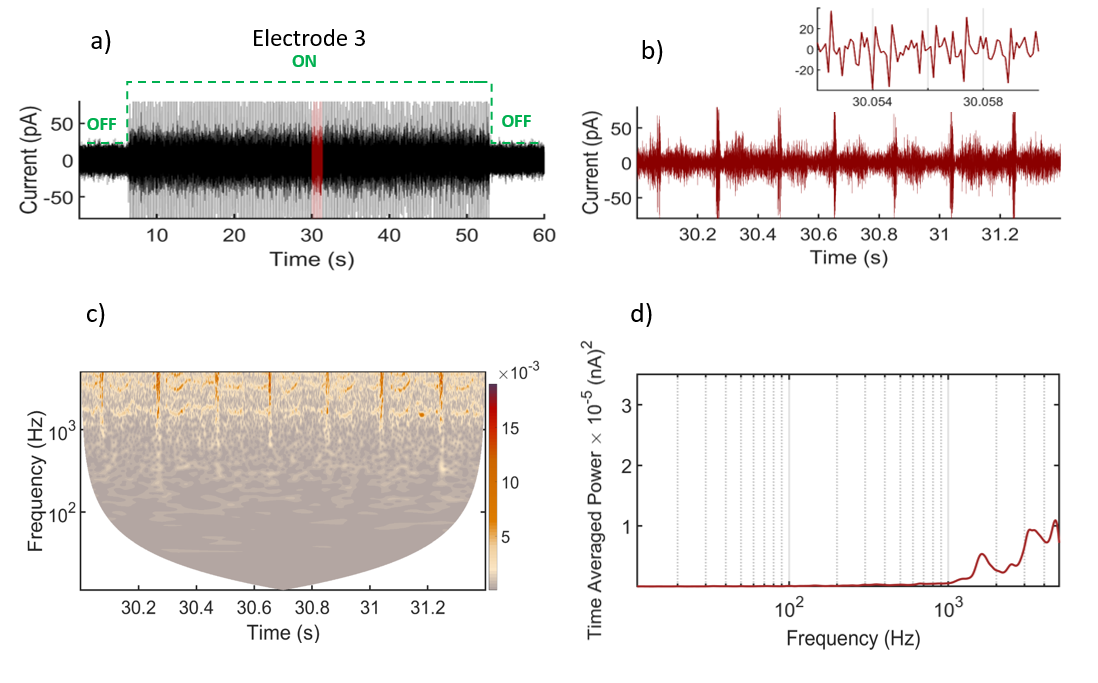}
\caption{(continued) Signals from electrodes E2 and E3, and their spectral analyses. (a) Signals were recorded for 60 seconds at a pressing voltage of $V\,=\,4.18$~V and electron density $n_e=2.2\times {10^{6}}$~\si{cm^{-2}}. The MW (139~GHz) is switched On/Off where indicated. (b) Enlarged reddened 1.4~s part of (a) used for analysis, where the inset shows a further zoom. (c) Wavelet transform with frequency resolution~3~Hz. (d) Time-averaged wavelet power of signals in (b) showing the fundamental frequency and higher harmonics. The $y$-axis of (c) and the $x$-axes of (d) are logarithmic.}
\end{figure}  

\addtocounter{figure}{-1}
\clearpage 

\begin{figure*}
%\centering
\includegraphics[scale=0.9]{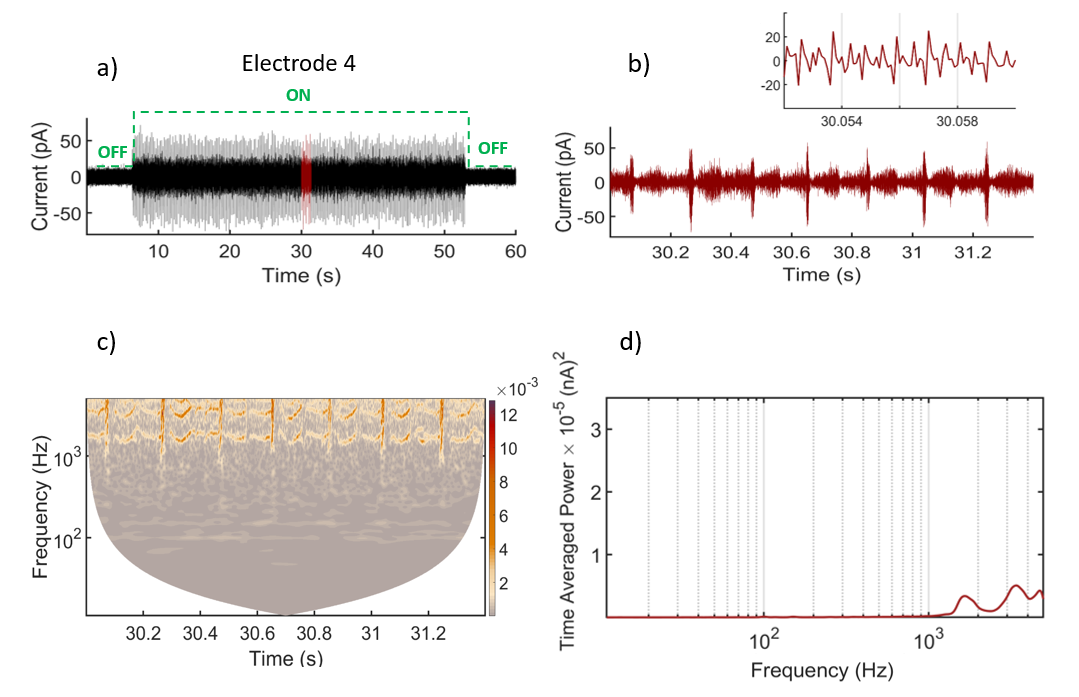}[c]
\caption{(continued) Signal from electrode E4, and its spectral analysis. (a) The signal was recorded for 60 seconds at a pressing voltage of $V\,=\,4.18$~V and electron density $n_e\,=\,2.2\times {10^{6}}$~\si{cm^{-2}}. The MW (139~GHz) is switched On/Off where indicated. (b) Enlarged reddened 1.4~s part of (a) used for analysis, where the inset shows a further zoom. (c) Wavelet transform with frequency resolution~3~Hz. (d) Time-averaged wavelet power of signals in (b) showing the fundamental frequency and higher harmonics. The $y$-axis of (c) and the $x$-axes of (d) are logarithmic.}
\label{fig:16}
\end{figure*}

\clearpage
\underbar{\bf Pressing voltage $V\,=\,4.19$~V}

\begin{figure}[b!]
\centering
\includegraphics[scale=0.85]{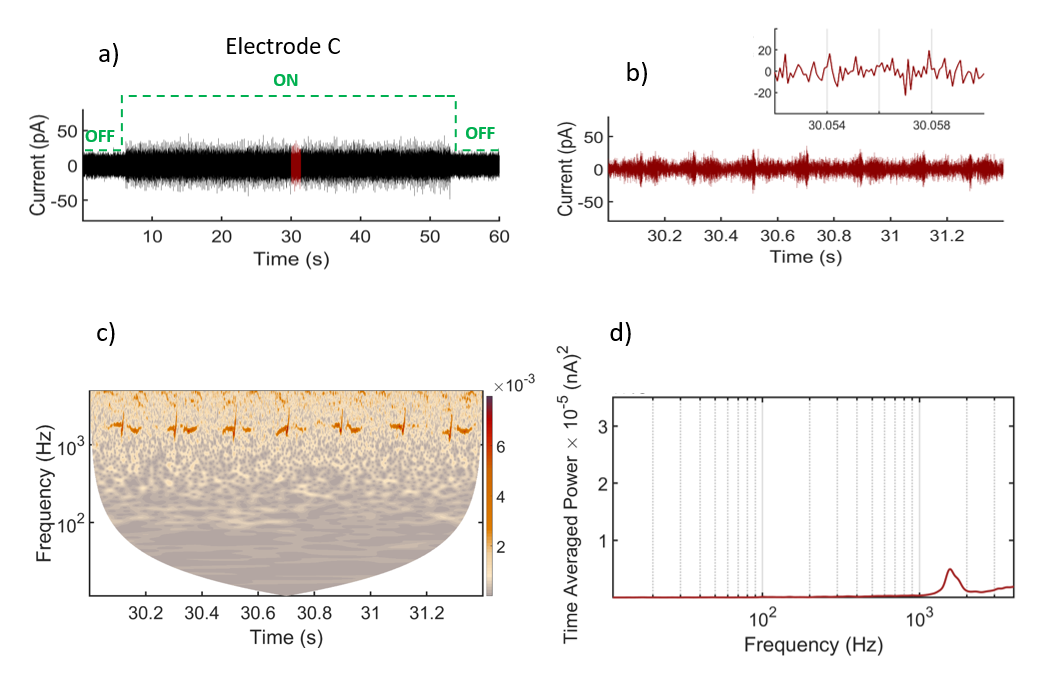}
\includegraphics[scale=0.85]{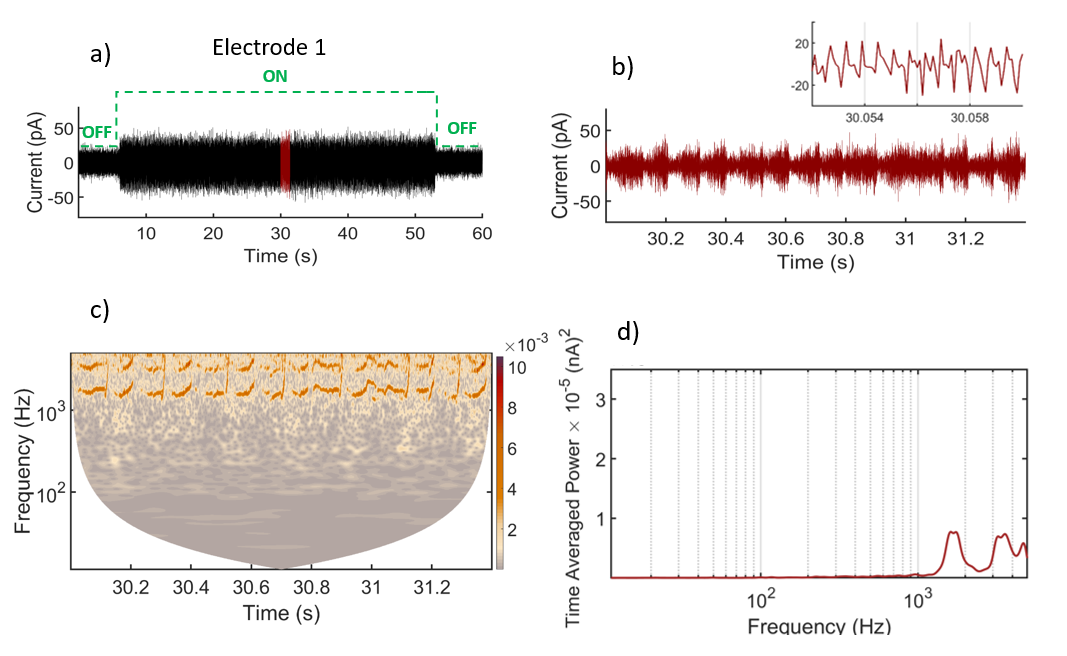}
\caption{ Signals from electrodes C and E1, and their spectral analyses. (a) Signals were recorded for 60 seconds at a pressing voltage of $V\,=\,4.19$~V and electron density $n_e\,=\,2.2\times {10^{6}}$~\si{cm^{-2}}. The MW (139\,GHz) is switched On/Off where indicated. (b) Enlarged reddened 1.4~s part of (a) used for analysis, where the inset shows a further zoom. (c) Wavelet transform with frequency resolution~3~Hz. (d) Time-averaged wavelet power of signals in (b) showing the fundamental frequency and higher harmonics. The $y$-axis of (c) and the $x$-axes of (d) are logarithmic.}
\end{figure}

\addtocounter{figure}{-1}
\clearpage 

\begin{figure}[b!]
\centering
\includegraphics[scale=0.85]{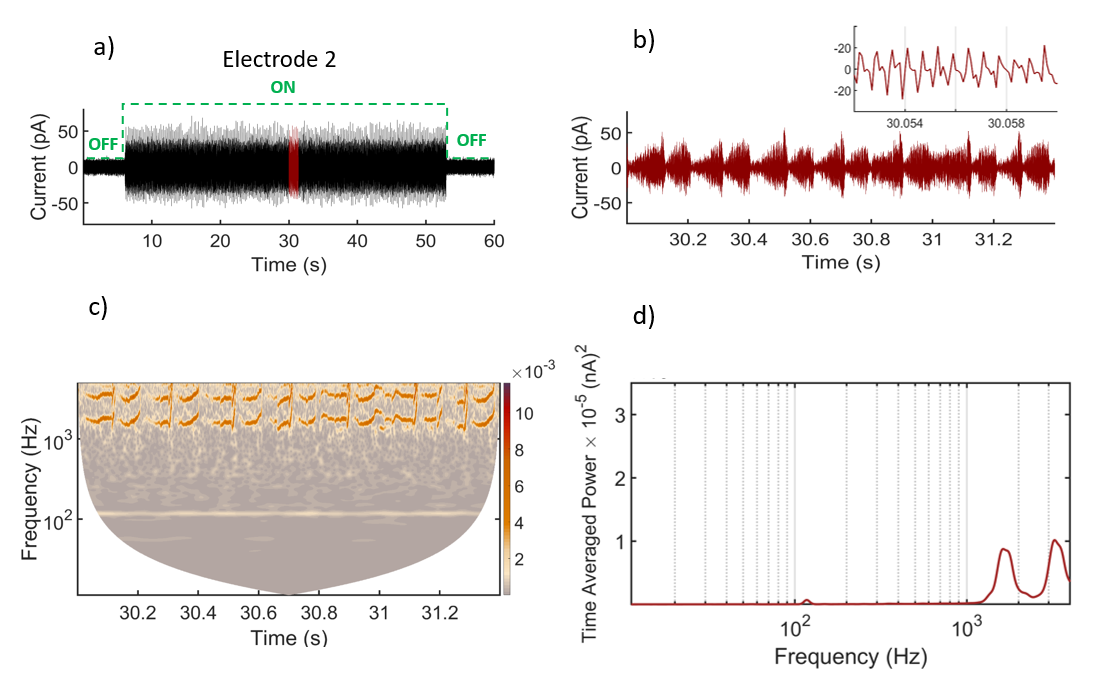}
\includegraphics[scale=0.85]{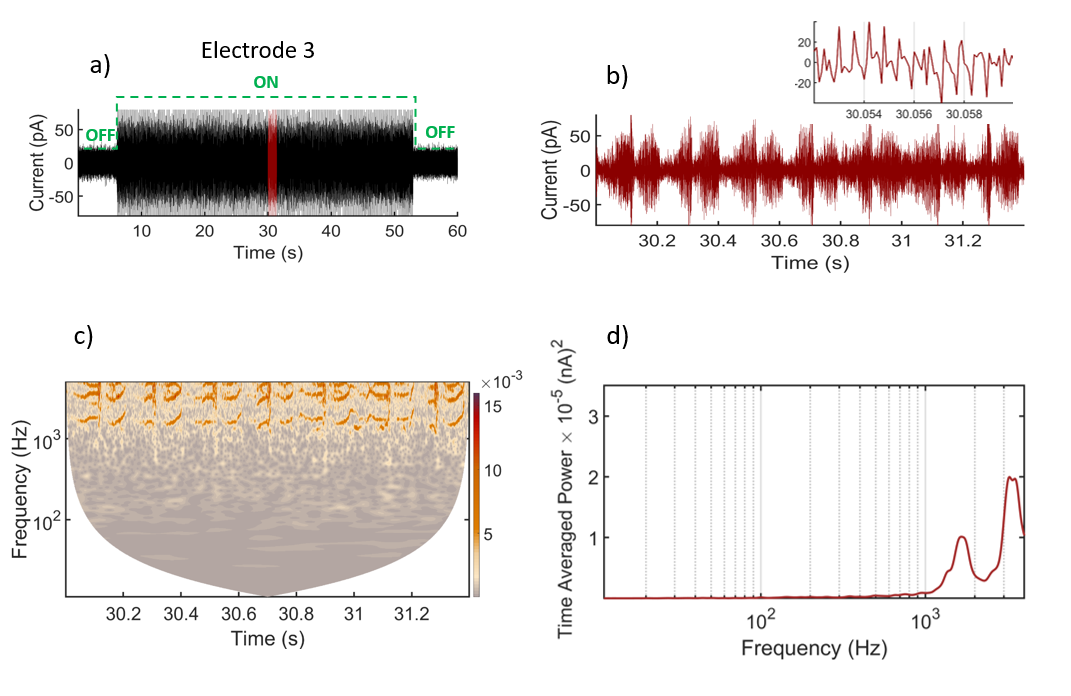}
\caption{(continued) Signals from electrodes E2 and E3, and their spectral analyses. (a) Signals were recorded for 60 seconds at a pressing voltage of $V\,=\,4.19$~V and electron density $n_e\,=\,2.2\times {10^{6}}$~\si{cm^{-2}}. The MW (139\,GHz) is switched On/Off where indicated. (b) Enlarged reddened 1.4~s part of (a) used for analysis, where the inset shows a further zoom. (c) Wavelet transform with frequency resolution~3~Hz. (d) Time-averaged wavelet power of signals in (b) showing the fundamental frequency and higher harmonics. The $y$-axis of (c) and the $x$-axes of (d) are logarithmic.}
\end{figure}  

\addtocounter{figure}{-1}
\clearpage 

\begin{figure*}
%\centering
\includegraphics[scale=0.9]{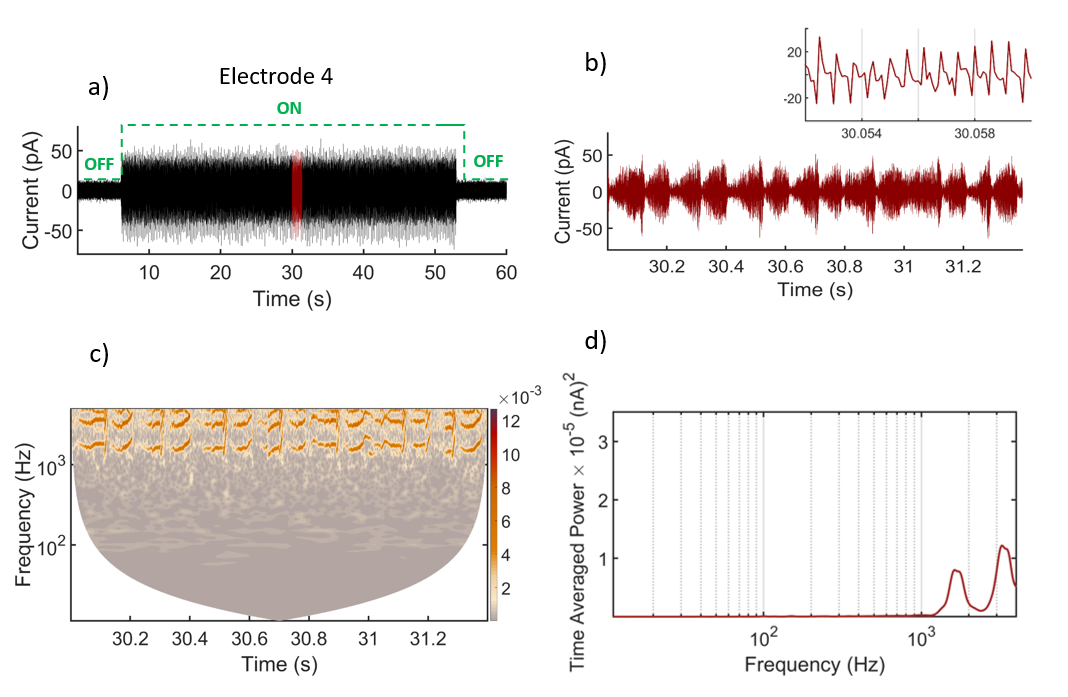}[c]
\caption{(continued) Signal from electrode E4, and its spectral analysis. (a) The signal was recorded for 60 seconds at a pressing voltage of $V\,=\,4.19$~V and electron density $n_e\,=\,2.2\times {10^{6}}$~\si{cm^{-2}}. The MW (139~GHz) is switched On/Off where indicated. (b) Enlarged reddened 1.4~s part of (a) used for analysis, where the inset shows a further zoom. (c) Wavelet transform with frequency resolution~3~Hz. (d) Time-averaged wavelet power of signals in (b) showing the fundamental frequency and higher harmonics. The $y$-axis of (c) and the $x$-axes of (d) are logarithmic.}
\label{fig:16}
\end{figure*}

\clearpage
\underbar{\bf Pressing voltage $V\,=\,4.20$~V}

\begin{figure}[b!]
\centering
\includegraphics[scale=0.70]{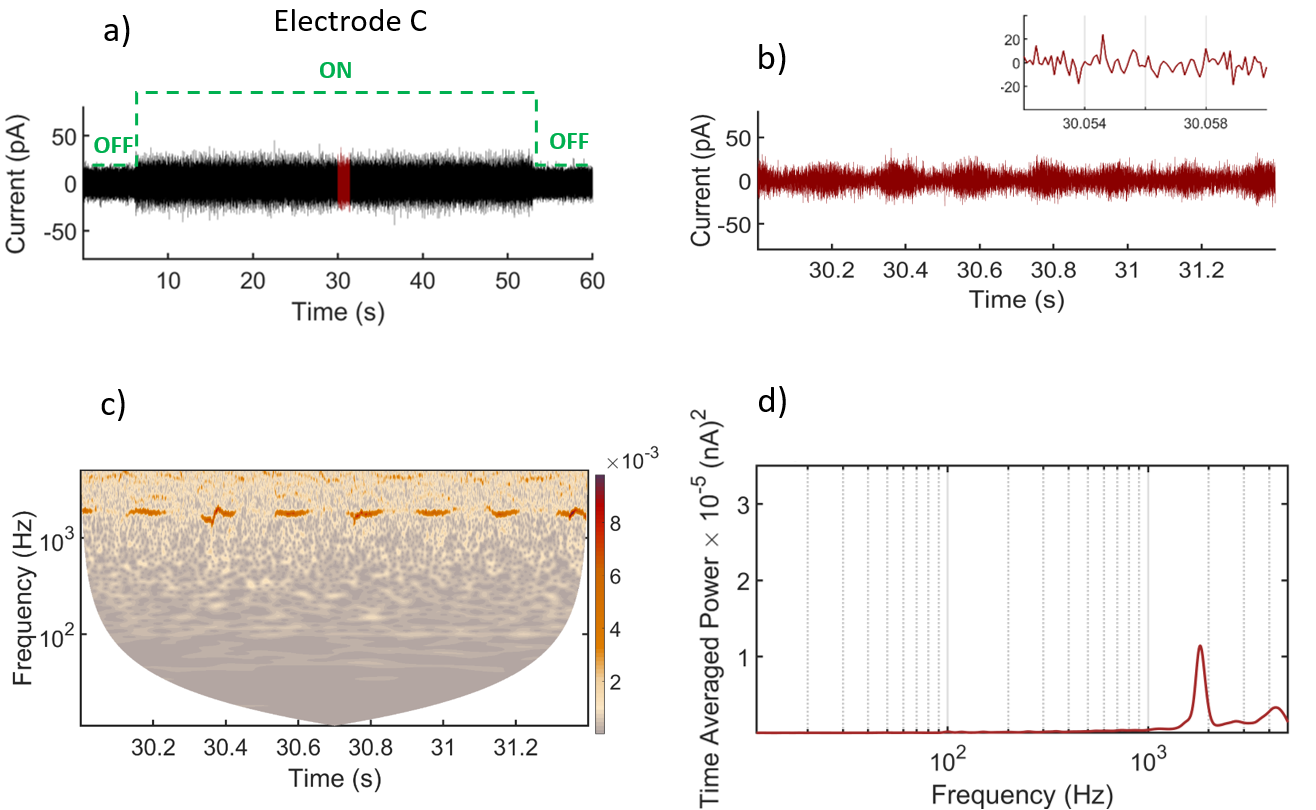}
\includegraphics[scale=0.70]{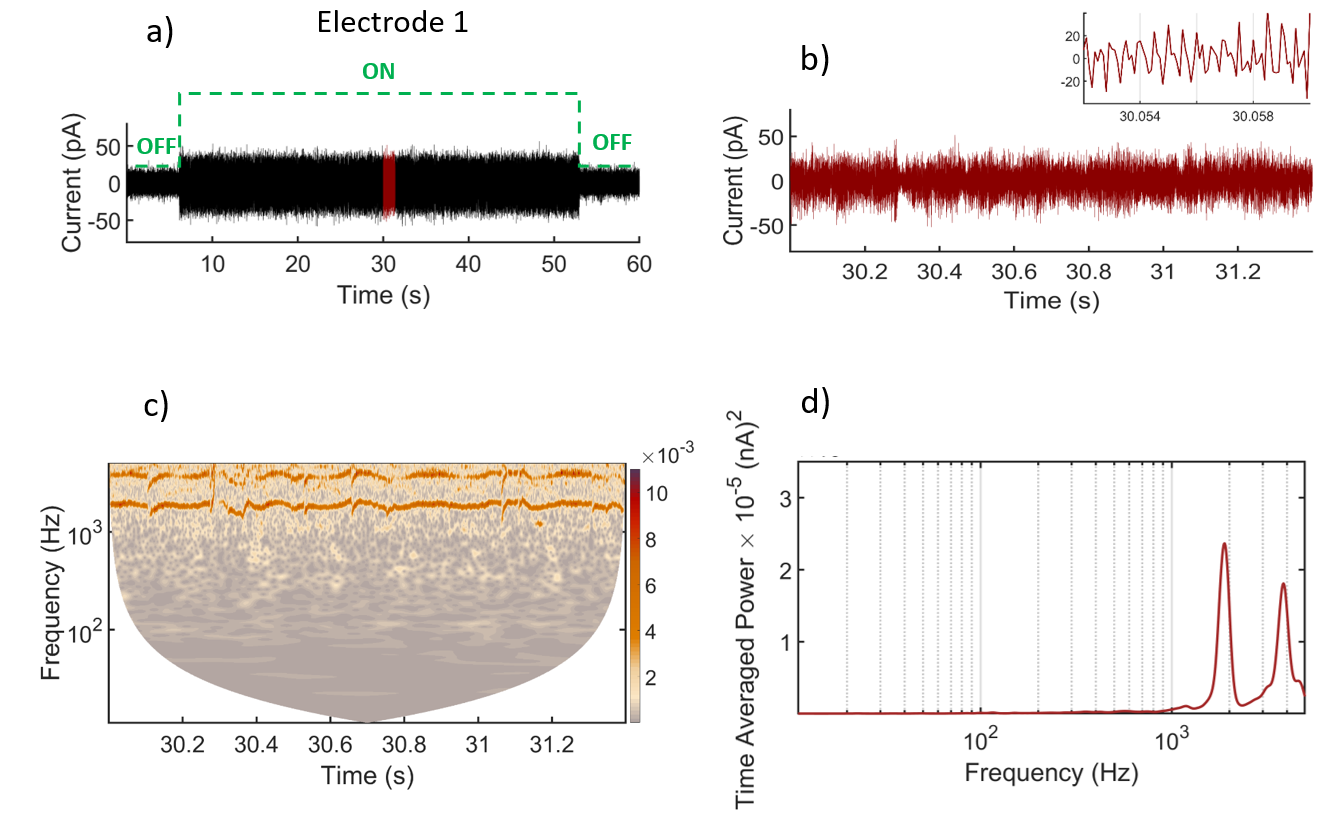}
\caption{ Signals from electrodes C and E1, and their spectral analyses. (a) Signals were recorded for 60 seconds at a pressing voltage of $V\,=\,4.20$~V and electron density $n_e\,=\,2.2\times {10^{6}}$~\si{cm^{-2}}. The MW (139~GHz) is switched On/Off where indicated. (b) Enlarged reddened 1.4~s part of (a) used for analysis, where the inset shows a further zoom. (c) Wavelet transform with frequency resolution~3~Hz. (d) Time-averaged wavelet power of signals in (b) showing the fundamental frequency and higher harmonics. The $y$-axis of (c) and the $x$-axes of (d) are logarithmic.}
\label{fig:220}
\end{figure}

\addtocounter{figure}{-1}
\clearpage 

\begin{figure}[b!]
\centering
\includegraphics[scale=0.7]{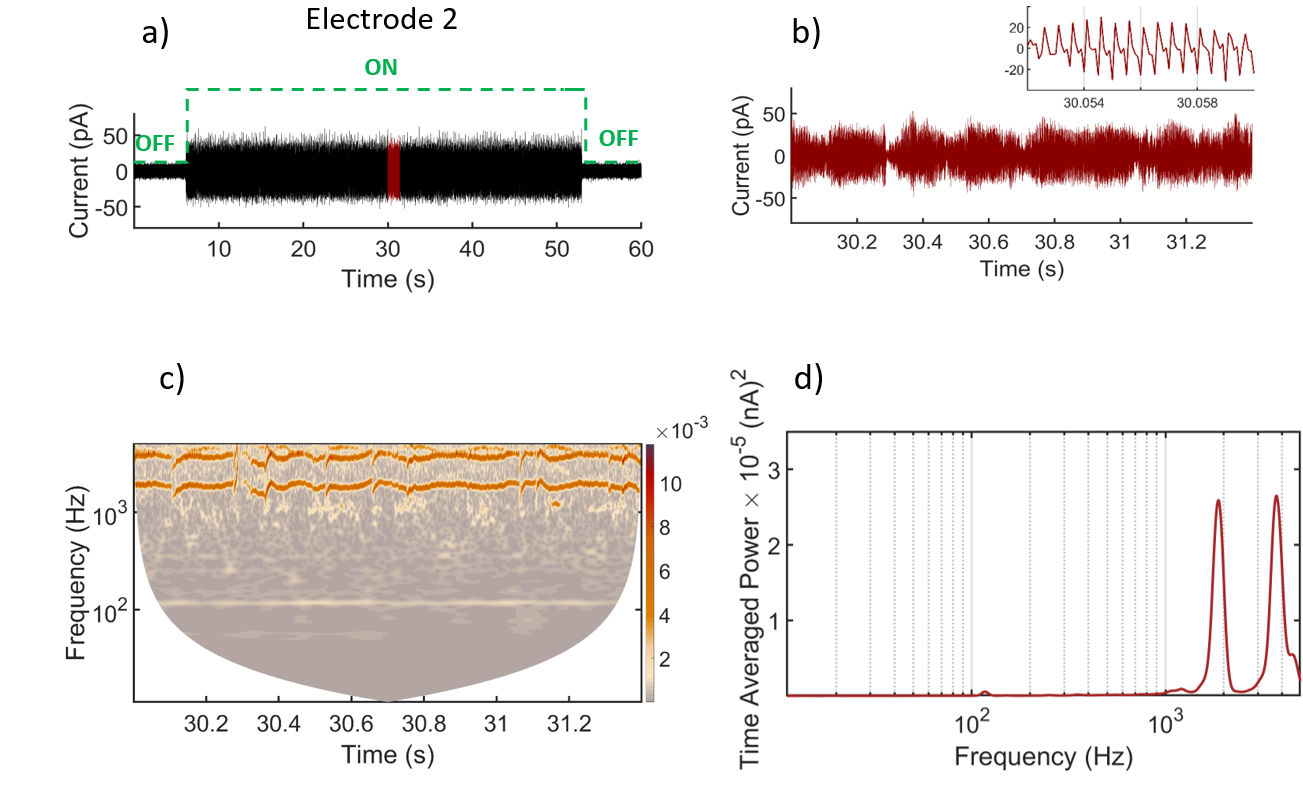}
\includegraphics[scale=0.7]{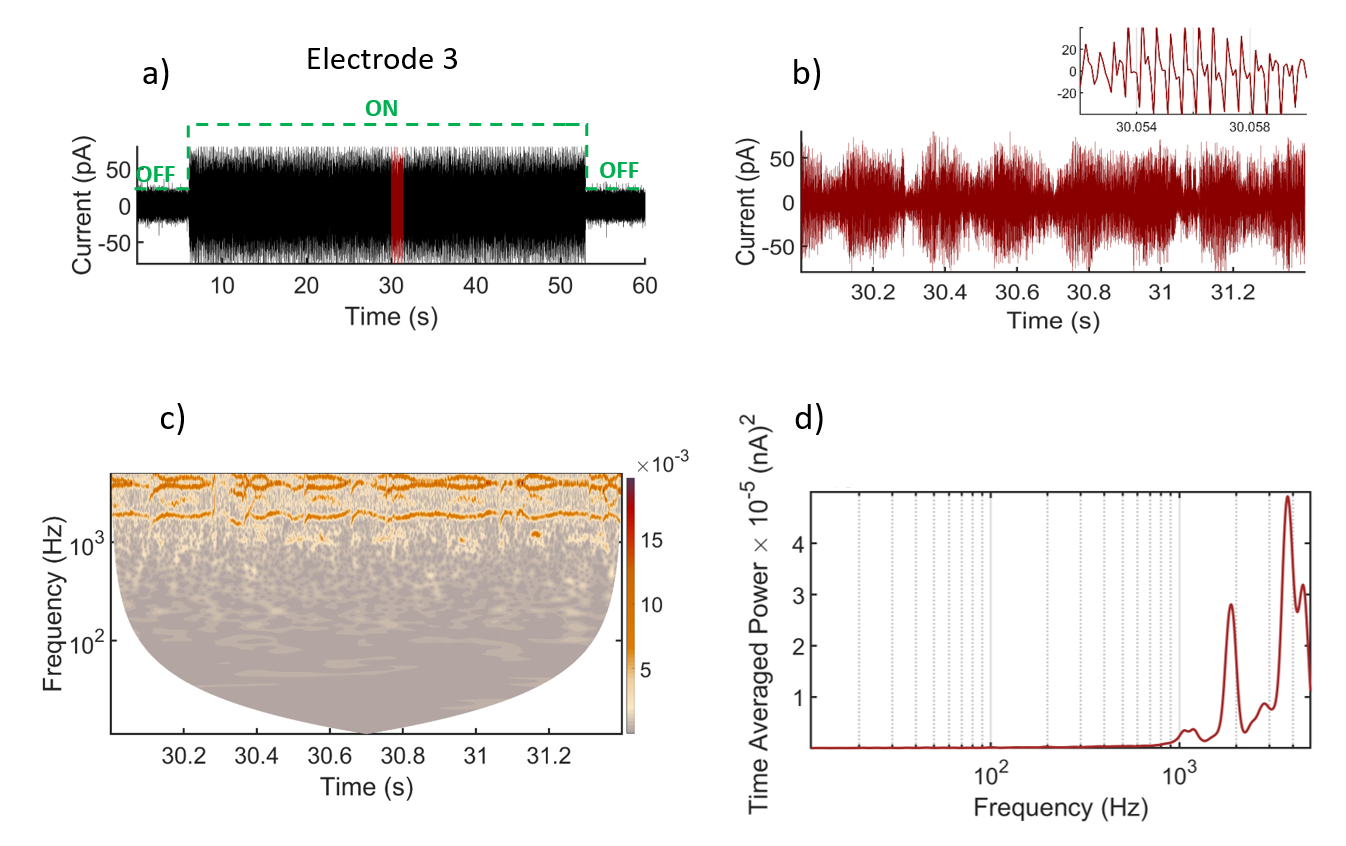}
\caption{(continued) Signals from electrodes E2 and E3, and their spectral analyses. (a) Signals were recorded for 60 seconds at a pressing voltage of $V\,=\,4.20$~V and electron density $n_e\,=\,2.2\times {10^{6}}$~\si{cm^{-2}}. The MW (139~GHz) is switched On/Off where indicated. (b) Enlarged reddened 1.4~s part of (a) used for analysis, where the inset shows a further zoom. (c) Wavelet transform with frequency resolution~3~Hz. (d) Time-averaged wavelet power of signals in (b) showing the fundamental frequency and higher harmonics. The $y$-axis of (c) and the $x$-axes of (d) are logarithmic.}
\end{figure}  

\addtocounter{figure}{-1}
\clearpage 

\begin{figure*}
%\centering
\includegraphics[scale=0.7]{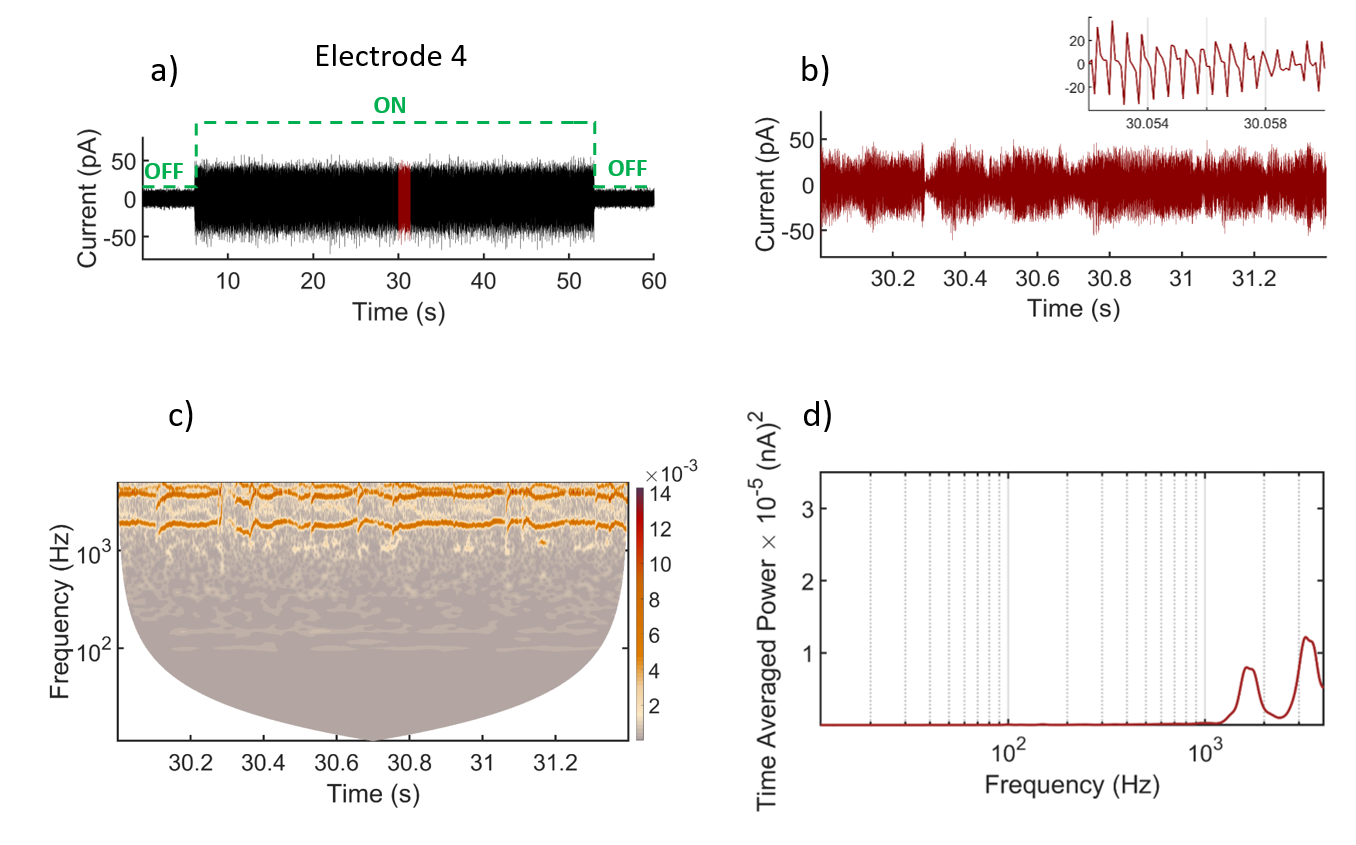}[c]
\caption{(continued) Signal from electrode E4, and its spectral analysis. (a) The signal was recorded for 60 seconds at a pressing voltage of $V\,=\,4.20$~V and electron density $n_e\,=\,2.2\times {10^{6}}$~\si{cm^{-2}}. The MW (139~GHz) is switched On/Off where indicated. (b) Enlarged reddened 1.4~s part of (a) used for analysis, where the inset shows a further zoom. (c) Wavelet transform with frequency resolution~3~Hz. (d) Time-averaged wavelet power of signals in (b) showing the fundamental frequency and higher harmonics. The $y$-axis of (c) and the $x$-axes of (d) are logarithmic.}
\label{fig:16}
\end{figure*}

\clearpage
\underbar{\bf Pressing voltage $V\,=\,4.21$~V}

\begin{figure}[b!]
\centering
\includegraphics[scale=0.80]{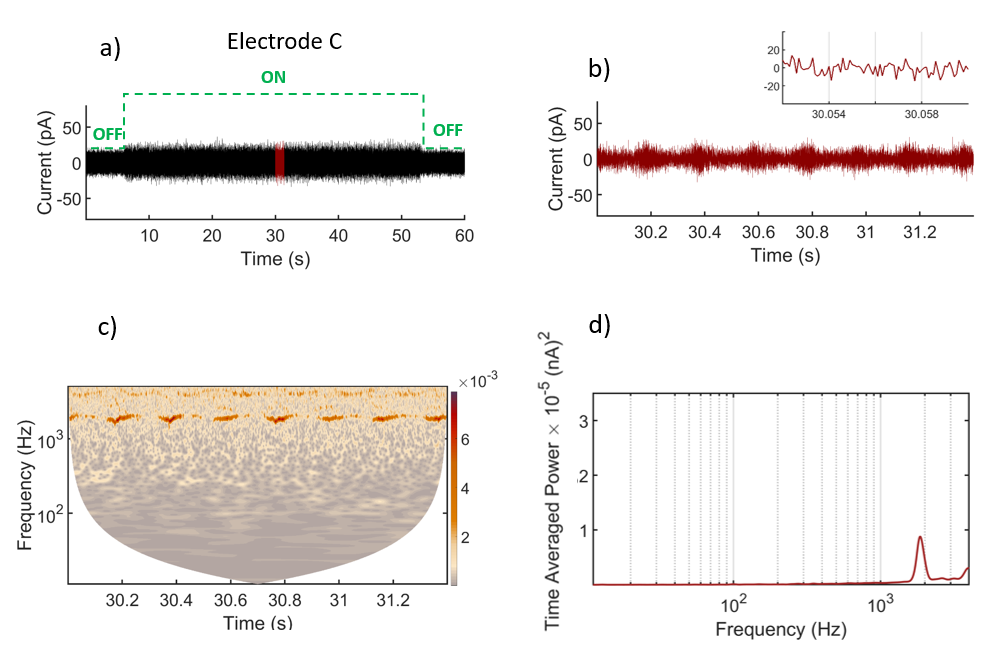}
\includegraphics[scale=0.80]{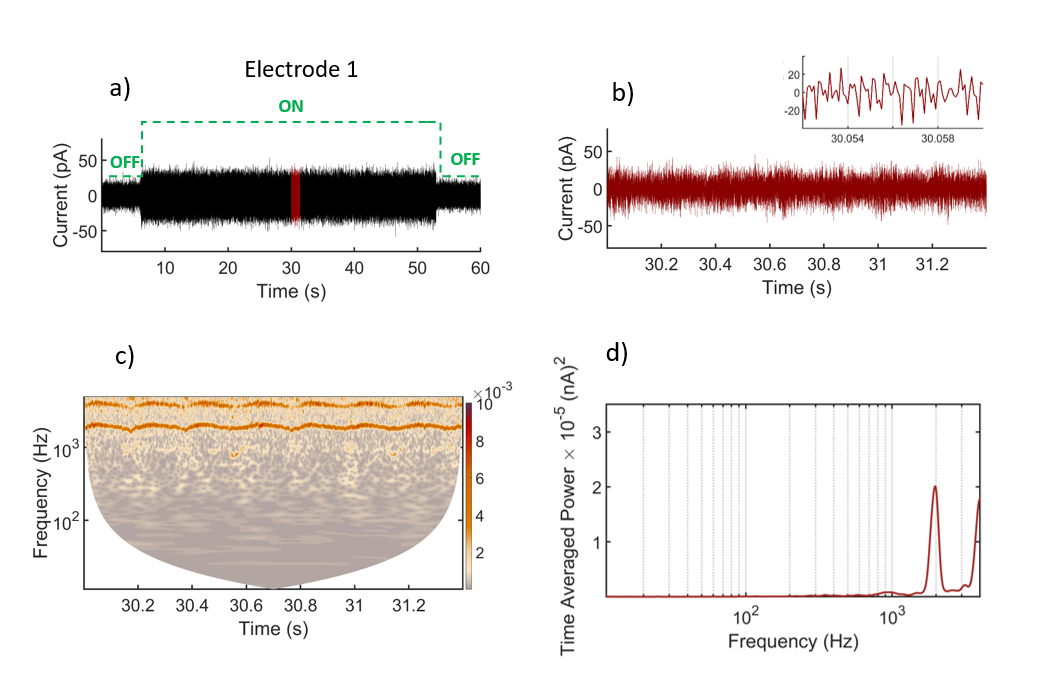}
\caption{ Signals from electrodes C and E1, and their spectral analyses. (a) Signals were recorded for 60 seconds at a pressing voltage of $V\,=\,4.21$~V and electron density $n_e\,=\,2.2\times {10^{6}}$~\si{cm^{-2}}. The MW (139~GHz) is switched On/Off where indicated. (b) Enlarged reddened 1.4~s part of (a) used for analysis, where the inset shows a further zoom. (c) Wavelet transform with frequency resolution~3~Hz. (d) Time-averaged wavelet power of signals in (b) showing the fundamental frequency and higher harmonics. The $y$-axis of (c) and the $x$-axes of (d) are logarithmic.}
\end{figure}

\addtocounter{figure}{-1}
\clearpage 

\begin{figure}[b!]
\centering
\includegraphics[scale=0.85]{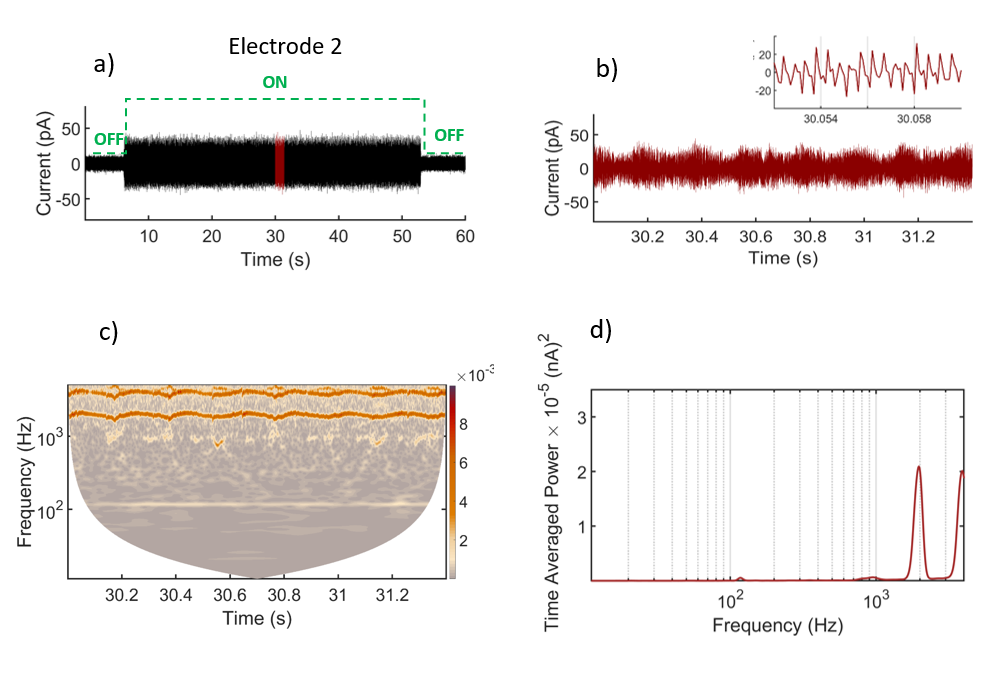}
\includegraphics[scale=0.85]{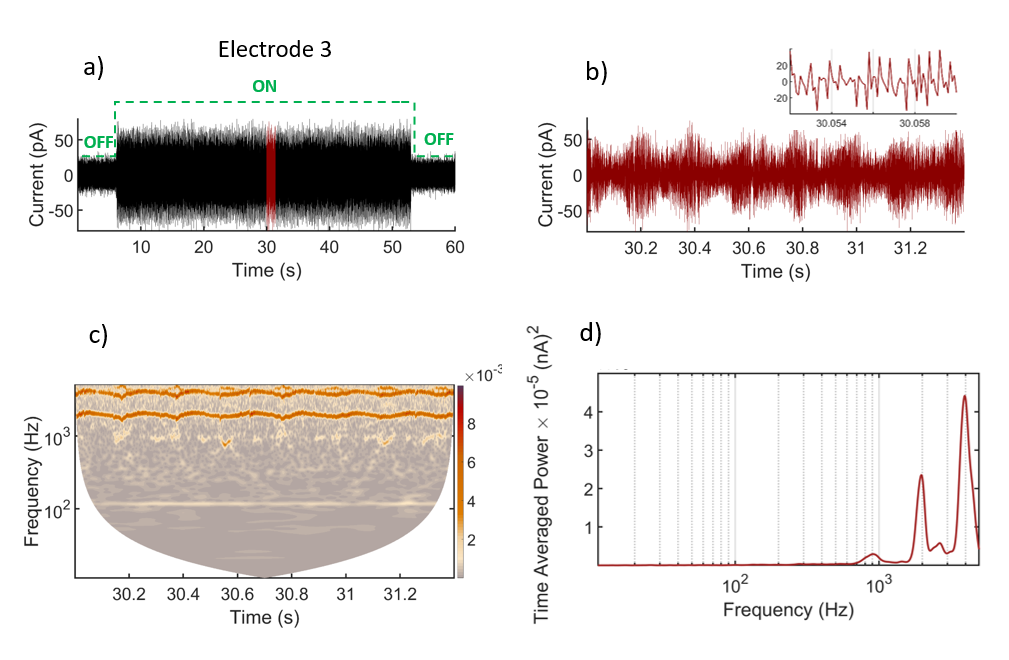}
\caption{(continued) Signals from electrodes E2 and E3, and their spectral analyses. (a) Signals were recorded for 60 seconds at a pressing voltage of $V\,=\,4.21$~V and electron density $n_e\,=\,2.2\times {10^{6}}$~\si{cm^{-2}}. The MW (139\,GHz) is switched On/Off where indicated. (b) Enlarged reddened 1.4~s part of (a) used for analysis, where the inset shows a further zoom. (c) Wavelet transform with frequency resolution~3~Hz. (d) Time-averaged wavelet power of signals in (b) showing the fundamental frequency and higher harmonics. The $y$-axis of (c) and the $x$-axes of (d) are logarithmic.}
\end{figure}  

\addtocounter{figure}{-1}
\clearpage 

\begin{figure*}
%\centering
\includegraphics[scale=0.9]{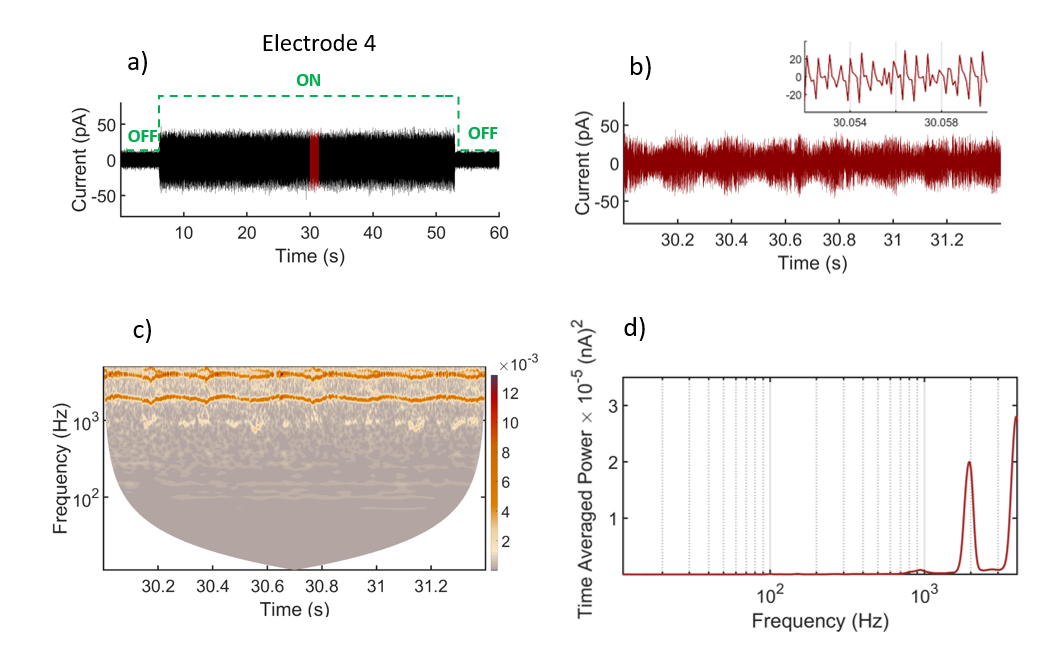}[c]
\caption{(continued) Signal from electrode E4, and its spectral analysis. (a) The signal was recorded for 60 seconds at a pressing voltage of $V\,=\,4.21$~V and electron density $n_e\,=\,2.2\times {10^{6}}$~\si{cm^{-2}}. The MW (139~GHz) is switched On/Off where indicated. (b) Enlarged reddened 1.4~s part of (a) used for analysis, where the inset shows a further zoom. (c) Wavelet transform with frequency resolution~3~Hz. (d) Time-averaged wavelet power of signals in (b) showing the fundamental frequency and higher harmonics. The $y$-axis of (c) and the $x$-axes of (d) are logarithmic.}
\label{fig:16}
\end{figure*}

\clearpage
\underbar{\bf Pressing voltage $V\,=\,4.22$~V}

\begin{figure}[b!]
\centering
\includegraphics[scale=0.85]{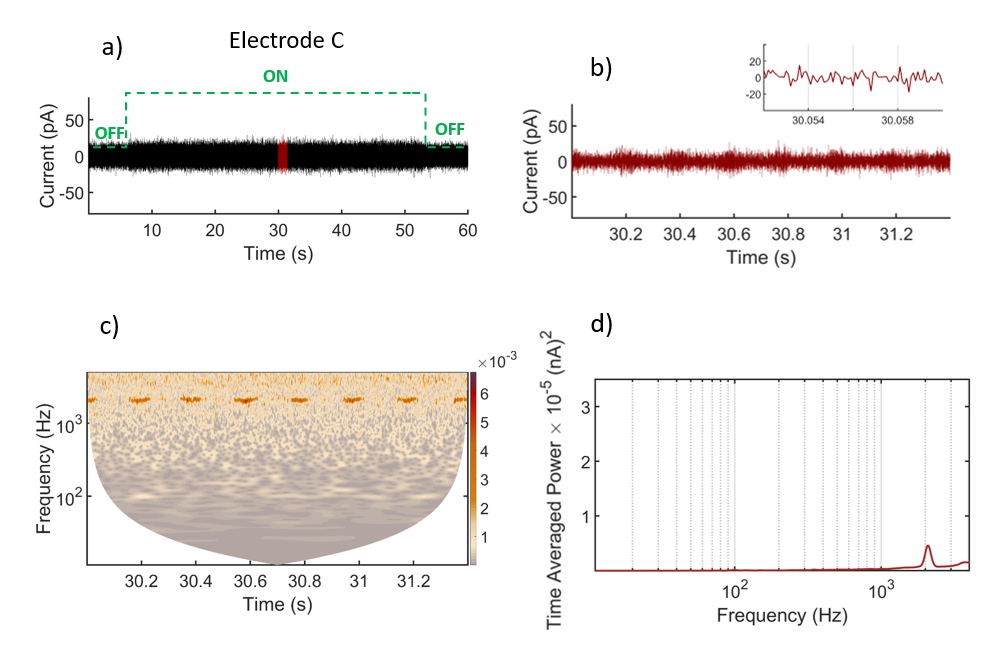}
\includegraphics[scale=0.85]{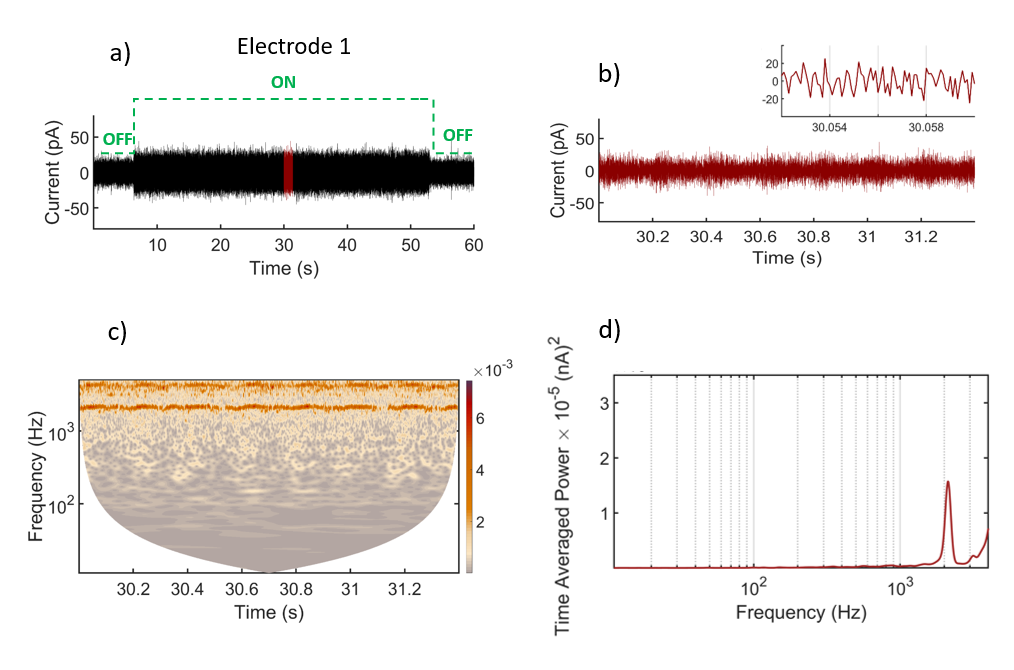}
\caption{ Signals from electrodes C and E1, and their spectral analyses. (a) Signals were recorded for 60 seconds at a pressing voltage of $V\,=\,4.22$~V and electron density $n_e\,=\,2.2\times {10^{6}}$~\si{cm^{-2}}. The MW (139\,GHz) is switched On/Off where indicated. (b) Enlarged reddened 1.4~s part of (a) used for analysis, where the inset shows a further zoom. (c) Wavelet transform with frequency resolution $f_{0}$\,=\,3~Hz. (d) Time-averaged wavelet power of signals in (b) showing the fundamental frequency and higher harmonics. The $y$-axis of (c) and the $x$-axes of (d) are logarithmic.}
\end{figure}

\addtocounter{figure}{-1}
\clearpage 

\begin{figure}[b!]
\centering
\includegraphics[scale=0.9]{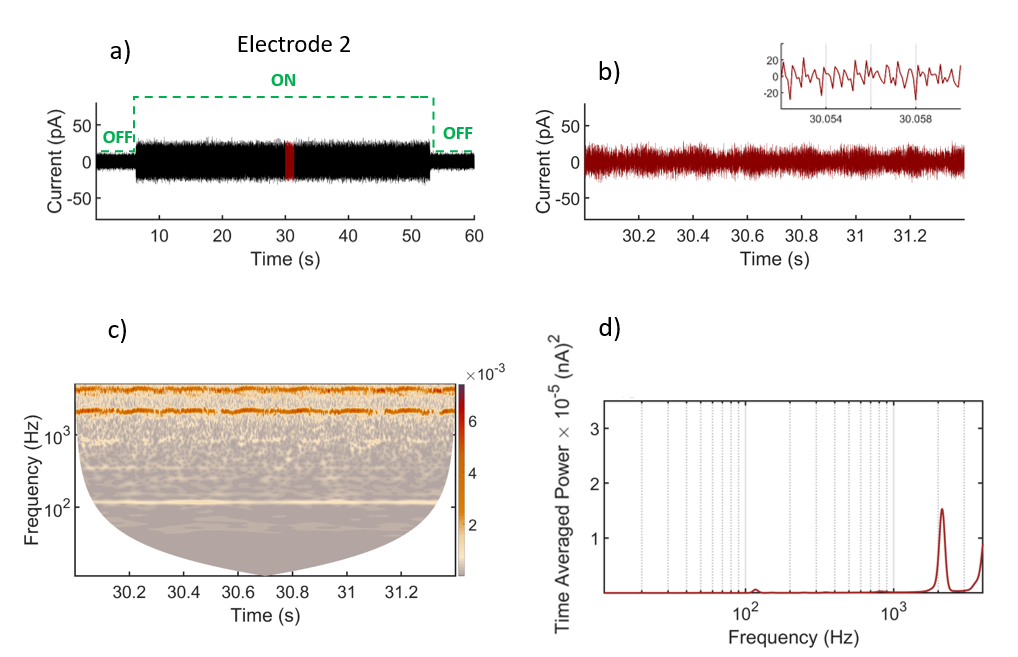}
\includegraphics[scale=0.9]{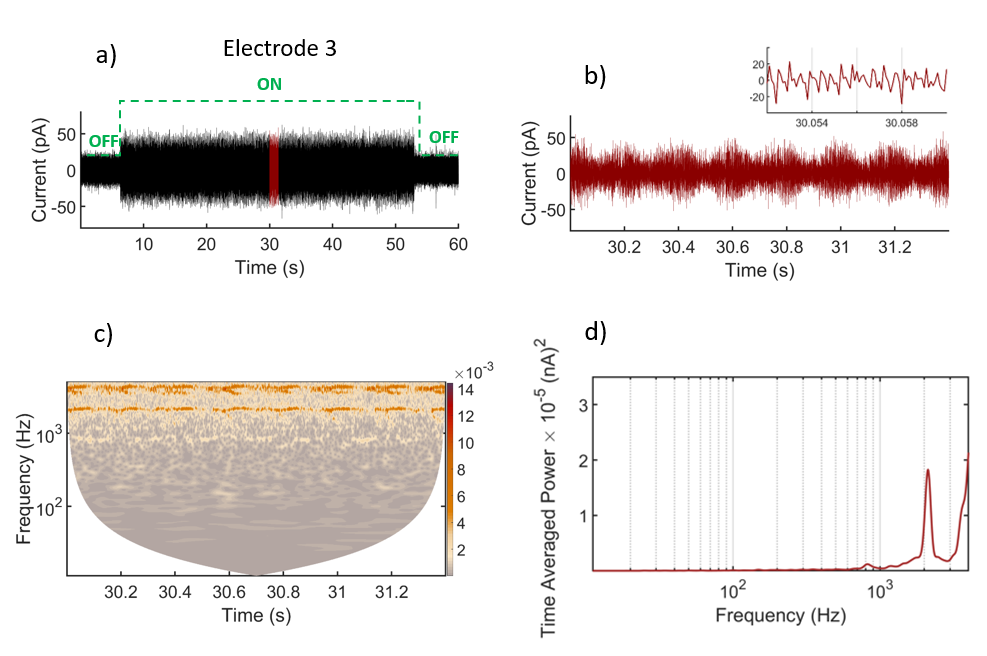}
\caption{(continued) Signals from electrodes E2 and E3, and their spectral analyses. (a) Signals were recorded for 60 seconds at a pressing voltage of $V\,=\,4.22$\,V and electron density $n_e\,=\,2.2\times {10^{6}}$~\si{cm^{-2}}. The MW (139~GHz) is switched On/Off where indicated. (b) Enlarged reddened 1.4~s part of (a) used for analysis, where the inset shows a further zoom. (c) Wavelet transform with frequency resolution~3~Hz. (d) Time-averaged wavelet power of signals in (b) showing the fundamental frequency and higher harmonics. The $y$-axis of (c) and the $x$-axes of (d) are logarithmic.}
\end{figure}  

\addtocounter{figure}{-1}
\clearpage 

\begin{figure*}
%\centering
\includegraphics[scale=0.9]{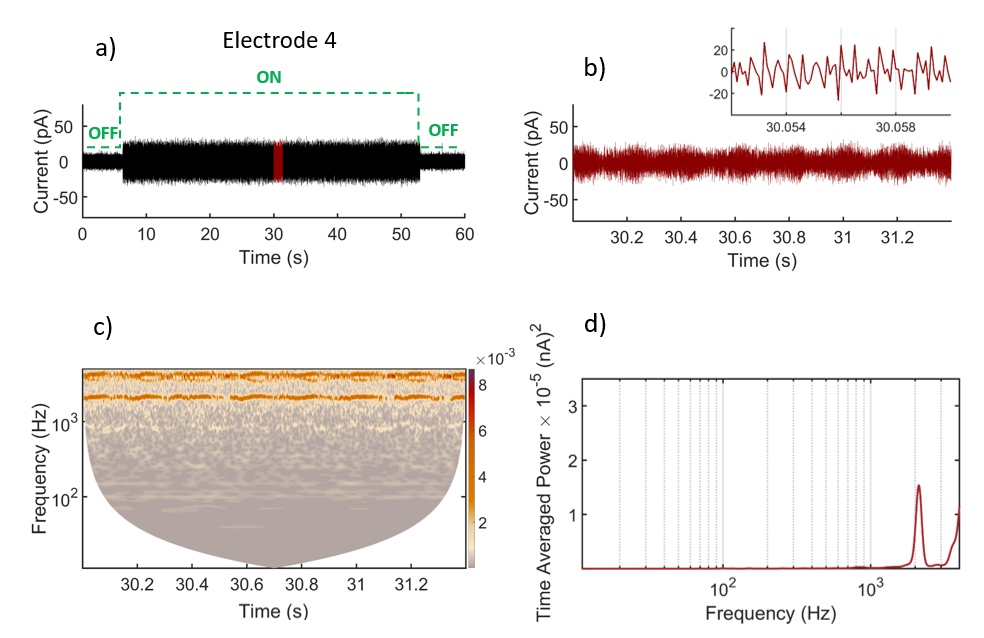}[c]
\caption{(continued) Signal from electrode E4, and its spectral analysis. (a) The signal was recorded for 60 seconds at a pressing voltage of $V = 4.22$\,V and electron density $n_e\,=\,2.2\times {10^{6}}$~\si{cm^{-2}}. The MW (139\,GHz) is switched On/Off where indicated. (b) Enlarged reddened 1.4~s part of (a) used for analysis, where the inset shows a further zoom. (c) Wavelet transform with frequency resolution~3~Hz. (d) Time-averaged wavelet power of signals in (b) showing the fundamental frequency and higher harmonics. The $y$-axis of (c) and the $x$-axes of (d) are logarithmic.}
\label{fig:222}
\end{figure*}

\clearpage

%\newpage
By implementing the wavelet transform on all the signals, for both low and high electron density as shown in figures~(\ref{fig:onesix}~-~\ref{fig:22}) and ~(\ref{fig:216}~-~\ref{fig:222}), respectively, while varying the pressing voltage from 4.16~V to 4.22~V, we observe that the the frequency varies with pressing voltage. For both electron densities, we found that the oscillations for all five electrodes (C, E1, E2, E3 and E4) occurred at the same frequency. We observe in the time-frequency representation that there is more than one frequency component for each electrode. We have therefore checked in each case whether the the second frequency is an independent mode, or is a higher harmonic of the first mode, by using the harmonic finder method (see section~\ref{sec:H}). This tool extracted the phases from the TFR of the time series, divided them into bins, and then computed the mutual information between two of the phases. High mutual information implies that the phases are mutually dependent. In each case it was found that the second frequency is a higher harmonic of the main mode, resulting from nonlinearity in the interaction between the electron and the surface of the liquid helium. At low electron density, the frequency-modulated oscillations are clearly seen at pressing voltages from 4.16 to 4.19 V. The frequencies are most well-defined for $V\,=\,4.20$~V. We therefore claim that a small electron density together with a pressing voltage of 4.20~V represent resonance conditions for this system. Data for high electron density appear to follow the trend that a larger current amplitude causes a higher frequency of oscillation, which means physically that the oscillations are more pronounced relative to the background. Some small peaks are clearly evident in the E3 signal at 1kHz in figures~\ref{fig:220} and fig~\ref{fig:222}. They may result from the intermittency (modulation). This modulation implies that the electron direction is changed.

\clearpage
\newpage

%%%%%%%%%%%%%%%%%%%%%%%%%%%%%%%%%%%%%5
%%%%%%%%%%%%%%%%%%%%%%%%%%%%%%%%%%%%%%%%%%%%%55555555555
%%%%%%%%%%%%%%%%%%%%%%%%%%%%%%%%%%%%%%%%%%%%%%%%%%%%%%%%%%%

\subsection{Time-averaged wavelet power analysis}

We now present the time-averaged wavelet power for both electron densities and the current from each of the five electrodes (C, E1, E2, E3 and E4) with pressing voltages in the range $4.16$ $\leq V \leq 4.22$~V. Fig~\ref{fig:TP}. (a) shows the time-averaged wavelet power for low electron density. It is immediately evident that significant changes occur in the spectra for all electrodes as the pressing voltage changes. We found that the most intense oscillation peak is at the centre electrode for pressing voltages $V$ in the range 4.16 to 4.19~V; the peak starts to decrease as $V$ is increased from 4.20 to 4.22~V; while, for E1, E2, E3 and E4, the most intense oscillation peak is at a pressing voltage of 4.20~V. This indicates that the most of the electrons are located under  electrode C, with some under electrodes 3 and 4 for $V \leq 4.20$~V. For higher $V$, the electrons tend to concentrate near the edge of the cell: see fig~\ref{fig:TP}(b).

 \begin{figure*}[h!]
   \centering
 \includegraphics[width=17.5cm]{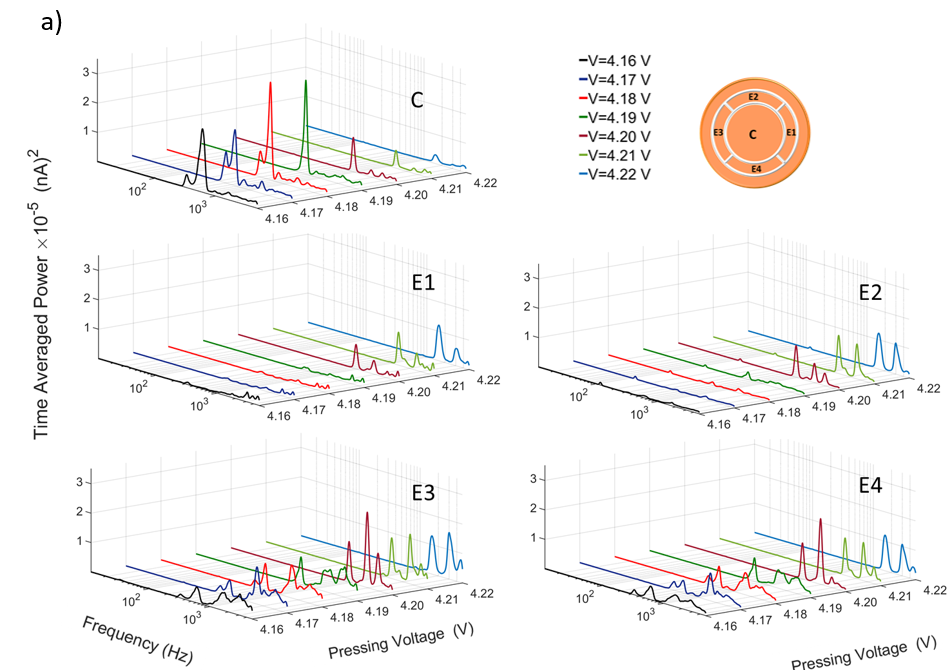} %
    \caption{(a) 3D time-averaged wavelet power of signals with MW On for each of the electrodes, for the lower $n_e$\,=\,1.4$\times {10^{6}}$~\si{cm^{-2}} electron density with different pressing voltages. The frequency axes are logarithmic. }%
    %\subfloat{{\includegraphics[width=16.5cm]{p22.PNG} }}%

   %\caption{Multiple figures in latex.}%

    \label{fig:L}%

\end{figure*}

\addtocounter{figure}{-1}

 \begin{figure*}[h!]
    \centering
    % \subfloat
    %\subfloat{{\includegraphics[width=16.5cm]{3d1.PNG} }}%

   {{\includegraphics[width=17.5cm]{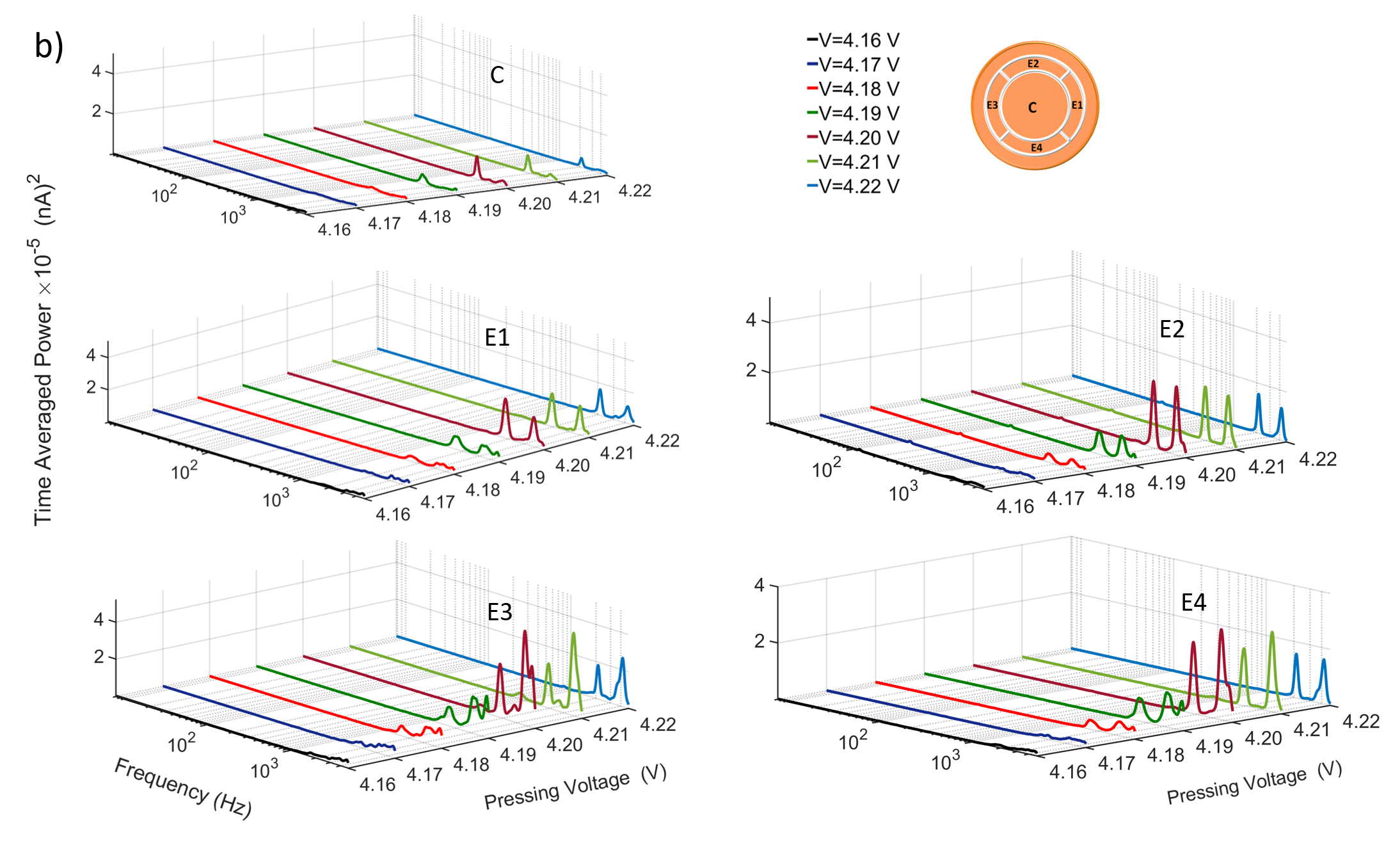} }}%

    \caption{(b) 3D time-averaged wavelet power of signals with MW On for the higher $n_e$\,=\,2.2$\times {10^{6}}$~\si{cm^{-2}} electron density and the same pressing voltages as in (a). The frequency axes are logarithmic. }%

    \label{fig:TP}%

\end{figure*}

\clearpage
\newpage

\subsection{Ridge Curve Extraction}\

\begin{figure*}[h!]
    \includegraphics[width=17.2cm]{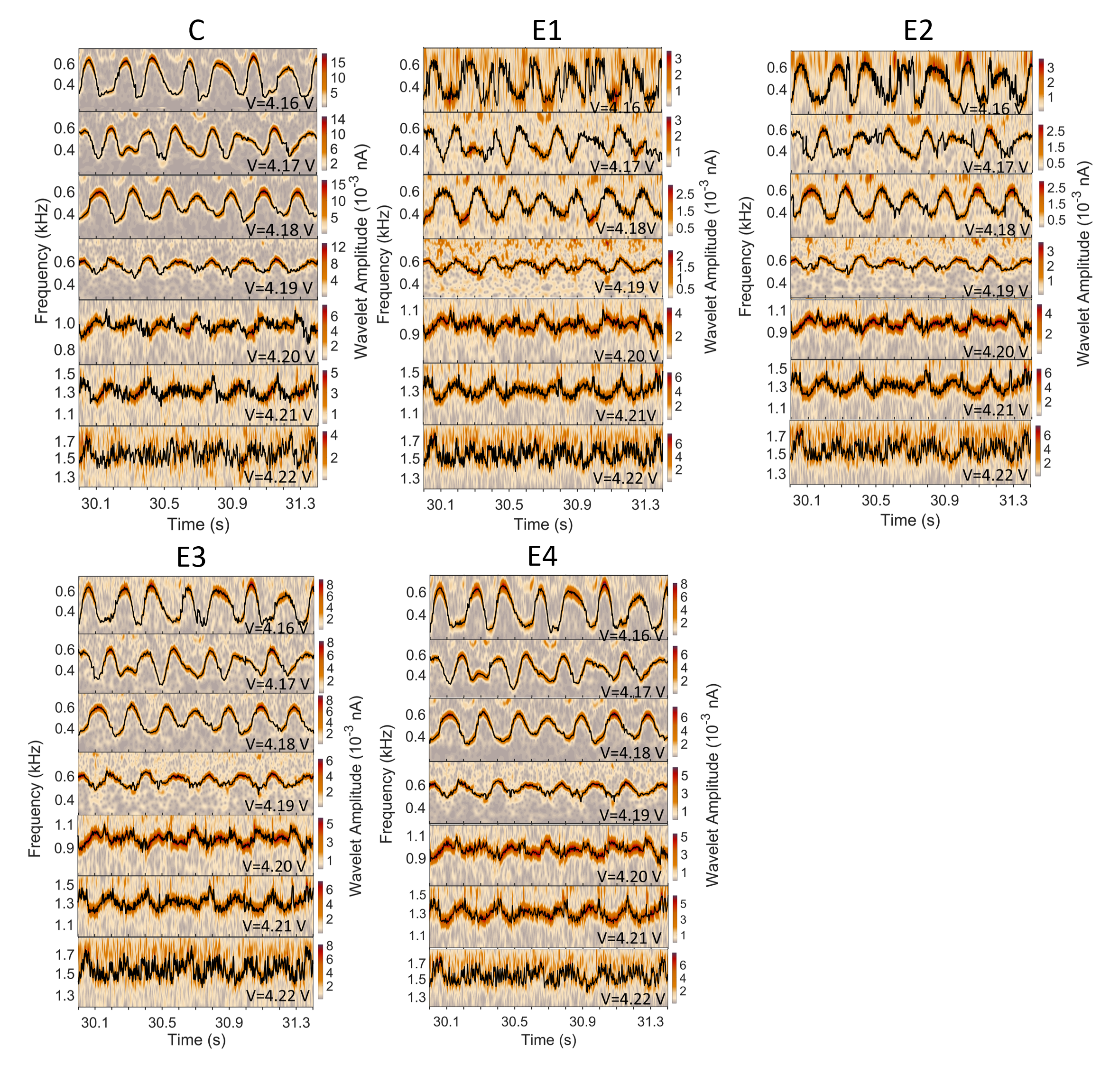}%
\caption{\label{fig:lo} Ridge extractions (black line), obtained from the time frequency representations for each of the five electrodes (C, E1, E2, E3 and E4), of the instantaneous frequency oscillations and for the main component with varying pressing voltage. The periodicity is lost for pressing voltage higher than 4.19~V. Colour bars show the intensity of the oscillations.}
\end{figure*}

\begin{figure*}[h!]
    \includegraphics[width=17.2cm]{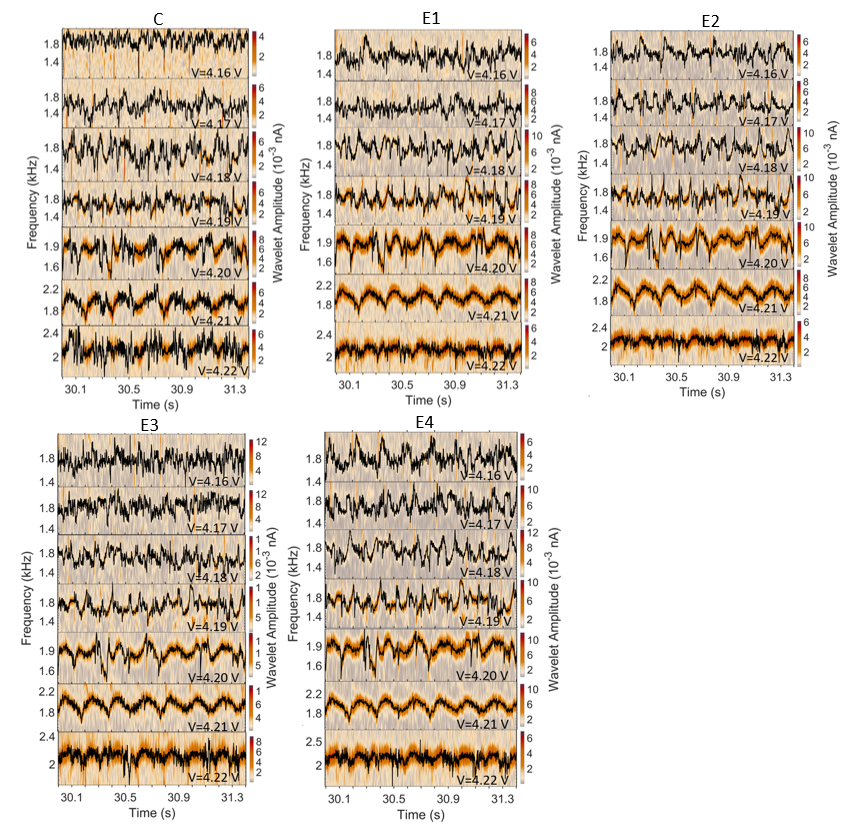}%
 \caption{\label{fig:l1o}Ridge extractions (black line), obtained from the time frequency representation method for all the five electrodes (C, E1, E2, E3 and E4), for the main component of the instantaneous frequency oscillations and  at various pressing voltages. Colour bars show the intensity of the oscillations.}
\end{figure*}

The oscillatory components are extracted by ridge extraction from the TFR. The results include the frequency and amplitude of the oscillatory components over time. Figures~\ref{fig:lo} and ~\ref{fig:l1o} show the ridge extractions (black lines), obtained from the time frequency representations for low and high electron densities, respectively, and computed for each of the five electrodes (C, E1, E2, E3 and E4). They show the instantaneous frequency of oscillation for the main component, at different pressing voltages.

\begin{figure}[H]
\centering
    \includegraphics[width=8cm]{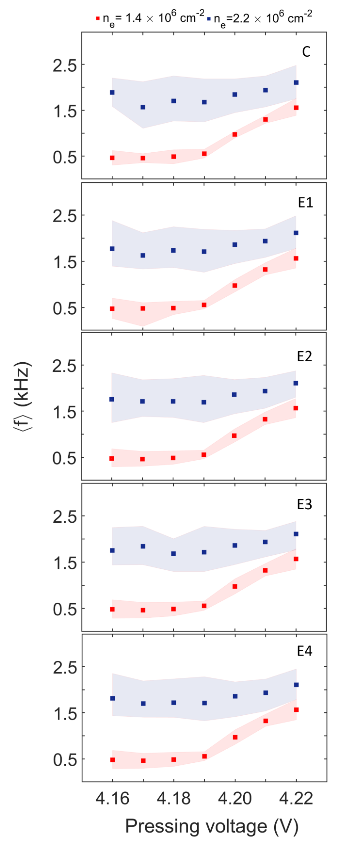}%
    \caption{\label{fig:me}The mean frequencies of oscillations at different pressing voltages for low electron density $n_e$\,=\,1.4 $\times10^{6}$~$\si{cm}^{-2}$ (red squares) and high electron density $n_e$\,=\,2.2 $\times 10^{6}$~ $\si{cm}^{-2}$ (blue squares). The shadows show the full range of frequencies of the oscillations for all electrodes. Increasing the electron density considerably decreases the dependence of frequency upon pressing voltage.
}
\end{figure}

Tables~\ref{table:T1} and~\ref{table:T2} show the mean, median, frequency modulation and standard deviation of the instantaneous oscillation frequency for different pressing voltage, calculated for each of the five electrodes for both low and high electron densities, respectively. Both tables show the maximum and minimum frequencies of the main frequency mode of oscillations. The median and the mean of the main frequency component increase with increasing pressing voltage. The frequency of modulation (number of cycles completed per second) is independent of pressing voltage and is the same for both electron densities. However, in table~\ref{table:T1}, the standard deviation of the instantaneous frequency oscillations  for the main component for all five electrodes decreases with rising pressing voltage until 4.20~V, and then starts to increase. Table ~\ref{table:T2} shows that, for all five electrodes, the standard deviation of the main component  decreases with rising pressing voltage. Figure ~\ref{fig:me} shows how the mean frequencies of the oscillations for the five electrodes change with increasing pressing voltage, at different values of electron density.  

We note that, when the electron density is $n_e$\,=\,1.4 $\times 10^{6}$~$\si{cm}^{-2}$, the mean frequency for all electrodes starts to increase at pressing voltage 4.19~V with small differences between the maxima and minima of the oscillations (red shadow). This may be due to changes in the electron distribution within the cell as the electrons start to occupy the area under the edge electrodes. 

In contrast, the mean frequency at higher electron density ($n_e$\,=\,2.2 $\times 10^{6}$~$\si{cm}^{-2}$) is higher than at the lower density and hardly depends at all on the pressing voltage applied with wide range of oscillations. Also, the change in the distribution of the electrons in the cell is weak, possibly due to the low probability of electrons moving at higher electron density. This indicates that the interaction of the electrons with the helium surface at lower electron density is larger than in the case of higher electron density. %Therefore, the ability to increase the mean frequency of the oscillations suggests that the surface of the liquid helium will be not flat due to electron gravity wave interaction.

\newpage

\newpage

\begin {table}[h!]
{\scriptsize
	%\caption {Parameters of example (3); $(.)_i$: represents $A_i$, $f_{1i}$ or $f_{2i}$} \label{tab:Parameters} 
	%\vspace*{10pt}
	\begin{center}
		\begin{tabular}{|p{1.4cm}|p{1.3cm}|p{1.3cm}|p{1.2cm}|p{2.8cm}|p{2cm}|p{2.4cm}|}
			\hline
 \multicolumn{7}{|c|}{C} \\
            \hline
			 Pressing Voltage\,(V)& $f_{max}$~(Hz) & $f_{min}$~(Hz) & Median $f$~(Hz) & Frequency modulation $f_m$~(Hz) & Mean Frequency $\langle f \rangle$~(Hz) & Standard deviation $f$~(Hz)\\
			\hline
			4.16 & 683 & 285 & 430 & 5.25 & 463 &116    \\
			\hline
			4.17 & 597 &  319 & 454 & 5.10 &  454 & 78   \\
			\hline
			4.18 & 643 & 331 & 453& 5.14 &  488 & 81   \\
			\hline
			4.19 & 656 &  474& 549& 5.22 &  553 & 47    \\
			\hline
			4.20 & 1102 &  875 & 965 & 5.25 &  975 & 44  \\
			\hline
			4.21 & 1466 & 1149& 1297& 5.25 &  1300 & 68  \\
			\hline
			4.22 & 1763 & 1358 & 1539& 5.33 & 1555 & 97  \\
			\hline
			\end {tabular}
			\label{table:1}
			\end{center}
			\vspace*{-1pt}}
\end{table}

\begin {table}[h!]
{\scriptsize
	%\caption {Parameters of example (3); $(.)_i$: represents $A_i$, $f_{1i}$ or $f_{2i}$} \label{tab:Parameters} 
	%\vspace*{10pt}
	\begin{center}
		\begin{tabular}{|p{1.4cm}|p{1.3cm}|p{1.3cm}|p{1.2cm}|p{2.8cm}|p{2cm}|p{2.4cm}|}
			\hline
 \multicolumn{7}{|c|}{E1} \\
            \hline
			 Pressing Voltage\,(V)& $f_{max}$~(Hz) & $f_{min}$~(Hz) & Median $f$~(Hz) & Frequency modulation $f_m$~(Hz) & Mean Frequency $\langle f \rangle$~(Hz) & Standard deviation $f$~(Hz)\\
			\hline
			4.16 & 701 & 261 & 349 & 5.88 & 472 &117    \\
			\hline
			4.17 & 602 & 290 & 413 & 5.00 &  479 & 80  \\
			\hline
			4.18 & 631 & 343 & 450 & 5.14 &  486 & 81   \\

			\hline
			4.19 & 650 &  468& 547& 5.18 &  555 & 46    \\
			\hline
			4.20 & 1101 &  833 & 968 & 5.25 &  971 & 44  \\
			\hline
			4.21 & 1468 & 1205& 1298& 5.25 &  1322 & 67  \\
			\hline
			4.22 & 1780 & 1351 & 1532& 5.14 & 1563 & 93  \\
			\hline
			\end {tabular}
			\label{table:parameters}
			\end{center}
			\vspace*{-1pt}}
\end{table}

\begin {table}[h!]
{\scriptsize
	%\caption {Parameters of example (3); $(.)_i$: represents $A_i$, $f_{1i}$ or $f_{2i}$} \label{tab:Parameters} 
	%\vspace*{10pt}
	\begin{center}
		\begin{tabular}{|p{1.4cm}|p{1.3cm}|p{1.3cm}|p{1.2cm}|p{2.8cm}|p{2cm}|p{2.4cm}|}
			\hline
 \multicolumn{7}{|c|}{E2} \\
           \hline
			 Pressing Voltage\,(V)& $f_{max}$~(Hz) & $f_{min}$~(Hz) & Median $f$~(Hz) & Frequency modulation $f_m$~(Hz) & Mean Frequency $\langle f \rangle$~(Hz) & Standard deviation $f$~(Hz)\\
			\hline
			4.16 & 684 & 294 & 445 & 5.10 & 470 &117    \\
			\hline
			4.17 & 627 & 306 & 480 & 5.07 &  462 & 74 \\
			\hline
			4.18 & 638 & 347 & 456 & 5.14 &  487 & 82   \\

			\hline
			4.19 & 653 &  466& 546& 5.18 &  555 & 45   \\
			\hline
			4.20 & 1111 &  826 & 967 & 5.25 &  968 & 44  \\
			\hline
			4.21 & 1479 & 1207& 1298& 5.25 &  1322 & 66  \\
			\hline
			4.22 & 1771 & 1358 & 1531& 5.25 & 1568 & 92  \\
			\hline
			\end {tabular}
			\label{table:parameters}
			\end{center}
			\vspace*{-1pt}}
\end{table}

\begin {table}[h!]
{\scriptsize
	%\caption {Parameters of example (3); $(.)_i$: represents $A_i$, $f_{1i}$ or $f_{2i}$} \label{tab:Parameters} 
	%\vspace*{10pt}
	\begin{center}
		\begin{tabular}{|p{1.4cm}|p{1.3cm}|p{1.3cm}|p{1.2cm}|p{2.8cm}|p{2cm}|p{2.4cm}|}
			\hline
 \multicolumn{7}{|c|}{E3} \\
            \hline
			 Pressing Voltage\,(V)& $f_{max}$~(Hz) & $f_{min}$~(Hz) & Median $f$~(Hz) & Frequency modulation $f_m$~(Hz) & Mean Frequency $\langle f \rangle$~(Hz) & Standard deviation $f$~(Hz)\\
			\hline
			4.16 & 685 & 285 & 442 & 5.00 & 477 &117    \\
			\hline
			4.17 & 627 & 290 & 455 & 5.18 &  457 & 79 \\
			\hline
			4.18 & 634 & 336 & 452 & 5.18 &  484 & 81   \\

			\hline
			4.19 & 656 &  456& 551& 5.18 &  557 & 45   \\
			\hline
			4.20 & 1135 &  810 & 965 & 5.25 &  971 & 45  \\
			\hline
			4.21 & 1469 & 1200& 1296& 5.25 &  1322 & 65  \\
			\hline
			4.22 & 1788 & 1347 & 1534& 5.30 & 1568 & 96  \\
			\hline
			\end {tabular}
			\label{table:parameters}
			\end{center}
			\vspace*{-1pt}}
\end{table}

\begin {table}[h!]
{\scriptsize
	%\caption {Parameters of example (3); $(.)_i$: represents $A_i$, $f_{1i}$ or $f_{2i}$} \label{tab:Parameters} 
	%\vspace*{10pt}
	\begin{center}
		\begin{tabular}{|p{1.4cm}|p{1.3cm}|p{1.3cm}|p{1.2cm}|p{2.8cm}|p{2cm}|p{2.4cm}|}
			\hline
 \multicolumn{7}{|c|}{E4} \\
             \hline
			 Pressing Voltage\,(V)& $f_{max}$~(Hz) & $f_{min}$~(Hz) & Median $f$~(Hz) & Frequency modulation $f_m$~(Hz) & Mean Frequency $\langle f \rangle$~(Hz) & Standard deviation $f$~(Hz)\\
			\hline
			4.16 & 684 & 289 & 433 & 5.01 & 477 &117    \\
			\hline
			4.17 & 620 & 295 & 457 & 5.18 &  459 & 78 \\
			\hline
			4.18 & 643 & 338 & 453 & 5.14 &  488 & 81   \\
			\hline
			4.19 & 656 &  462& 547& 5.18 &  557 & 47   \\
			\hline
			4.20 & 1131 &  815 & 966 & 5.25 &  969 & 43  \\
			\hline
			4.21 & 1471 & 1207& 1299& 5.25 &  1321 & 66  \\
			\hline
			4.22 & 1786 & 1347 & 1534& 5.25 & 1567 & 96  \\
			\hline
			\end {tabular}
			\caption {Maximum, minimum, median frequency, frequency modulation, mean frequency and deviation of the instantaneous frequency oscillations  for the main component with varying pressing voltage calculated to each of the five electrodes at low electron density.}
			\label{table:T1}
			\end{center}
			\vspace*{-1pt}}
\end{table}

%\begin{figure*}[h!]
    %\includegraphics[width=7.2cm]{Ec.PNG}%
%\end{figure*}\\

%\begin{figure*}[h!]
    %\includegraphics[width=7cm]{E1.PNG}%
%\end{figure*}\\

%\begin{figure*}[h!]
    %\includegraphics[width=7cm]{E2.PNG}%
%\end{figure*}\\

%\begin{figure*}[h!]
   % \includegraphics[width=7cm]{E3.PNG}%
%\end{figure*}\\

\newpage
\clearpage
\begin {table}[h!]
{\scriptsize
	%\caption {Parameters of example (3); $(.)_i$: represents $A_i$, $f_{1i}$ or $f_{2i}$} \label{tab:Parameters} 
	%\vspace*{10pt}
	\begin{center}
		\begin{tabular}{|p{1.4cm}|p{1.3cm}|p{1.3cm}|p{1.2cm}|p{2.8cm}|p{2cm}|p{2.4cm}|}
			\hline
 \multicolumn{7}{|c|}{C}\\
             \hline
			 Pressing Voltage\,(V)& $f_{max}$~(Hz) & $f_{min}$~(Hz) & Median $f$~(Hz) & Frequency modulation $f_m$~(Hz) & Mean Frequency $\langle f \rangle$~(Hz) & Standard deviation $f$~(Hz)\\
			\hline
			4.16 & 2203 & 1583 & 1888 & 5.00 & 1888 &143    \\
			\hline
			4.17 & 2122 &  1106 & 1563 & 5.00 &  1567 & 162   \\
			\hline
			4.18 & 2248 & 1265 & 1669& 5.80 &  1704 & 180   \\
			\hline
			4.19 & 2184 &  1246& 1669& 5.00 &  1675 & 143    \\
			\hline
			4.20 & 2184 &  1447 &1845 & 5.00 &  1842 & 121  \\
			\hline
			4.21 & 2247 & 1572& 1932& 5.10 &  1939 & 108 \\
			\hline
			4.22 & 2380 & 1750 & 2116& 5.50 & 2105 & 68 \\
			\hline
			\end {tabular}
			\label{table:parameters}
			\end{center}
			\vspace*{-1pt}}
\end{table}

\begin {table}[h!]
{\scriptsize
	%\caption {Parameters of example (3); $(.)_i$: represents $A_i$, $f_{1i}$ or $f_{2i}$} \label{tab:Parameters} 
	%\vspace*{10pt}
	\begin{center}
		\begin{tabular}{|p{1.4cm}|p{1.3cm}|p{1.3cm}|p{1.2cm}|p{2.8cm}|p{2cm}|p{2.4cm}|}
			\hline
 \multicolumn{7}{|c|}{E1} \\
            \hline
			 Pressing Voltage\,(V)& $f_{max}$~(Hz) & $f_{min}$~(Hz) & Median $f$~(Hz) & Frequency modulation $f_m$~(Hz) & Mean Frequency $\langle f \rangle$~(Hz) & Standard deviation $f$~(Hz)\\
			\hline
			4.16 & 2380 & 1392 & 1759 & 4.60 & 1770 &162    \\
			\hline
			4.17 & 2117 &  1331& 1610 & 5.20 &  1623 & 129   \\
			\hline
			4.18 & 2247& 1363 & 1726& 5.00 &  1735 & 162   \\

			\hline
			4.19 & 2190&  1260& 1687& 5.00 &  1706 & 161    \\
			\hline
			4.20 & 2168 &  1448 & 1866 & 5.00 &  1858 & 111  \\
			\hline
			4.21 & 2200 & 1583& 1938 & 5.10 &  1935 & 90 \\
			\hline
			4.22 & 2484 & 1775 & 2105& 5.70 & 2114 & 123 \\
			\hline
			\end {tabular}
			\label{table:parameters}
			\end{center}
			\vspace*{-1pt}}
\end{table}

\begin {table}[h!]
{\scriptsize
	%\caption {Parameters of example (3); $(.)_i$: represents $A_i$, $f_{1i}$ or $f_{2i}$} \label{tab:Parameters} 
	%\vspace*{10pt}
	\begin{center}
		\begin{tabular}{|p{1.4cm}|p{1.3cm}|p{1.3cm}|p{1.2cm}|p{2.8cm}|p{2cm}|p{2.4cm}|}
			\hline
 \multicolumn{7}{|c|}{E2} \\
              \hline
			 Pressing Voltage\,(V)& $f_{max}$~(Hz) & $f_{min}$~(Hz) & Median $f$~(Hz) & Frequency modulation $f_m$~(Hz) & Mean Frequency $\langle f \rangle$~(Hz) & Standard deviation $f$~(Hz)\\
			\hline
			4.16 & 2329& 1251 & 1740 & 5.70 & 1758 &171    \\
			\hline
			4.17 & 2180 &  1382& 1669 & 5.00 &  1712 & 155   \\
			\hline
			4.18 & 2204& 1363 & 1701& 5.30 &  1712 & 149   \\

			\hline
			4.19 & 2273&  1251& 1674& 4.80 &  1690 & 159    \\
			\hline
			4.20 & 2184 &  1481 & 1864 & 4.80 &  1856 & 111  \\
			\hline
			4.21 & 2231 & 1561& 1937 & 5.10 &  1934 & 89\\
			\hline
			4.22 & 2373 & 1779 & 2114& 5.70 & 2100& 57 \\
			\hline
			\end {tabular}
			\label{table:parameters}
			\end{center}
			\vspace*{-1pt}}
\end{table}

\begin {table}[h!]
{\scriptsize
	%\caption {Parameters of example (3); $(.)_i$: represents $A_i$, $f_{1i}$ or $f_{2i}$} \label{tab:Parameters} 
	%\vspace*{10pt}
	\begin{center}
		\begin{tabular}{|p{1.4cm}|p{1.3cm}|p{1.3cm}|p{1.2cm}|p{2.8cm}|p{2cm}|p{2.4cm}|}
			\hline
 \multicolumn{7}{|c|}{E3} \\
           \hline
			 Pressing Voltage\,(V)& $f_{max}$~(Hz) & $f_{min}$~(Hz) & Median $f$~(Hz) & Frequency modulation $f_m$~(Hz) & Mean Frequency $\langle f \rangle$~(Hz) & Standard deviation $f$~(Hz)\\
			\hline
			4.16 & 2247 & 1437& 1742& 4.60 & 1754 &122    \\
			\hline
			4.17 & 2274&  1443& 1845 & 5.30 &  1838 & 146   \\
			\hline
			4.18 & 2005& 1295 & 1668& 5.20 &  1680 & 139   \\

			\hline
			4.19 & 2272&  1295& 1703& 5.00 &  1715 & 156    \\
			\hline
			4.20 & 2210 &  1445 & 1868& 5.00 &  1860 & 111  \\
			\hline
			4.21 & 2184 & 1611& 1938 & 5.10 &  1937 & 90 \\
			\hline
			4.22 & 2380 & 1763 & 2115& 5.70 & 2106 & 74\\
			\hline
			\end {tabular}
			\label{table:parameters}
			\end{center}
			\vspace*{-1pt}}
\end{table}

\begin {table}[h!]
{\scriptsize
	%\caption {Parameters of example (3); $(.)_i$: represents $A_i$, $f_{1i}$ or $f_{2i}$} \label{tab:Parameters} 
	%\vspace*{10pt}
	\begin{center}
		\begin{tabular}{|p{1.4cm}|p{1.3cm}|p{1.3cm}|p{1.2cm}|p{2.8cm}|p{2cm}|p{2.4cm}|}
			\hline
 \multicolumn{7}{|c|}{E4} \\
            \hline
			 Pressing Voltage\,(V)& $f_{max}$~(Hz) & $f_{min}$~(Hz) & Median $f$~(Hz) & Frequency modulation $f_m$~(Hz) & Mean Frequency $\langle f \rangle$~(Hz) & Standard deviation $f$~(Hz)\\
			\hline
			4.16 & 2359 & 1436& 1792& 5.70 & 1811 &162    \\
			\hline
			4.17 & 2189&  1402& 1666 & 4.70 &  1700 & 142   \\
			\hline
			4.18 & 2234& 1400 & 1703& 5.10 &  1715 & 163   \\

			\hline
			4.19 & 2280&  1324& 1703& 4.80 &  1713 & 162    \\
			\hline
			4.20 & 2168 &  1413 & 1865& 5.00 &  1858 & 113  \\
			\hline
			4.21 & 2231 & 1539& 1939 & 5.10 &  1935 & 88 \\
			\hline
			4.22 & 2449 & 1788 & 2115& 5.70 & 2105 & 69\\
			\hline
			\end {tabular}
				\caption {Maximum, minimum, median frequency, frequency modulation, mean frequency and deviation of the instantaneous frequency oscillations  for the main component with varying pressing voltage calculated to each of the five electrodes at high electron density.}
			\label{table:T2}
			\end{center}
			\vspace*{-1pt}}
\end{table}

\newpage

%\endfoot

\newpage

\subsection {Phase coherence and phase difference and motion of the electrons}
We now present the results of the phase coherence analysis and the phases differences for both electron densities (lower and higher). We also present the inferred motion of the electrons inside the cell, as obtained from the phase difference. We calculated the phase coherence and the phase difference as we described in section~(\ref{sec:1}). Next, we computed the values of the maximum coherence and the corresponding frequencies as functions of the pressing voltage, which is provided in the tables for low and high electron density.
  \begin{itemize}
       \item [a)] low electron density
   \end{itemize}

\begin{figure}[h!]%
    \includegraphics[width=17cm]{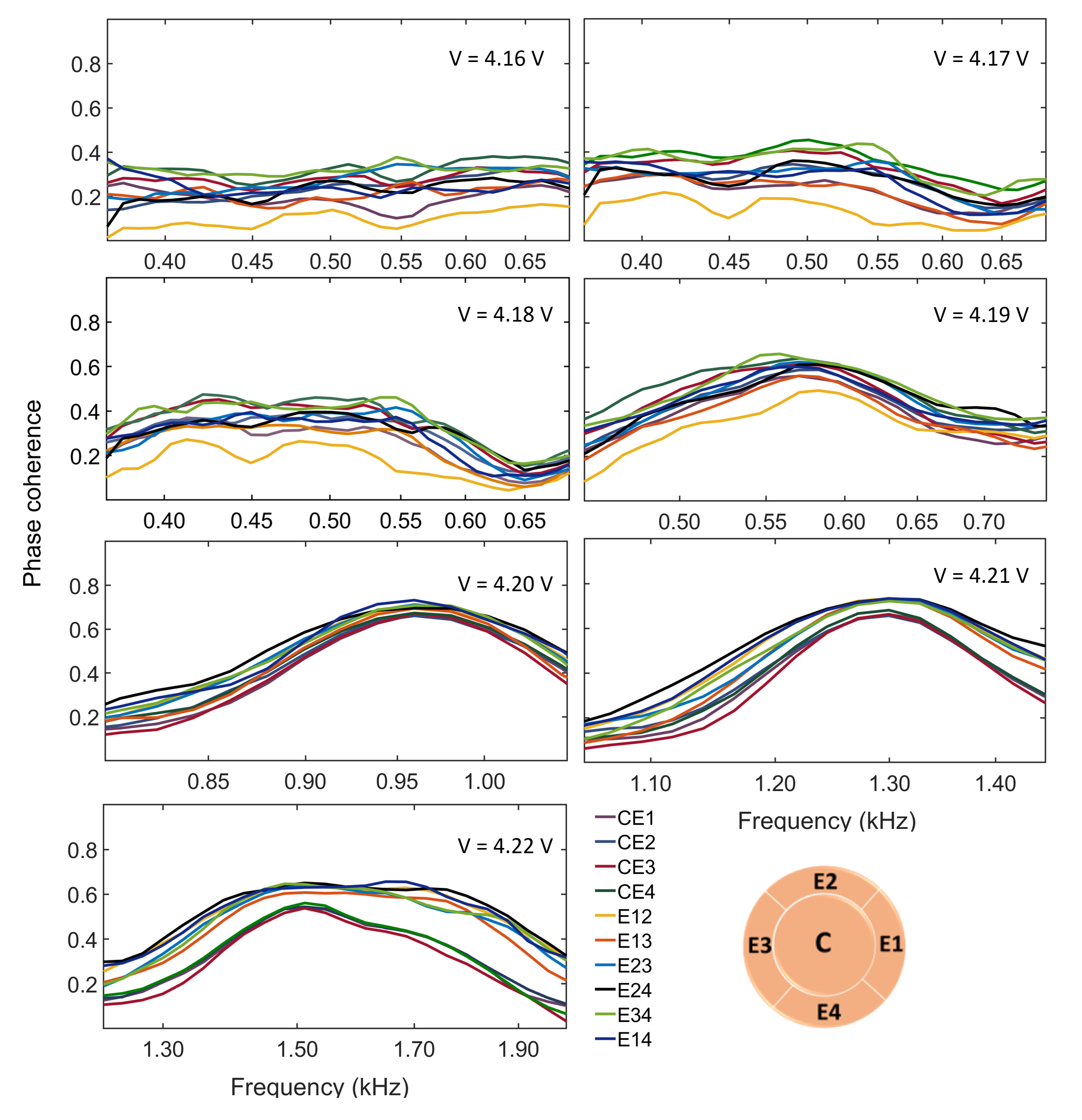} %
    \caption{ Significant wavelet phase coherence of pairs of electrodes for an electron density of $n_e$\,=\,1.4 $\times 10^{6}$~\si{cm^{-2}} at different pressing voltages. The lines are color-coded to indicate the particular electrode pairs.}%
    \label{fig:coh}%
\end{figure}

\begin{figure}[h!]%
    \includegraphics[width=17.2cm]{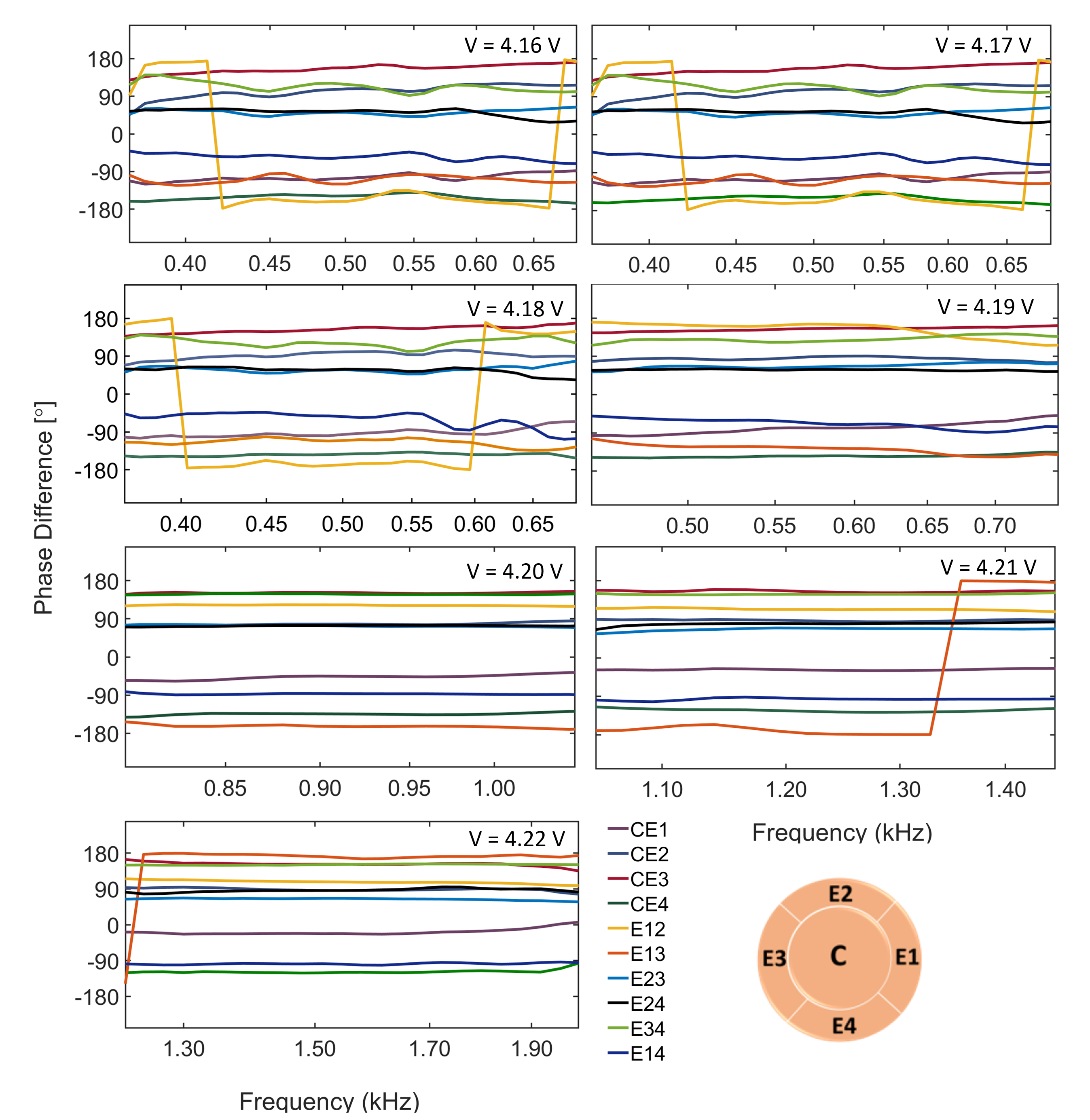} %
    \caption{ The phase difference of pairs of electrodes for electron density of $n_e$\,=\,1.4 $\times 10^{6}$ ~\si{cm^{-2}} at different pressing voltages. The lines are color-coded to indicate particular electrode pairs.}%
    \label{fig:phas}%
\end{figure}

%\endfoot
\newpage

\begin{figure}[h!]%
    \includegraphics[width=12cm]{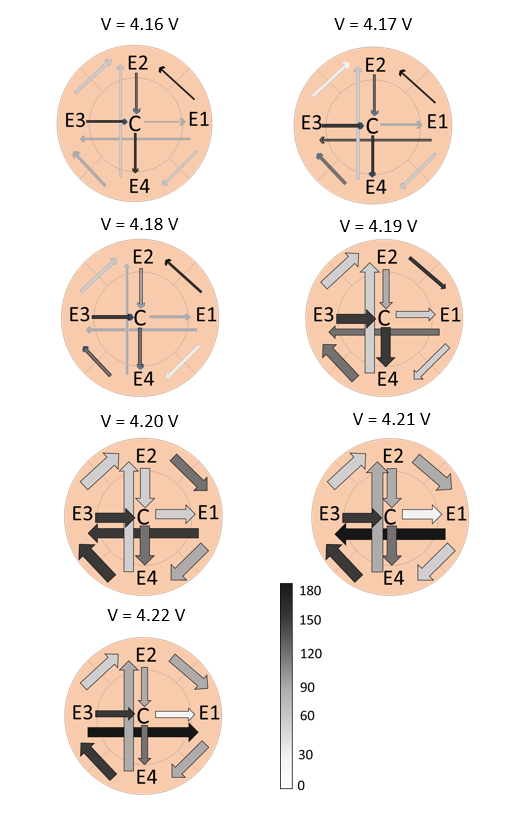} %
    \caption{ The circular schematic shows the movement of the electrons below the electrodes for different pressing voltages with electron density $n_e$\,=\,1.4 $\times 10^{6}$~\si{cm^{-2}}. It summarises the coherences and phases shifts between the currents. The thickness of the arrows indicates the magnitude of the coherence (fig~\ref{fig:coh}) and the white/black shading the size of the phase shift (fig~\ref{fig:phas})}%
    \label{fig:mov}%
\end{figure}

%\afterpage
\clearpage

\newpage

\begin {table}[!h]
{\scriptsize
	%\caption {Parameters of example (3); $(.)_i$: represents $A_i$, $f_{1i}$ or $f_{2i}$} \label{tab:Parameters} 
	\vspace*{25pt}
	\begin{center}
		\begin{tabular}{|p{1.1cm}|p{1.3cm}|p{1.3cm}|p{1.3cm}|p{1.3cm}|p{1.3cm}|p{1.3cm}|p{1.3cm}|p{1.3cm}|p{1.3cm}|p{1.3cm}|}
			\hline
 \multicolumn {11} {|c|} {Max Coherence}\\
            \hline
			 Pressing Voltage{ }(V) & CE1 & CE2 & CE3{} & CE4{} & E12{} & E13{} & E23{} & E24{} & E34{} & E14{}\\
			\hline
			4.16 & 0.17 & 0.25& 0.27& 0.3 & 0.11 & 0.25 & 0.33 &0.24 & 0.35 & 0.23\\
			\hline
			4.17 & 0.27 & 0.37& 0.40& 0.44 & 0.18 & 0.26& 0.36 & 0.40 &0.40& 0.31\\
			\hline
			4.18 & 0.32 & 0.36& 0.42& 0.45 & 0.24 & 0.31 & 0.41 &0.38 &0.46&0.35\\
			\hline
		     4.19 & 0.54& 0.58& 0.60& 0.63 & 0.48 & 0.54 & 0.61 &0.60 &0.65&0.59\\
			\hline
			4.20 & 0.66 &0.66& 0.66&0.67 & 0.69 & o.71 & 0.70 &0.70 &0.70&0.73\\
			\hline
			4.21 & 0.64 & 0.64& 0.65& 0.67& 0.72 & 0.71 & 0.72 &0.72 & 0.71 &0.73\\
			\hline
		    4.22 & 0.53&0.54& 0.53& 0.56& 0.63 &0.60 & 0.64 &0.65&0.64&0.63\\
			\hline
			\end {tabular}
				%\caption {The mean, median, frequency modulation and the stander deviation of the instantaneous frequency oscillations and for the main component with varying pressing voltage calculated to each five electrodes at high electron density ($n_e$=2.2 $\times 10^{6}$ $\si{cm}^{-2}$)}
			%\label{table:prs}
			\end{center}
			\vspace*{1pt}}
\end{table}

\begin {table}[h!]
{\scriptsize
	%\caption {Parameters of example (3); $(.)_i$: represents $A_i$, $f_{1i}$ or $f_{2i}$} \label{tab:Parameters} 
	\vspace*{25pt}
	\begin{center}
		\begin{tabular}{|p{1.1cm}|p{1.3cm}|p{1.3cm}|p{1.3cm}|p{1.3cm}|p{1.3cm}|p{1.3cm}|p{1.3cm}|p{1.3cm}|p{1.3cm}|p{1.3cm}|}
			\hline
 \multicolumn {11} {|c|} {Frequency of Max coherence}\\
            \hline
			 Pressing Voltage{ }(V) & $f$CE1{}(Hz) &  $f$CE2{}(Hz) & $f$CE3{}(Hz) & $f$CE4{}(Hz) & $f$E12{}(Hz) & $f$E13{}(Hz) & $f$E23{}(Hz) & $f$E24{}(Hz) & $f$E34{}(Hz) & $f$E14{}(Hz)\\
			\hline
			4.16 & 490 & 523& 523& 512 & 501 & 558 & 558 &501 & 546 & 570\\
			\hline
			4.17 & 512 & 490& 490& 501 & 490 & 523 & 546 & 490 &512& 501\\
			\hline
			4.18 & 523 & 523& 423& 523 & 512 & 534 & 546 &512 &534&523\\
			\hline
		     4.19 & 570 & 570& 570& 570 & 570 & 583 & 570 &570 &570&570\\
			\hline
			4.20 & 959 &959& 959& 959 & 959 & 959 & 959 &959 &959&959\\
			\hline
			4.21 & 1300 & 1300& 1300& 1300 & 1300 & 1300 & 1300 &1300 & 1300 &1300\\
			\hline
		    4.22 & 1512&1512& 1512& 1512 & 1512 & 1512 & 1512 &1512 &1512&1512\\
			\hline
			\end {tabular}
				\caption {The maximum coherence and the frequency of the maximum coherence for the lower electron density, calculated from the phase coherence shown in figure~\ref{fig:coh}.}
				
			\label{table:m1}
			\end{center}
			\vspace*{1pt}}
\end{table}

%\afterpage{
\clearpage

\newpage
 \begin{itemize}
       \item [b)] High electron density
   \end{itemize}

\begin{figure}[h!]%
    \includegraphics[width=17.2cm]{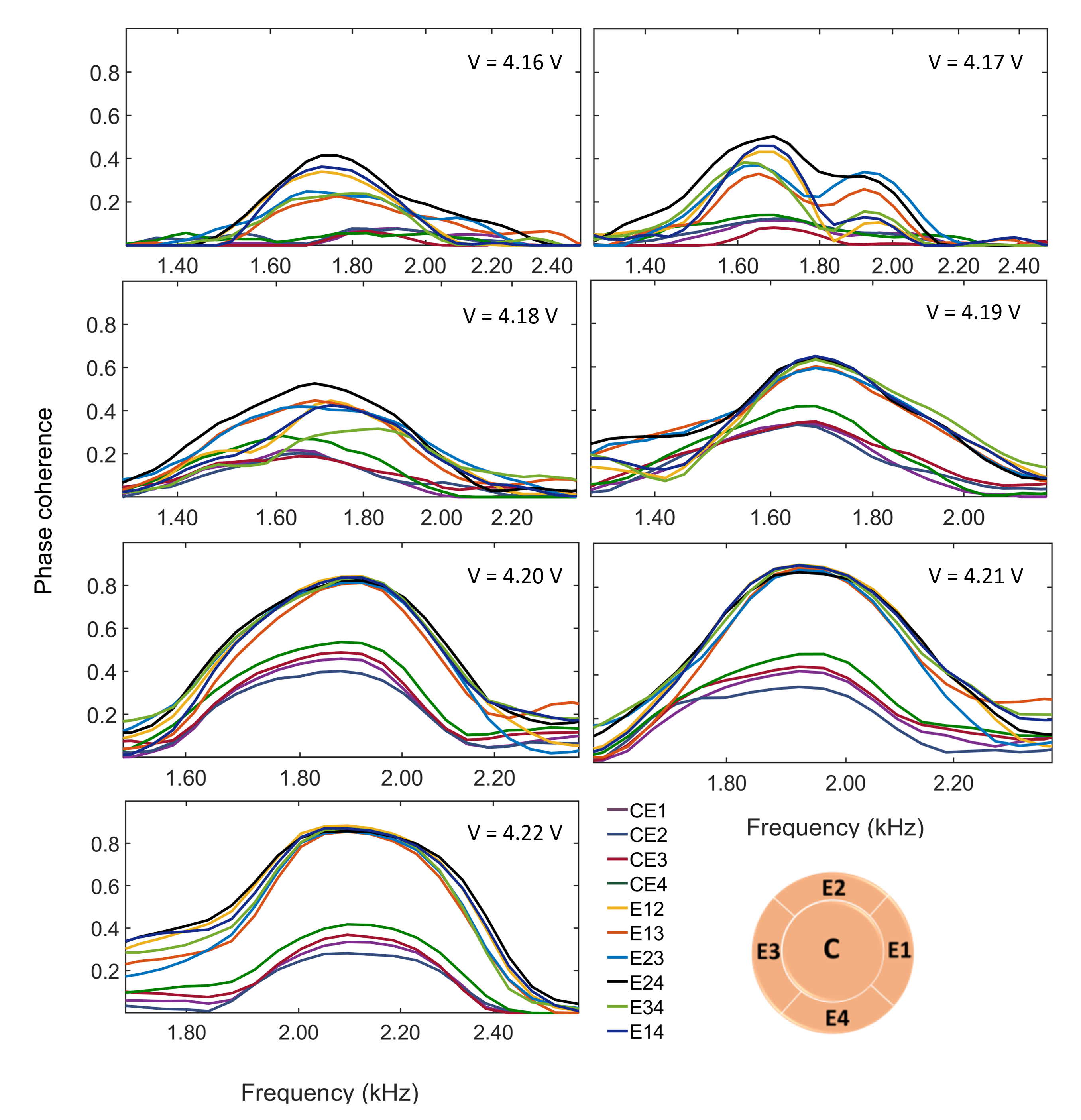} %
    \caption{ Significant wavelet phase coherence of pairs of electrodes for electron density $n_e$\,=\,2.2 $\times 10^{6}$~\si{cm^{-2}} at different pressing voltages. The lines are color-coded to indicate particular electrode pairs.}%
    \label{fig:coh2}%
\end{figure}

   \begin{figure}[h!]%
    \includegraphics[width=17.2cm]{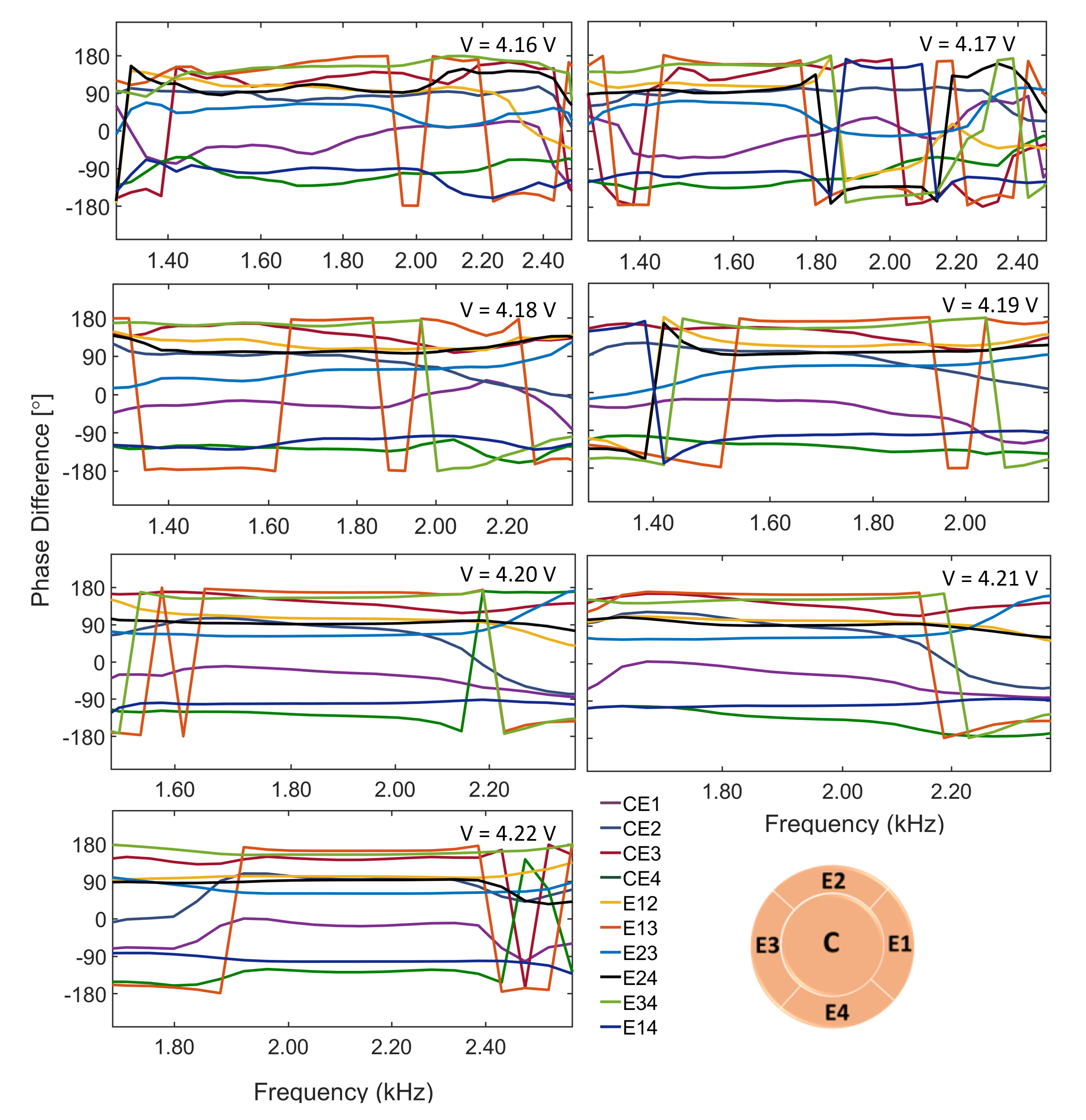} %
    \caption{ The phase difference of pairs of electrodes for electron density $n_e$\,=\,2.2 $\times 10^{6}$ ~\si{cm^{-2}} at different pressing voltages. The lines are color-coded to indicate particular electrode pairs. }%
    \label{fig:phas2}%
\end{figure}

\begin{figure}[h!]%
    \includegraphics[width=12cm]{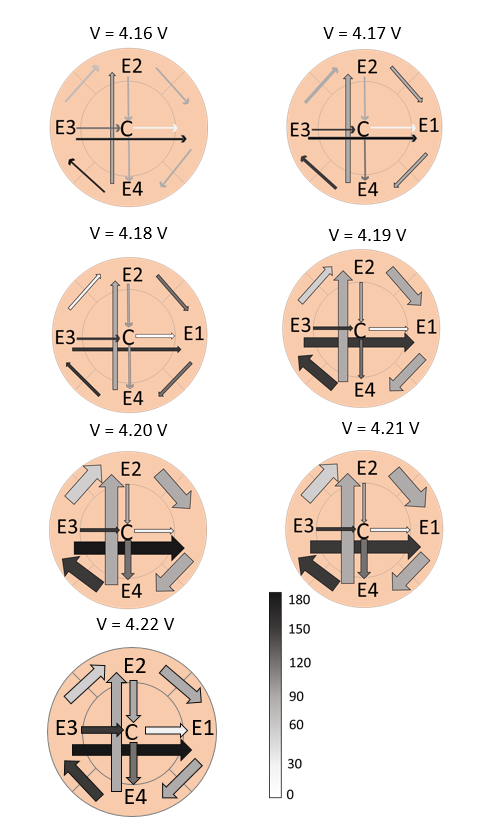} %
    \caption{ The circular schematics of the motion of the electrons below the electrodes at different pressing voltages and at electron density $n_e$\,=\,2.2 $\times 10^{6}$~\si{cm^{-2}}. It summarises the coherences and phases shifts between the currents. The thickness of the arrows indicates the magnitude of the coherence (fig~\ref{fig:coh2}) and the white/black shading the size of the phase shift (fig~\ref{fig:phas2})}%
    \label{fig:mov2}%
\end{figure}

\clearpage

\begin {table}[h!]
{\scriptsize
	%\caption {Parameters of example (3); $(.)_i$: represents $A_i$, $f_{1i}$ or $f_{2i}$} \label{tab:Parameters} 
	\vspace*{25pt}
	\begin{center}
		\begin{tabular}{|p{1.1cm}|p{1.3cm}|p{1.3cm}|p{1.3cm}|p{1.3cm}|p{1.3cm}|p{1.3cm}|p{1.3cm}|p{1.3cm}|p{1.3cm}|p{1.3cm}|}
			\hline
 \multicolumn {11} {|c|} { Max Coherence}\\
            \hline
			 Pressing Voltage{ }(V) & CE1 & CE2 & CE3{} & CE4{} & E12{} & E13{} & E23{} & E24{} & E34{} & E14{}\\
			\hline
			4.16 & 0.076 & 0.077& 0.065 & 0.057 & 0.34 & 0.22 & 0.25 & 0.41 & 0.24 & 0.36\\
			\hline
			4.17 & 0.11 & 0.12& 0.08 & 0.14 & 0.43 & 0.33& 0.37 & 0.5 &0.38& 0.46\\
			\hline
			4.18 & 0.20 & 0.22 & 0.20& 0.28 & 0.44 & 0.44 & 0.41 &0.52 & 0.31 & 0.42\\
			\hline
		     4.19 & 0.33& 0.33& 0.34& 0.41 & 0.64 & 0.60 & 0.59 & 0.63 &0.63 & 0.65\\
			\hline
			4.20 & 0.46 & 0.40 & 0.48 & 0.53 & 0.84 & 0.81 & 0.81 &0.82 &0.83 & 0.84\\
			\hline
			4.21 & 0.41 & 0.34& 0.43& 0.5& 0.90 & 0.88 & 0.87 &0.86 & 0.89 &0.89\\
			\hline
		    4.22 & 0.33 &0.28& 0.36& 0.41& 0.88 &0.85 & 0.85 &0.85&0.87&0.87\\
			\hline
			\end {tabular}
			%	\caption {The maximum coherence and the frequency value of the maximum coherence for the high electron density that we calculated from the phase coherence figure~\ref{fig:coh2}}
			%\label{table:prs}
			\end{center}
			\vspace*{1pt}}
\end{table}

\begin {table}[h!]
{\scriptsize
	%\caption {Parameters of example (3); $(.)_i$: represents $A_i$, $f_{1i}$ or $f_{2i}$} \label{tab:Parameters} 
	\vspace*{25pt}
	\begin{center}
		\begin{tabular}{|p{1.1cm}|p{1.3cm}|p{1.3cm}|p{1.3cm}|p{1.3cm}|p{1.3cm}|p{1.3cm}|p{1.3cm}|p{1.3cm}|p{1.3cm}|p{1.3cm}|}
			\hline
 \multicolumn {11} {|c|} {Frequency of the Max coherence}\\
            \hline
			 Pressing Voltage (V) & $f$CE1\,(Hz) &  $f$CE2\,(Hz) & $f$CE3\,(Hz) & $f$CE4\,(Hz) & $f$E12\,(Hz) & $f$E13\,(Hz) & $f$E23\,(Hz) & $f$E24\,(Hz) & $f$E34\,(Hz) & $f$E14\,(Hz)\\
			\hline
			4.16 & 1878 & 1798& 1838& 1798& 1722 & 1760 & 1685 &1722 & 1708 & 1722\\
			\hline
			4.17 & 1685 & 1685& 1685& 1685 & 1685 & 1649 & 1649 &1685 &1614& 1685\\
			\hline
			4.18 & 1614 & 1649& 1649& 1714 & 1722 & 1685 & 1649 & 1685 & 1798 &   1722\\
			\hline
		     4.19 & 1685 & 1649 & 1685& 1685 & 1685 & 1685 & 1685 & 1685 & 1684 &1685\\
			\hline
			4.20 & 1878 & 1878 & 1878 & 1878 & 1878& 1919 & 1878 & 1919 & 1878 & 1778\\
			\hline
			4.21 & 1919 & 1919& 1919& 1961 & 1919 & 1919 & 1919 & 1919 & 1919 &1919\\
			\hline
		    4.22 & 2093 & 2093 & 2093 & 2093 & 2048 & 2048 & 2048 &2048 & 2048 & 2048\\
			\hline
			\end {tabular}
				\caption {The maximum coherence and the frequency of the maximum coherence for the high electron density, calculated from the phase coherence shown in figure~\ref{fig:coh2}.}
			\label{table:m2}
			\end{center}
			\vspace*{1pt}}
\end{table}

Significant coherence was found between the signals from paired electrodes recorded for different  pressing voltages at both low and high electron density, as shown in figs~\ref{fig:coh} and~\ref{fig:coh2}. At low electron density (Fig.~\ref{fig:coh}), clear coherence peaks were well-defined and mostly constant for all electrode pairs with pressing voltage 4.19 and 4.20~V. Therefore, we can say that the resonance condition is satisfied at pressing voltage V\,=\,4.20~V which means that the oscillations are uniform over the electrodes that have been measured.  At high electron density (Fig.~\ref{fig:coh2}), the peaks in phase coherence are less dependent on the pressing voltage. The phase coherence is lower in the case when the central electrode is paired with one of the four edge electrodes, whilst the coherence is higher for pairings of the four edge electrodes. 

For the phase difference, we can see in Fig.~\ref{fig:phas} that the phase difference in the case of low electron density is constant at pressing voltage 4.19 and 4.20~V. However, the phase difference for the high electron density (intervals Fig.~\ref{fig:phas2} exhibits more slipping and they complete 360 degrees compared to the Fig.~\ref{fig:phas}). They happen when the coherence is quite low (although still nonzero), and that may result of a phase transitions due to external perturbation.

Studying the phase difference helps us to reveal the direction of electron motion inside the cell. The circular schematics in Figs.~\ref{fig:mov} and~\ref{fig:mov2} show the movement of the electrons below the electrodes inside the cell at different pressing voltages for two values of electron densities. It is clear when changing the electron density and pressing voltage it becomes possible to change the direction of electron motion below all five electrodes.

We provide in table (\ref{table:m1},~\ref{table:m2}) the values of the maximum coherence and their corresponding frequencies for different pressing voltages and for both electron densities. All these values are calculated from the phase coherences presented in figure~\ref{fig:coh},~\ref{fig:coh2} for the low and high electron density, respectively. The values of the maximum of the coherence for the both electron densities change with pressing voltages for all pairs of electrodes. However, the values of the frequency at which the maximum coherence is observed at the low electron density increase with increasing pressing voltage, but they hardly increase when the when the electron density is high: see tables ~\ref{table:m1}, ~\ref{table:m2}.

\subsection{Summary}
 
We have presented the significant results and information that we obtained by applying nonlinear dynamics
methods. We have explored the characteristic features of the current oscillations induced in the five electrodes for different electron densities and pressing voltages, but with fixed magnetic field, depth of liquid helium and temperature. 

We have revealed oscillatory electron motion with varying frequency but with a constant modulation frequency that was found by extracting the instantaneous frequencies from ridges in time-frequency representations. We find that a high electron density significantly decreases the dependence of frequency on pressing voltage. The constant modulation frequency may result from the interaction between the electrons and gravity eaves on the surface of the liquid helium. %This results in a non-linearity in the dynamics which leads to the existence of high frequency harmonics.

The time-averaged wavelet power provides additional information about the distribution of electrons inside the cell, and how it changes under different conditions of electron density and pressing voltage. At low electron density, it seems that the electrons are mostly at the center for low pressing voltages of 4.16-4.19\,V, but with some at the edge electrode E3 sand E4. Increasing the pressing voltage displaces the electrons from the centre towards the edge. Whilst, with high electron density and low pressing voltage, the electrons are unaffected. With increasing pressing voltage, the electrons seem to crawl along the edge of the liquid surface.

There is a significant phase coherence between the oscillations of current in the different electrodes for both electron densities. We have shown the motion of electrons inside the cell between all the electrodes by using the information of the phase difference.

\FloatBarrier

%%%%%%%%%%%%%%%%%%%%%%%%%%%%%%%%%%%%%%%%%%%%%%%
%%%%%%%%%%%%%%%%%%%%%%%%%%%%%%%%%%%%%%%%%%%%%%
\bibliography{supplemental}